\documentclass[10pt]{article}
\usepackage{fullpage}
\usepackage{graphicx}
\usepackage{url}
\usepackage{listings}
\usepackage{algpseudocode}
\usepackage{amsmath}
\usepackage[super]{nth}
\usepackage{color}
\usepackage[english]{babel}
\usepackage[autostyle]{csquotes}
\usepackage{epstopdf}
\DeclareGraphicsExtensions{.eps}
\usepackage{lmodern}
\usepackage{tikz}
\usepackage{pgfplots}
\usepackage{subcaption}

\newcommand{\blue}[1] 	{\color{blue} #1 \color{black}}

\begin{document}
\title{Hardening Cassandra Against Byzantine Failures}


\author{
	Roy Friedman \ \ \ \ \ Roni Licher\\
	Computer Science Department\\
	Technion - Israel Institute of Technology\\
	32000 Haifa, Israel\\
	\texttt{\{roy,ronili\}@cs.technion.ac.il}
}

\newcommand{\ZooKeeper} {\textsc{ZooKeeper}}
\newcommand{\HDFS}      {\textsc{HDFS}}
\newcommand{\MongoDB}   {\textsc{MongoDB}}
\newcommand{\Cassandra} {\textsc{Cassandra}}
\newcommand{\Riak}  	{\textsc{Riak}}
\newcommand{\Dynamo}  	{\textsc{Dynamo}}
\newcommand{\e}[1]  	{\emph{#1}}

\maketitle

\begin{abstract}
%With the continuous increase in the amount of published information, a variety of data-stores have been developed.
%These aspire to support high-availability, scalability, low-latency operations, fault-tolerance and more.
%While part of these preserve traditional strong consistency semantics, some are willing to relax this requirement in order to achieve better performance.
%The data models have been divided as well into multiple categories, in favor of better compliance with the applications needs.

Cassandra is one of the most widely used distributed data stores these days.
Cassandra supports flexible consistency guarantees over a \emph{wide-column} data access model and provides almost linear scale-out performance.
This enables application developers to tailor the performance and availability of Cassandra to their exact application's needs and required semantics.
Yet, Cassandra is designed to withstand benign failures, and cannot cope with most forms of Byzantine attacks.
%It was developed by Facebook, combining Google's BigTable~data model with Amazon's Dynamo~distributed structure.
%Cassandra~is a highly popular distributed data-store, it is a top-level Apache project, used by more than 1,500 companies.

In this work, we present an analysis of Cassandra's vulnerabilities and propose protocols for hardening Cassandra against Byzantine failures.
We examine several alternative design choices and compare between them both qualitatively and empirically by using the Yahoo! Cloud Serving Benchmark (YCSB) performance benchmark.
We include incremental performance analysis for our algorithmic and cryptographic adjustments, supporting our design choices.
%We report on our findings and insights and draw some conclusions for future work.
\end{abstract}

%%%%%%%%%%%%%%%%%%%%%%%%% Section 1 : Introduction %%%%%%%%%%%%%%%%%%%%%%%
\section{Introduction}
\label{sec:intro}
Distributed data stores are commonly used in data centers and cloud hosted applications, as they provide fast, reliable, and scalable access to persistently stored data.
Such data stores enable developers to treat scalable persistent storage as a service.
While persistent storage is a fundamental aspect of almost any application, developing an effective one is notoriously difficult.
Hence, the existence of such data stores relieves developers from the burden of creating and maintaining one themselves.

Due to inherent tradeoffs between semantics and performance~\cite{BF96,Br00} as well as the desire to offer various flexible data management models, a plethora of products has been developed.
These differ in the data access model, which can range from traditional relational databases, to wide-columns~\cite{bigtable,cassandraPaper}, key-value stores~\cite{Riak,dynamo}, as well as graph databases and more.
Another axis by which such systems differ is the consistency guarantees, which can range from strong consistency~\cite{lamport1978time} to eventual consistency~\cite{petersen1997flexible} and a myriad of intermediate options.

In our work, we focus on Cassandra~\cite{cassandraPaper}.
Cassandra follows the wide-column model, and offers very flexible consistency guarantees.
Among open source data stores, it is probably the most widely used; according to the Cassandra Apache project page~\cite{cassandraapache}, more than 1,500 companies are currently using Cassandra, including, e.g., Apple, CERN, Comcast, eBay, Easou, GitHub, GoDaddy, Hulu, Instagram, Intuit, Microsoft, Netflix, Reddit, The Weather Channel and more.

Like many distributed data stores, Cassandra has very effective protection against benign failures, but was not designed to withstand Byzantine attacks, in which some nodes in the system may act arbitrarily, including in a malicious manner.
Overcoming Byzantine failures requires sophisticated protocols and more resources.
However, ever since the seminal PBFT work of Castro and Liskov~\cite{pBFTcastro}, the practicality of building Byzantine fault tolerant replicated state machines has been demonstrated by multiple academic projects, e.g.,~\cite{upright,thenext700BFT} to name a few.
Interestingly, storage systems offer weaker semantics than general replicated state machines, and therefore it may be possible to make them resilient to Byzantine failures using weaker timing and failure detection assumptions, as been proposed in~\cite{cachin2014separating,liskov2005byzantine,ByzantineQuorumSystems,malkhi1998secure}.
Yet, to the best of our knowledge, we are the first to offer an extension of Cassandra that can withstand Byzantine failures.

Specifically, we analyze Cassandra's structure and protocols to uncover their vulnerabilities to Byzantine behavior.
We then propose alterations to Cassandra's existing protocols that overcome these failures.
In particular, we examine several alternative solutions and compare between them qualitatively and quantitatively.
Let us emphasize that one of our main design principles is to maintain Cassandra's basic interaction model as much as possible, to increase the likelihood of adoption and in order to minimize the number of lines of code we need to change.
After all, our goal in this study is to harden the existing system, not to create a new one.

We have benchmarked both the original Cassandra and our hardened versions of Cassandra using the standard YCSB benchmark~\cite{cooper2010benchmarking}.
We were able to demonstrate that the best performing configuration of the hardened Cassandra was only twice worse than the original Cassandra in the settings we measured.
Interestingly, we discovered that a key factor to obtaining reasonable performance is in the type of cryptography used.
That is, using traditional RSA signatures dramatically lowers the performance.
In contrast, our noval combination of vectors of MACs with the more modern Elliptic Curve Digital Signature Algorithm (ECDSA)~\cite{johnson2001elliptic} can yield a significant performance boost.

The rest of this paper is organized as follows:
We survey related works in Section \ref{sec:related}.
The system model and assumptions are presented in Section~\ref{sec:model}.
A brief overview of \Cassandra{} is presented in Section~\ref{sec:cassandra}.
In Section~\ref{sec:hardened}, we identify Byzantine vulnerabilities in \Cassandra{} and suggest ways to overcome them.
Section~\ref{sec:perf} details the performance evaluation.
We conclude with a discussion in Section~\ref{sec:disscussion}. 

%%%%%%%%%%%%%%%%%%%%%%%%% Section 2 : Related Work %%%%%%%%%%%%%%%%%%%%%%%
\section{Related Work}
\label{sec:related}
\label{sec:related}

Castro \& Liskov~\cite{pBFTcastro} were the first to show a practical BFT protocol using replicated state machine.
Based on their work, Clement et al.~\cite{upright} introduced \textsc{UpRight}, a modular library to support BFT using replicated state machine.
They have shown results for integrating the library with \ZooKeeper~\cite{zookeeper} and \HDFS~\cite{hadoop}, two popular open-source systems.
%\ZooKeeper~is a distributed coordination service while \HDFS~is a distributed filesystem with a single point of failure.
\textsc{BFT-SMaRt}~\cite{bessani2014state} and \textsc{Prime}~\cite{amir2011prime} have improved these algorithms in order to produce better performance even when facing Byzantine behaviour.
\textsc{Abstract}~\cite{thenext700BFT} is the state of the art in BFT replicated state machine.
It adds the ability to abort a client request when faults occur.
Then it can dynamically switch to a different BFT protocol that produces better performance under the new system conditions.

Replicating existing transactions-oriented databases using a middleware solution have been studied both in the context of benign failure~\cite{correia2007gorda} and Byzantine failures~\cite{garcia2011efficient, luiz2014mitra}.

\e{Quorum systems}~\cite{herlihy1984replication} are common tools for ensuring consistency and availability of replicated data in spite of benign faults.
In these protocols, each read request must be processed by a quorum (set) of nodes that intersects with all quorums of nodes that were used for earlier writes~\cite{ABD90}.
%Quorums sizes are configured according to the number of benign faults the system was configured to mask.
Quorum systems are employed in many distributed storage systems such as \Cassandra~\cite{cassandraPaper}, \Dynamo~\cite{dynamo} and \Riak~\cite{Riak}.

Malkhi \& Reiter~\cite{ByzantineQuorumSystems,malkhi1998secure} were the first to discuss \e{Byzantine} quorum systems, i.e., using read and write quorums such that any two quorums intersects in at least one \e{correct} node.
Furthermore, the system remains available in spite of having up to \e{f} Byzantine nodes.

Aguilera \& Swaminathan~\cite{aguilera2009remote} explored BFT storage for slow client-server links. 
In their solution, clients communicate with the system through a proxy and rely on a synchronized clock. 
Their goal was to implement a \e{linearizable} \e{abortable} register that provides the \e{limited effect} property.
That is, partial writes due to benign client failures do not have any effect. 
To do so, they used unique timestamps and timestamp promotion when conflicts appear.
Their work did not show an actual implementation nor performance analysis. 
As our work preserves \Cassandra's~semantics, we are able to design faster operations requiring lighter cryptography measures even when conflicts occur.

Byzantine clients in quorum systems might try to perform \e{split-brain-writes}.
A split-brain-write is a write performed to different servers using the same timestamp but not the same values.
There are two main approaches for handling split-brain-writes in quorum systems.
In both of them, the idea is to get a commitment from a quorum to bind a timestamp and a value on every write.
In Malkhi \& Reiter's approach~\cite{ByzantineQuorumSystems}, on every write, the servers exchange inter-servers messages agreeing on the binding.
In Liskov \& Rodrigues's approach~\cite{liskov2005byzantine}, the servers transmit signed agreements to the client that are later presented to the servers as a proof for the quorum agreement.
%In our work, we followed previous works in the goal to make sure that correct clients observe values in a consistent way under a refined read schematics~\cite{ByzantineQuorumSystems,pBFTcastro,liskov2005byzantine}.
%However, we do not prevent split-brain-writes rather repair the object state on a read request (or in the background).
In our work, we do not prevent split-brain-writes, but rather repair the object state on a read request (or in the background).

Basescu et al.~\cite{buasescu2012robust} investigated how to build robust storage systems using multiple key-value stores generating a \e{cloud-of-clouds}, but focusing on benign failures.
%Their work did not focus on Byzantine failures.

Several BFT cloud storage systems provide \e{eventual consistency} semantics~\cite{petersen1997flexible}.
\textsc{Zeno}~\cite{singh2009zeno} requires at least $f+1$ correct servers and guarantees \e{causal order consistency}~\cite{lamport1978time} while \textsc{Depot}~\cite{mahajan2011depot} can tolerate any number of Byzantine clients and servers and guarantees \e{Fork-Join-Causal order consistency}.

Aniello et al.~\cite{CassandraGossip} showed how Byzantine nodes can launch DoS attacks in distributed systems that use a gossip based membership protocol.
In their paper, they have demonstrated their attack on \Cassandra~\cite{cassandraPaper} and presented a way to prevent it by using signatures on the gossiped data.
Other more general solutions for BFT gossip membership algorithms were shown in \textsc{Fireflies}~\cite{johansen2015fireflies} and \textsc{Brahms}~\cite{BFTgossip}.
The first uses digital signatures, full membership view and a pseudorandom mesh structure and the latter avoids digital signatures by sophisticated sampling methods.

%Considering data distributed across multiple decentralized nodes, a method to map each value and the host that is responsible to store it should be defined.
%One method to accomplish this is by using a Distributed Hash Table (DHT): each value is represented by a unique key using a hash function; each node in the system is responsible for storing keys of a certain range.
Sit \& Morris~\cite{securityP2PHash} mapped classic attacks in \emph{Distributed Hash Tables} (DHT) systems.
Some of the attacks can be disrupted by using SSL communication.
According to the documentation of recent versions of \Cassandra~\cite{DataStaxCassandra}, it supports inter-nodes and client-node SSL communication.
Other attacks described in \cite{securityP2PHash}, such as storage and retrieval attacks, are addressed in our work.

Okman et al.~\cite{SecurityNoSQL} showed security concerns in NoSQL systems, focusing on \Cassandra~\cite{cassandraPaper} and \MongoDB~\cite{MongoDB}.
Their work concentrated on implementation issues while our we focus on architectural concepts and algorithms that add BFT resilience.

%%%%%%%%%%%%%%%%%%%%%%%%% Section 3 : Model and assumptions %%%%%%%%%%%%%%%%%%%%%%%
\section{Model and Assumptions}
\label{sec:model}
We assume a Cassandra system consisting of \e{nodes} and \e{clients}.
Each of the entities may be \e{correct} or \e{faulty} according to the Byzantine failure model \cite{lamport1982byzantine}.
A correct entity makes progress only according to its specification while a faulty entity can act arbitrarily, including colluding with others.
%We assume that faulty entities can collude in order to achieve their goals.

In our proposed solutions, we assume that the maximal number of faulty nodes is bounded by \e{f}.
We initially assume that all clients are \e{correct}, but later relax this assumption.
%We continue by relaxing this assumption allowing clients to fail.
When handling Byzantine clients, we do not limit the number of faulty clients nor change the assumption on the maximal number of \e{f} faulty nodes.
Yet, we assume that clients can be authenticated so correct nodes only respond to clients that are allowed to access the system according to some verifiable \emph{access control list} (ACL).
Let us emphasize that we use the terms nodes and processes interchangeably and only to refer to Cassandra nodes.

We assume a partially synchronous distributed system that is fully connected.
Every node can directly deliver messages to every other node and every client can directly contact any system node.
We also assume that each message that is sent from one correct entity to another will eventually arrive exactly once and without errors.
That can be implemented, e.g., on top of fair lossy networks, using retransmission and error detection codes.
We do not assume any bound on messages delay or computation time in order to support our safety and liveness properties.
However, efficiency depends on the fact that most of the time messages and computation steps do terminate within bounded time~\cite{DLS84}.

Every system entity has a verifiable PKI certificate \cite{PKIrfc5280}.
We assume a trusted \e{system administrator}.
The system administrator can send signed membership configuration messages.

The system shares a loosely synchronized clock which enables detection of expired PKI certificates in a reasonable time but is not accurate enough to ensure coordinated actions.
We discuss this clock in Chapter~\ref{clock}.

%%%%%%%%%%%%%%%%%%%%%%%%% Section 4 : Cassandra overview %%%%%%%%%%%%%%%%%%%%%%%
\section{Brief Overview of Cassandra}
\label{sec:cassandra}
\Cassandra{} stores data in tables with varying number of columns.
Each node is responsible for storing a range of rows for each table.
Values are replicated on multiple nodes according to the configurable \e{replication factor}.

Specifically, mapping of data to nodes follows the \e{consistent hashing} principle~\cite{karger1997consistent}, where nodes are logically placed on a virtual ring by hashing their ids.
To be precise, on each node installation, multiple \e{virtual nodes}~\cite{dynamo} are created.
Each virtual node generates a randomized key on the ring, called a \e{token}, which we refer to as its place.
This virtual node takes responsibility for hashed keys that fall in the range from its place up to the next node on the ring, known as its \e{successor}.
Additionally, the node also stores keys in the ranges of the $N-1$ preceding nodes that require replication, where $N$ is the replication factor parameter.
The $N$ nodes responsible for storing a given value are called its \e{replication set}.

\Cassandra{} uses a \e{full membership view}, where every node knows about every other node.
A node that responds to communication is considered \e{responsive} and otherwise it is \e{suspected}.
In order to ensure that the nodes' views are consistent, nodes exchange their views via \e{gossip}~\cite{van1998gossip}.
The gossip is disseminated periodically and randomly; every second, each node tries to exchange views with up to three other nodes: one responsive, one suspected, and a \e{seed}~\cite{cassandraPaper}.
On node installation, seed nodes can be configured to be the first point of contact.
As these nodes are part of the system, they are constantly being updated about the membership changes and can provide an updated membership view.

\Cassandra{} provides tunable consistency per operation.
On every operation, the client can specify the \e{consistency level} that determines the number of replicas that have to acknowledge the operation.
Some of the supported consistency levels are: \e{one} replica, a \e{quorum}~\cite{herlihy1984replication} of replicas and \e{all} of the replicas.
According to the consistency level requested in the write and in the respectively read of a value, \e{eventual consistency}~\cite{petersen1997flexible} or \e{strong consistency} can be achieved.

\begin{figure}[t!]
	\centering
	\includegraphics[scale=0.53]{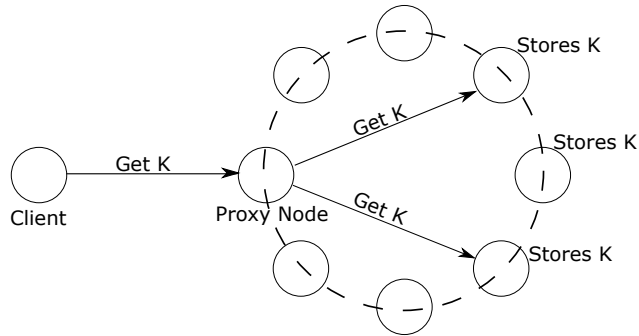}
	\caption[The read operation in \Cassandra.]{The read operation in \Cassandra. Replication factor is 3. A client connects to a system node (proxy) and requests a read quorum (in this case, majority, satisfied with 2 responses). The proxy contacts the relevant nodes using its view of the system.}
	\label{fig:ClientToRing}
\end{figure}

On each operation, a client connects to any node in the system in order to perform the operation.
This selected node acts as a \e{proxy} on behalf of the client and contacts the relevant nodes using its view of the system as illustrated in Figure~\ref{fig:ClientToRing}.
In the common configuration, the client selects a proxy from all of the system nodes in a \e{Round Robin} manner.
The proxy node may contact up to $N$ nodes that are responsible for storing the value according to the requested consistency level.
If the required threshold of responses is satisfied, the proxy will acknowledge the write or forward the latest value, according to the stored timestamp, to the client.
If the proxy fails to contact a node on a write, it stores the value locally and tries to update the suspected node at a later time.
The stored value is called \e{hinted handoff} \cite{DataStaxCassandra}.
If a proxy receives multiple versions on a read query, it performs a \e{read repair}, a method to update nodes that hold a stale version with the most updated one.

If a node is unresponsive for a long time, hinted handoffs that were saved for this node may be deleted.
Similarly, a hinted handoff may not get to its targeted node is if the node that stores it fails.
\Cassandra{} provides a manual \e{anti-entropy} tool for these cases.
This tool can sync a node's data by asking nodes that hold replicas for its range to compute and exchange \e{Merkle trees}~\cite{merkle1980protocols} for their values and sync the outdated values.

%\Cassandra~provides \e{eventual consistency} \cite{petersen1997flexible}. Although it lets a client request acknowledgments from a \e{quorum} \cite{herlihy1984replication} or even from all of the replicas, this is not enough to ensure \e{strong consistency}. Instead, the current implementation provides a \e{regular register} semantics \cite{lamport1986interprocess}, i.e., a write that is taking place in parallel with two read queries can result in the first read returning the new value while the second read returning the old value. This is due to the fact that in read operations, if inconsistency is detected, the proxy only schedules a read repair for a later time and does not block the read operation till it completes the repair \cite{attiya1995sharing}.

The primary language for communicating with \Cassandra~is the \e{Cassandra Query Language} (CQL)~\cite{DataStaxCassandra}.
CQL is similar to \e{SQL} with adjustments to the NoSQL concepts.
For example, \e{join} queries are not available.
In our work, we ignore the wide selection of options and focus on put and get commands as available in standard NoSQL key-value databases.

Previous versions of \Cassandra{} supported \e{sloppy quorums}~\cite{dynamo}, by which responsive nodes outside the replication set were used instead of failed ones.
This was deprecated in version 1.0 by switching the responsibility of storing the replica value to the proxy node.
In both cases, only nodes of the true replication set count for the consistency level requirement.

While \Cassandra{} can handle benign failures, it is unable to detect nor mask \e{Byzantine} failures.
In our work, we suggest solutions that improve the Byzantine robustness of the system.
We have analyzed the system mechanisms and extended them with the ability to mask up to \emph{f} (configurable) \e{Byzantine} nodes.

%One of the consistency levels supported by \Cassandra~is \e{quorum} \cite{herlihy1984replication}, implemented by requesting majority acknowledgments in each operation. In order to increase the availability

% In addition, nodes that are placed on the same physical server along with at least another node that was already used for replicating the same value should be skipped.

% On a node failure, the other nodes detect it and the succeeding available node is taking its range responsibility. If the node recovers, it queries the system nodes and gets updated with newer values. If replication is available, it will query multiple nodes in order to load balance this heavy operation.

%\todo{Add about the snitch, there are about 4 types}

%%%%%%%%%%%%%%%%%%%%%%%%% Section 5 : Solution %%%%%%%%%%%%%%%%%%%%%%%
\section{Hardened Cassandra}
\label{sec:hardened}
\label{sec:hardened}

% Overview of our solution and each part of its mechanisms
In this section, we identify Byzantine vulnerabilities in \Cassandra{} and suggest ways to overcome them.

\subsection{Impersonating}
\Cassandra{} supports the use of SSL and enables each message to be authenticated by each party.
In some cases, messages are required to be authenticated by a third party, e.g., a read response sent from a node to a client using a proxy node.
In order to support such authentication, we use digital signatures.
When using SSL or digital signatures, we depend on PKI.

Digital signatures are divided into two main categories according to the type of keys they use: \e{public/private keys} vs \e{MAC tags}.
Public key signatures are more powerful than MAC tags as they enable anyone to verify messages without being able to sign them.
MAC tags are mostly useful when there are exactly two entities that have to prove to each other that they have generated the messages.
In the last case, the receiver should also be able to identify that received messages were not actually generated by itself.
The trade-off for using public key signatures is the compute time, which is about two to three orders of magnitude slower than MAC tags and these signatures are significantly larger, e.g., RSA 2048b versus AES-CBC MAC 128b.

\subsection{Consistency Level}
Recall that in \Cassandra{} the user can configure the replication factor $N$ (the number of nodes that have to store a value) and in addition on each read and each write to require how many nodes ($R$ and $W$, respectively) must acknowledge it.
This required threshold can be one node or a quorum (in \Cassandra, always configured as majority) or all $N$ nodes.
When up to $f$ nodes may be Byzantine, querying fewer than $f+1$ nodes may retrieve old data (signed data cannot be forged), violating the consistency property.
On the other hand, querying more than $N - f$ nodes may result in loss of availability.
In our work, we present two approaches: (1) using Byzantine quorums for obtaining Strong Consistency and (2) using \Cassandra{} quorums with a scheduled run of the anti-entropy tool for obtaining \e{Byzantine Eventual Consistency}.

%  (1) using Byzantine quorums for obtaining near Strong Consistency (we preserve a regular register behaver)

\subsubsection{Byzantine Quorums}
By requesting that each read and each write will intersect in at least $f+1$ nodes, we ensure that every read will intersect with every write in at least one \e{correct} node.
That is, $R+W\geq N+f+1$.
As for liveness, to be able to ensure that Byzantine nodes will not be able to block a write or a read, we must require that $R\leq N-f,W\leq N-f$.
By combining the above 3 requirements, we obtain: $N\geq 3f+1$.

The last bound was formally proved by Malkhi \& Reiter~\cite{ByzantineQuorumSystems}.
Cachin et al.~\cite{cachin2014separating} have lowered this bound to $2f+1$ by separating between the actual data and its \e{metadata}; storing the medadata still requires $3f+1$ nodes.
The above separation was presented under the assumptions of benign writers and Byzantine readers.

The last solution is beneficial for storing large data as it uses less storage space and network load.
However, when storing small values, the method of~\cite{cachin2014separating} only increases the overhead.
A system may offer either solution according to the system usage, or use them both in a hybrid way, according to each value's size.

\subsubsection{Byzantine Eventual Consistency}
As mentioned earlier, eventual consistency offers faster operations and higher availability in exchange for weakened semantics.
To satisfy eventual consistency, all replication set nodes must eventually receive every update.
Further, every writes order conflict should be resolved deterministically.
In this model, there is no bound on the propagation time of a write, but it should be finite.
In particular, if no additional writes are made to a row, eventually all reads to that row will return the same value.

Byzantine eventual consistency can be obtained through majority quorums.
In this approach, the replication set is of size $2f+1$ nodes while write and read quorums are of size of $f+1$.
Hence, each write operation acknowledged by $f+1$ nodes is necessarily executed by at least one correct node.
This node is trusted to update the rest of the nodes in the background.
As this node is correct, it will eventually use the anti-entropy tool to update the rest of the replication set.
Recall that the client request is signed so the servers will be able to authenticate this write when requested.

Every read is sent to $f+1$ nodes and thus reaches at least one correct node.
This correct node follows the protocol and accepts writes from proxy nodes and from the anti-entropy tool.
So, eventually, it retrieves the latest update.
Due to the cryptographic assumptions, a Byzantine node can only send old data and cannot forge messages.
Hence, on receiving a value from the anti-entropy tool that does not pass the signature validation, we can use it as a Byzantine failure detector and notify the system administrator about a Byzantine behavior.

\subsection{Proxy Node} \label{sec:proxy}
\begin{figure}[t!]
	\centering
	\includegraphics[width=0.90\columnwidth]{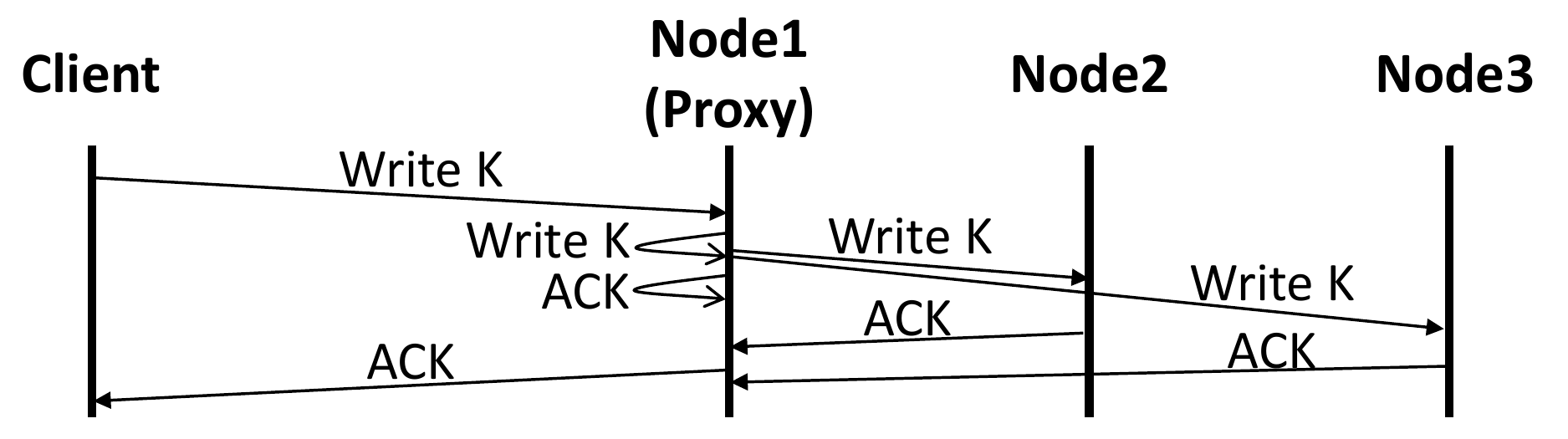}
	%\caption{Diagram for the algorithem from Figure~\ref{fig:clean-cassandra-write-flow}, this is the current write algorihem. Configuration:  N=4 and W=3.}
	\caption[The write algorithm in original \Cassandra.]{The write algorithm in original \Cassandra. Configuration: N=3 and W=2.}
	\label{fig:WriteBase}
\end{figure}

\begin{figure}[t!]
	\centering
	\includegraphics[width=0.90\columnwidth]{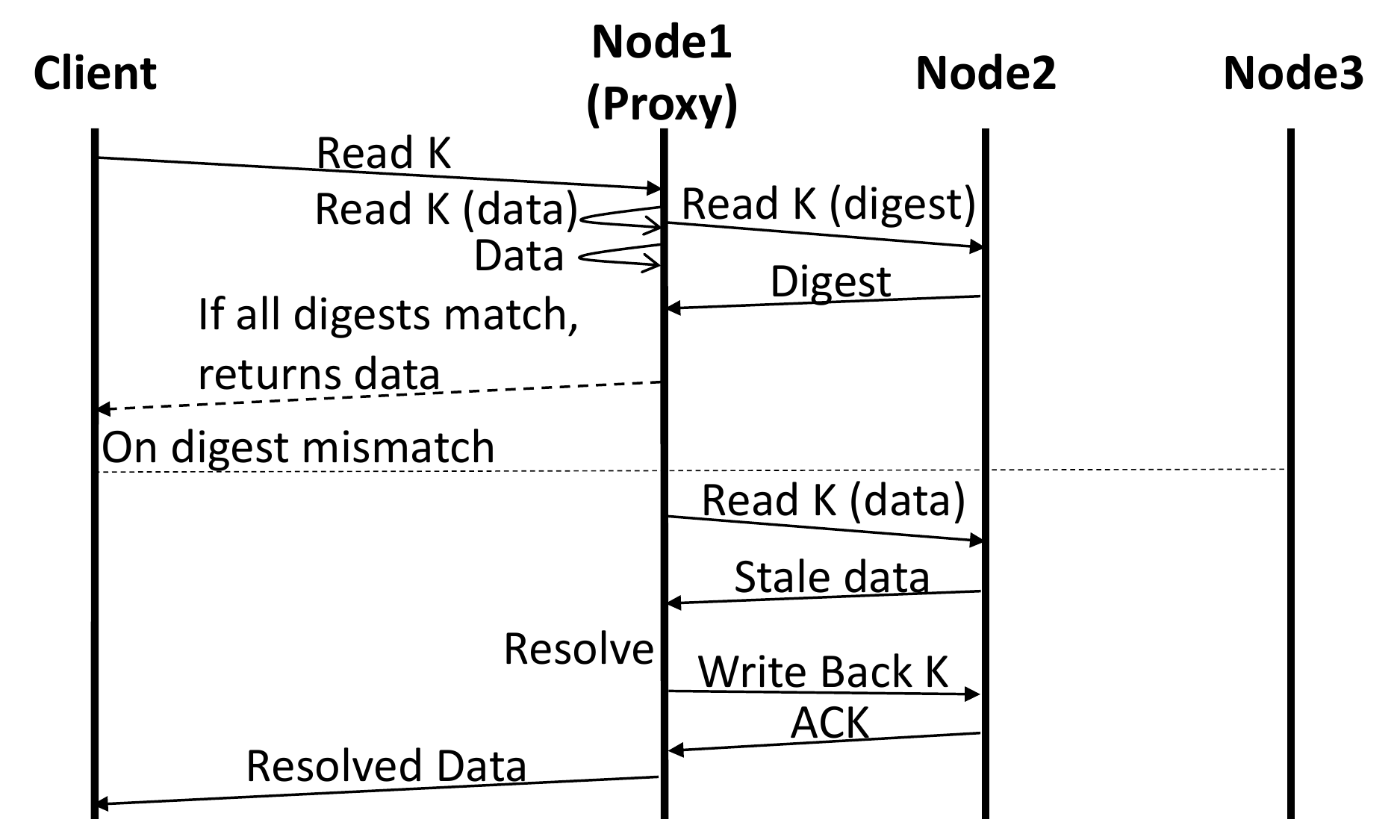}
	\caption[The read algorithm in original \Cassandra.]{The read algorithm in original \Cassandra. Configuration: N=3 and R=2.}
	\label{fig:ReadBase}
\end{figure}

Figures~\ref{fig:WriteBase} and~\ref{fig:ReadBase} present the current write and read flows in \Cassandra, including the role of proxies.
A Byzantine proxy node can act in multiple ways, such as (1) respond that it has successfully stored the value without doing so, (2) perform a split-brain-write, and (3) respond that the nodes are not available while they are.
We augment the existing flows of writing and reading in \Cassandra{} to overcome these vulnerabilities below.

\subsubsection{Write Operation in Details}
\begin{figure}[t!]
	\centering
	\includegraphics[width=\columnwidth]{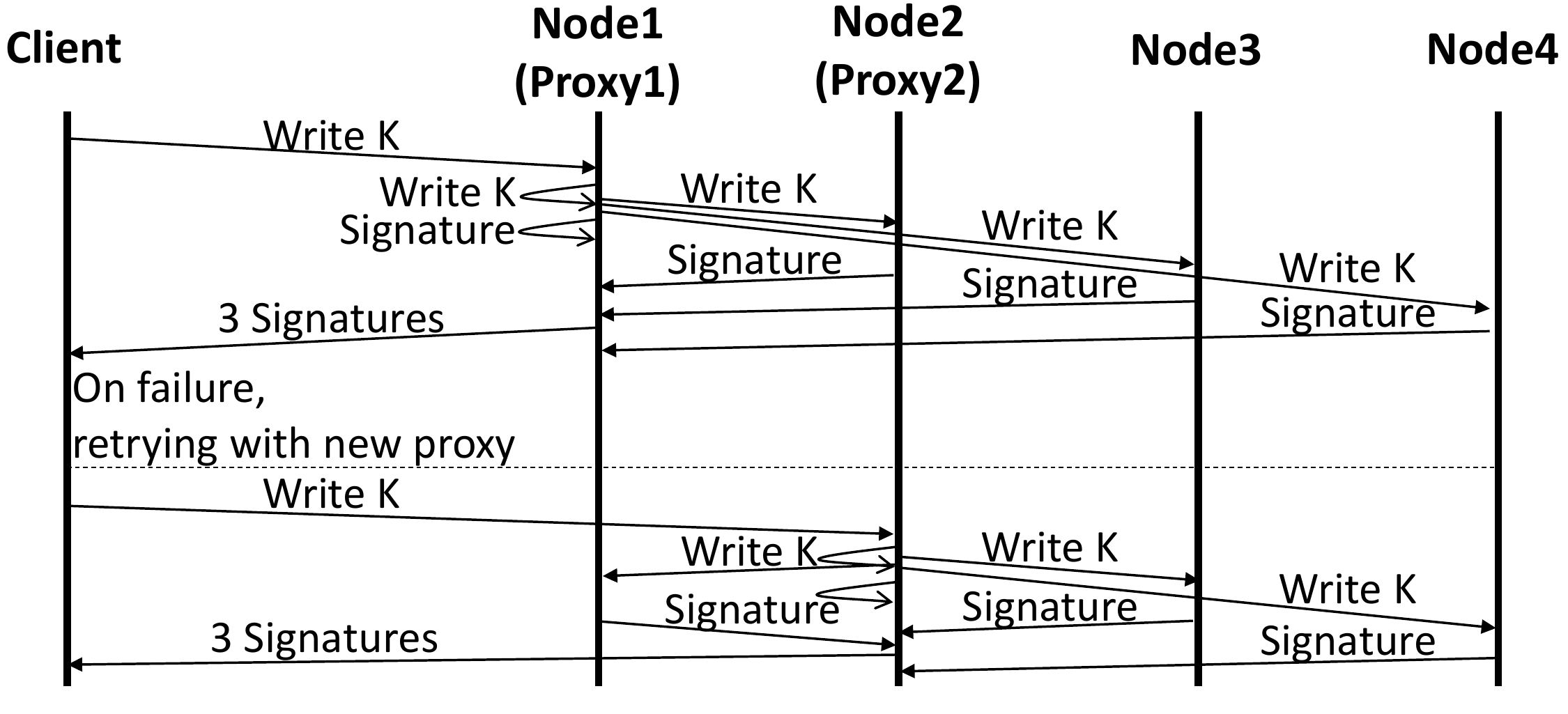}
	\caption[Illustrating our write algorithm from Figure~\ref{fig:write-flow} where the proxy verifies each store acknowledgment.]{Illustrating our write algorithm from Figure~\ref{fig:write-flow} where the proxy verifies each store acknowledgment. Configuration: N=4 and W=3.}
	\label{fig:write1}
\end{figure}

We present our modified write algorithm in Figures~\ref{fig:write1} and~\ref{fig:write-flow}.
In this solution, when storing a new value, the client signs the value and a node will store it only if it is signed by a known client according to the ACL and with a timestamp that is considered fresh (configurable).
On each store, the storing node signs an acknowledgment so that the client can verify it.
In addition, the signed acknowledgment covers the timestamp provided by the client, preventing replay attacks by the proxy.
A client completes a write only after obtaining the required threshold of signed responses, which now the proxy cannot forge.
If one proxy fails to respond with enough signed acknowledgments in a configurable reasonable time, the client contacts another node and asks it to serve as an alternative proxy for the operation.
After contacting at most $f+1$ proxy nodes when needed, the client knows for sure that at least one \e{correct} proxy node was contacted.

\begin{figure*}[t!]
	\captionsetup{font=scriptsize}
	\begin{algorithmic}[1] 		
		\scriptsize
		\Function{OnNodeToNodeWriteRequest}{$key, value, ts, clientSignature, clientID$}
			\If {clientSignature is valid}
				\State $nodeSignature \gets ComputeSignature(clientSignature) $ \Comment{The client signature covers a fresh ts}
				\State Store locally $<key, value, ts, clientSignature, clientID>$
				\State \Return $nodeSignature$ \Comment{A verifiable acknowledgment}
			\EndIf
		\EndFunction
		\\
		\Function{OnClientToNodeWriteRequest}{$key, value, ts, clientSignature, clientID$}
		\For {each node $n$ that is responsible for the $key$} \Comment{N nodes}
		\State Send write request with \e{$<$key, value, ts, clientSignature, clientID$>$} to $n$
		\EndFor
		\State Wait for $2f+1$ \textbf{verified} acknowledgements OR tmeout 
		\State \Comment{Verified in the manner of correct node signature}
		\State \Return responses
		\EndFunction
		\\
		\Function{ClientWriteRequest}{$key,value$}
		\State $ ts \gets $ Current timestamp \Comment{From a secure synchronized clock}
		\State $ clientSignature \gets $ ComputeSignature($key$ $||$ $value$ $||$ $ts$)
		\State $ p \gets $ Some random system node
		\State Send write request with \e{$<$key, value, ts, clientSignature, clientID$>$} to $p$ \label{lst:line:sendToPP}
		\State Wait for acknowledgments OR timeout 
		\If {$|valid Responses| \geq 2f+1$}
		\State \Return Success
		\EndIf
		
		\State $ p \gets $ Some random system node that was not used in this function invocation
		\If {$p = \bot$ OR $contacted Nodes > f$ }
		\State \Return Failure
		\EndIf
		\State goto line~\ref{lst:line:sendToPP} \Comment{Use another node as proxy}

		\EndFunction
\end{algorithmic}
\normalsize
	\caption[Our hardened write algorithm.]{Our hardened write algorithm. ClientWriteRequest is invoked by the client for each write. OnClientToNodeWriteRequest is invoked on the proxy node by the client. OnNodeToNodeWriteRequest is invoked on a node that has the responsibility to store the value. \emph{Store locally} appends the write to an append log without any read. When \emph{key} is queried, the latest store (according to timestamp) is retrieved. }
		% $W$ is the required quorum size for the write opertaion based on the consistency level
	\label{fig:write-flow}
\end{figure*}

\subsubsection{Read Operation in Details}
The read algorithm of a proxy has three parts:
(1) Reading data from one node and only a digest from the rest of the nodes.
In some cases, as an optimization, the read will target only a known live quorum instead of to all relevant nodes.
(2) On digests mismatch, a full read is initiated to all contacted nodes from the first phase, retrieving the data instead of a digest.
(3) The proxy resolves the conflict by creating a row with the most updated columns according to their timestamps, using the lexicographical order of the values as tie breakers when needed.
The resolved row is written back to out-dated nodes.

\begin{figure}[t!]
	\centering
	\includegraphics[width=\columnwidth]{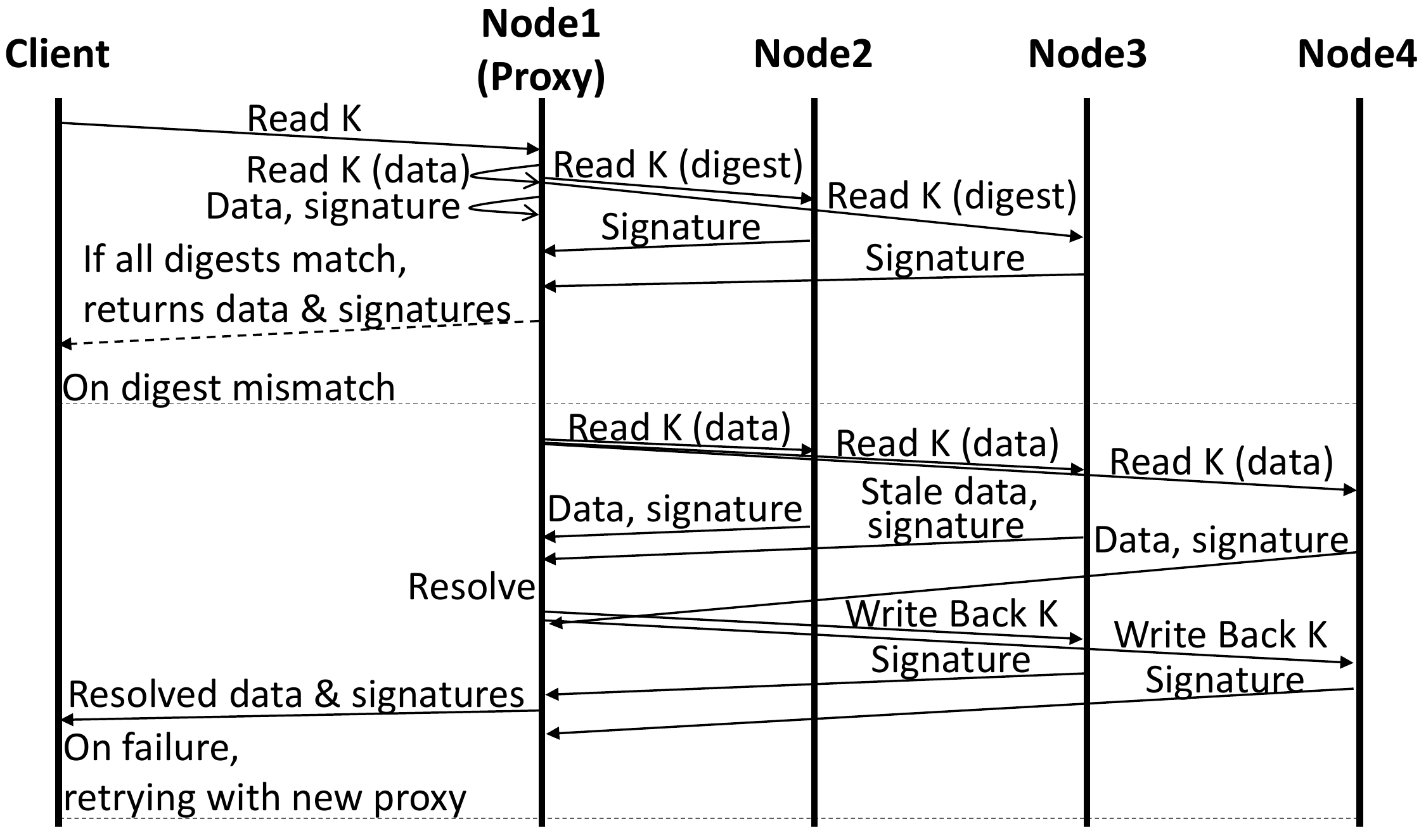}
	\caption[Illustrating our read algorithm from Figure~\ref{fig:read-flow} where the proxy verifies each answer.]{Illustrating our read algorithm from Figure~\ref{fig:read-flow} where the proxy verifies each answer. Configuration: N=4 and R=3.}
	\label{fig:read1}
\end{figure}

Figures~\ref{fig:read1} and~\ref{fig:read-flow} present our modified read algorithm, which consists of the following changes:
(1) In case the first phase is optimized by addressing only a known live quorum of nodes, if a failure occurs, we do not fail the operation but move to a full read from all nodes.
Thus, if a Byzantine node does not respond correctly, we do not fail the operation.
(2) If there is a digest mismatch in the first phase, we do not limit the full read only to the contacted nodes from the first phase but rather address all replication set nodes.
Hence, Byzantine nodes cannot answer in the first phase and fail the operation by being silent in the second phase.
(3) During resolving, the nodes issue a special signature, notifying the client about the write back.
The proxy then supplies the client with the original answers from the first phase, all are signed by the nodes.
This way, the client is able to authenticate that the resolving was executed correctly.

\begin{figure*}[t!]
	\captionsetup{font=scriptsize}
	\begin{algorithmic}[1]
		\scriptsize
			\Function{OnNodeToNodeReadRequest}{$key, client-ts$}
			\If {$key$ is sored in the node}
			\State $<value, ts, clientSignature, clientID> \gets $ The newest associated timestamp and value with $key$
			\Else
			\State $clientSignature \gets EMPTY$
			\EndIf
			
			\State $nodeSignature \gets ComputeSignature(key || hash(value) || clientSignature || client-ts) $
			
			\If {isDigestQuery}
			\State \Return $<hash(value), ts, clientSignature, clientID, nodeSignature>$ 
			\State \Comment{The hash is matched in the proxy}
			\Else
			\State \Return $<value, ts, clientSignature, clientID, nodeSignature>$
			\EndIf
			\EndFunction
			\\
			\Function{OnClientToNodeReadRequest}{$key, client-ts$}
			
			\State {$target Endpoints \gets all Relevant Nodes$ for $key$ OR a subset of $2f+1$ fastest relevant nodes} 
			\State \Comment{Optimization}
			\State {$dataEndpoind \gets$ One node from $target Endpoints$}
			\State Send read request for data to $ dataEndpoind $
			\State Send read request for digest to $target Endpoints \setminus \{dataEndpoind\} $
			\State Wait for $2f+1$ \textbf{verified} responses OR timeout
			\If {timeout AND all relevant nodes were targeted at the first phase}		
				\State \Return $\bot$
			\EndIf
			\If {got response from $dataEndpoind$ AND all responses agree on the digest}
				\State \Return $<value, nodesSignatures>$
			\EndIf
			\State Send read request for data from all nodes in $ all Relevant Nodes $ \Comment {N nodes} \label{lst:line:readFull}
			\State Wait for  $2f+1$ \textbf{verified} responses OR timeout
			\If {timeout}
			\State \Return $\bot$
			\EndIf
			
			\State $resolvedValue \gets$ Latest response from $responses$ that is client-signature \textbf{verified}.
			\State Send write-back with $resolvedValue$ to $ all Relevant Nodes $ except those that are known to be updated
			\State Wait for responses till we have knowledge about $2f+1$ \textbf{verified} updated nodes OR timeout
			\State \Comment{Responded before with updated data or for the write back}
			\If {timeout}
			\State \Return $\bot$
			\EndIf
			
			\State \Return $<resolvedValue, nodesSignatures, originalValuesUsedForTheResolve>$
			\EndFunction
			\\
			
			\Function{ClientReadRequest}{$key$}
			\State $ client-ts \gets $ Current timestamp \Comment{Fresh timestamp}
			\State $ p \gets $ Some random system node
			\State Send read request with $<key, client-ts>$ to $p$ \label{lst:line:startWithNode}
			\State Wait for responses OR timeout
			\If {$|valid NodesSignatues| \geq 2f+1$}	
			\State \Comment {If a write-back is observed, the resolved row is verified with the original read answers}
			\State \Return data
			\EndIf
			
			\State $ p \gets $ Some random system node that was not used in this function invocation
			\If {$p = \bot$ OR $contacted Nodes > f$ }
			\State \Return Failure
			\EndIf
			\State goto line~\ref{lst:line:startWithNode}
			
			\EndFunction
    \end{algorithmic}
\normalsize
	\caption[Our hardened read algorithm.]{Our hardened read algorithm. ClientReadRequest is invoked by the client for each read. OnClientToNodeReadRequest is invoked on the proxy by the client. OnNodeToNodeReadRequest is invoked on a node that has the responsibility to store the value. $R$ is the quorum size for the read operation based on the consistency level. The read can be also sent to only $R$ nodes (a subset of the $N$ nodes) and only if some of them do not respond in the timeout interval, \e{p} will send the request to the remaining $N-R$ nodes.}
	\label{fig:read-flow}
\end{figure*}

Without supplying the set of original answers in the last case, a Byzantine proxy that has an old value could fool the client into accepting this old value.
This exploit is originated in the fast write optimization of \Cassandra~where new writes are appended to a commit log and reconciled in the background or during a following read request.
In our write solution, we follow this architecture and only verify the signature, letting old values be stored but preventing them from reaching clients.
We would like to emphasize that if there is already a newer value stored for that key, the stale value would not be served by any read.
Otherwise, the proxy could exploit this by requesting a quorum of nodes to store an old value, obtaining correct signatures that can be presented to the client.
When providing the client with the original answers, it can verify that the write back was necessary.

\subsubsection{Targeting Irrelevant Nodes} \label{irrelevant}

Another possible attack by a Byzantine proxy is directing read requests to irrelevant nodes.
These nodes will return a verifiable empty answer.

To overcome this, we have considered three options:
(1) Using clients that have full membership view, which is supported by \Cassandra.
This way, a client knows which nodes have to respond.
(2) Using an authentication service that is familiar with the membership view and can validate each response.
We do not elaborate on how this service should be implemented.
A client can use this service to authenticate answers. %each answer or only a sample.
(3) Configure the nodes to return a signed \enquote{irrelevant} message when requested a value that they are not responsible for.
A client then only counts as valid answers correctly signed messages that are not marked as \enquote{irrelevant}.

Using any of these solutions, a Byzantine proxy can always try to update the minimum number of nodes required for a successful write operation.
This performance attack can decrease the availability.
To overcome this attack and make sure that every value eventually gets updated to every correct node, we use the anti-entropy tool periodically.
As this action is costly, nodes should not run it too often.
The anti-entropy tool can be run correctly in a Byzantine environment as long as each value that is detected as new is delivered along with a correct client signature that can be authenticated.

\subsubsection{Proxy Acknowledgments Verification} \label{Proxy-Acknowledgments}

Our proposed hardened algorithms for read and write rely on digitally signed acknowledgments for later authenticating the actual completion of the operation by the nodes.
These acknowledgments provide a verifiable proof to the client that the nodes indeed treated the operation.
In our proposed solution as presented so far, we have requested the proxy to verify the nodes acknowledgments and accept a node response only if it is signed correctly.
In this section, we discuss the motivation for verifying the signatures in the proxy node and suggest an alternative of only verifying them at the client.
Specifically, when attempting to eliminate verification at the proxy, we have identified the following two problematic scenarios:
\begin{enumerate}
\item Consider a correct proxy and $f$ Byzantine nodes.
The Byzantine nodes manage to answer an operation faster (they have the advantage as they do not have to verify signatures nor sign) with bad signatures.
The proxy then returns to the client $f+1$ good signatures and $f$ bad signatures.
In this case, contacting an alternative proxy might produce the same behavior.

\item Consider a Byzantine proxy, which is also responsible to store data itself and it is colluding with $f-1$ Byzantine nodes.
On a write operation, the proxy asks the Byzantine nodes to produce a correct signature without storing the value.
The proxy also asks one correct node to store the data and in addition produces false, non-verifiable, $f$ signatures for some nodes.
The client will get $f+1$ correct signatures and $f$ bad signatures, while only one node really stored the value.
\end{enumerate}

%These behaviors are indistinguishable at the client side.
%When the proxy does not verify the acknowledgments, $f$ acknowledgments might be incorrect and a Byzantine proxy can exploit this assumption by supplying the client with $f$ misleading verifiable responses and only one correct response.

%We now present an alternative, where the proxy does not verify the signatures, but the client can still overcome a malicious proxy.
To enable the client to overcome Byzantine behavior without proxy acknowledgement verification, we let the client contact the proxy again in case it is not satisfied with the $2f+1$ responses it obtained.
On a write, the client requests the proxy to fetch more acknowledgments from new nodes.
On a read, the client requests the proxy to read again without contacting the nodes that supplied false signatures.

We would like to emphasize that if a client receives a bad signature, both the proxy and the node might be Byzantine.
In this case, we do not have evidence for the real Byzantine entity as one can try to frame the other.

The motivation for this alternative is that signatures verification is a heavy operation.
In the proxy verification option, on every write, the proxy is required to perform at least $2f+1$ signature verifications.
In the alternative client only verification option, the latency penalty will be noticed only when Byzantine failures are present and could be roughly bounded by the time of additional \e{RTT} (round-trip-time) to the system and $f$ parallel RTT's inside the system (counted as one), multiplying it all by $f$ (the number of retries with alternative proxies).
Assuming that in most systems Byzantine failures are rare, speeding the common correct case is a reasonable choice.

Another significant motivation for using the client only verification option is that it enables using MAC tags instead of public signatures, since only the client verifies signatures.
To that end, a symmetric key for each pair of system node and client should be generated.
Every client has to store a number of symmetric keys that is equal to the number of system nodes.
Every node has to store a number of symmetric keys that is equal to the number of (recently active) clients.
These keys can be pre-configured by the system administrator or be obtained on the first interaction through a secure SSL line.
%On each operation that is invoked on a node, the node can issue a dedicated MAC tag for the requesting client.
This produces significant speedups both for the node signing and for the client verification.

The exact algorithms appear in Appendix \ref{appendix:detailed_algos}.
Figures~\ref{fig:write-flow-v2} and~\ref{fig:read-flow-v2} describe the algorithms and Figures~\ref{fig:write2} and~\ref{fig:read2} illustrate their execution timelines.

\subsubsection{Proxy Resolving vs. Client Resolving} \label{Client-Resolving}
Recall that when \Cassandra's read operation detects an inconsistent state, a resolving process is initiated to update outdated replicas.
This way, the chance for inconsistency in future reads decreases.
In plain \Cassandra{} as well as in our solution as presented so far, the proxy is in charge of resolving such inconsistent states.
An alternative option is to let the client resolve the answers and write back the resolved value using a write request that specifies to the proxy which replicas are already updated.

As we wish to prevent Byzantine nodes from manipulating the resolver with false values, the resolver requires the ability to verify the client signature on each version.
When using the client resolving option in combination with using a proxy that is not verifying nodes acknowledgments (as discusses in Section~\ref{Proxy-Acknowledgments}), the proxy is released from all obligations of verifying client signatures, improving its scalability.

The exact details of the this algorithm appear in Appendix \ref{appendix:detailed_algos}. Figure \ref{fig:read-flow-v3} describes the algorithm and Figure~\ref{fig:read3} illustrates its execution timeline

\subsubsection{Switching From Public Key Signatures to MAC Tags}
The use of public key signatures has a major performance impact while switching to MAC tag is not trivial.
In Section~\ref{Proxy-Acknowledgments}, we have described a way to switch from public key signatures to MAC tags on messages sent from nodes to clients.

Supporting MAC tags on messages sent from clients to nodes present interesting challenges for certain \Cassandra{} features.
Such features are:
(1) Joining new nodes to \Cassandra{}.
These nodes have to fetch stored values for load-balancing.
As the client does not know who these future nodes are, it cannot prepare MAC tags for them.
A solution for this could be that a new node will only store values that were proposed by at least $f+1$ nodes.
Alternatively, have the client re-store all of relevant values that the new node has to store.
(2) Using the anti-entropy tool (exchanging Merkle trees and updating stale values) and resolving consistency conflicts need to ensure the correctness of the values by contacting at least $f+1$ nodes that agree on the values.
Alternatively, every node will have to store a vector of MAC tags for each responsible node.
Storing a signature vector poses another challenge: a Byzantine proxy can manipulate the signatures vector sent to each node, leaving only the node's signature correct and corrupting all other nodes' signatures (turning the stored vector useless).
This challenge can be overcome by adding another MAC tag on the signatures vector, proving to the node that the tags vector was not modified by the proxy.

\begin{figure}[t!]
	\centering
	\includegraphics[width=\columnwidth]{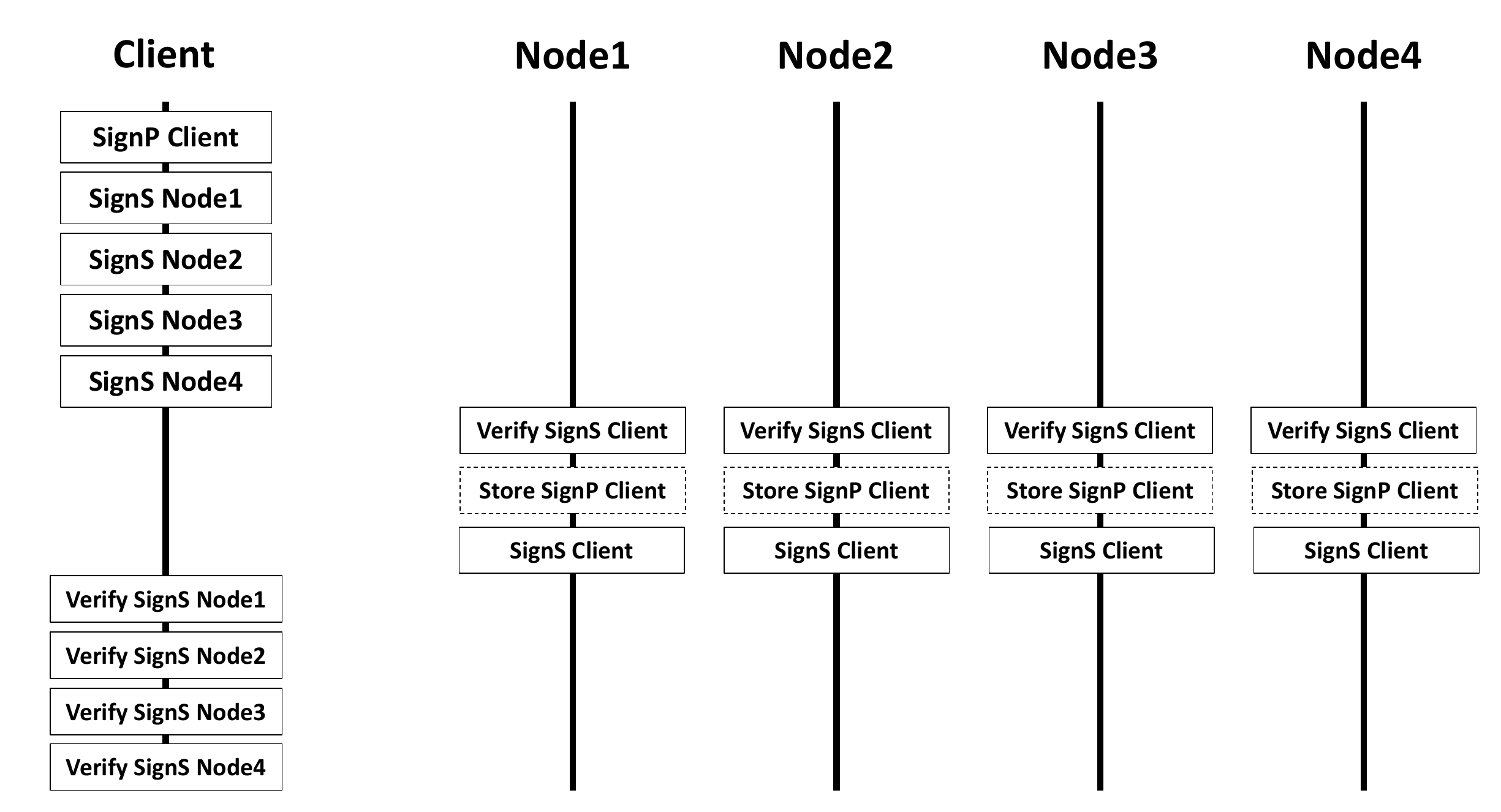}
	\caption[Illustration of our hybrid signing solution.]
	   {Illustration of our hybrid signing solution.
		The \emph{SignP} stands for public key signature, using the private key of the signing entity.
		The \emph{SignS} stands for MAC tag, using the shared key with the destination entity.}
	\label{fig:write-sign}
\end{figure}

Due to these limitations and in order to speed up the write path, we suggest a hybrid solution as presented in Figure~\ref{fig:write-sign}.
A write is signed with a public key signature and that signature is covered by MAC tags, one for each node.
A node then verifies only the MAC tag and stores only the public key signature.
Hence, in the common case, we will use a public key signature only once and will not use public key verifications at all.
When things go bad, we fall back to the public key signature.
Furthermore, some public key signature algorithms have better performance when signing, sacrificing their verification time.
For example, the \emph{Elliptic Curve Digital Signature Algorithm} (ECDSA)~\cite{johnson2001elliptic} in comparison with \e{RSA}~\cite{rivest1978method}.
In this case, ECDSA can greatly boost performance.

Finally, when using MAC tags on the client to node path, there is a need for the client to know what are the relevant nodes for that key.
One solution is to ensure clients are updated about the nodes tokens.
This way, on every write, the client knows what keys to use.
Since our solution has a fall back option, even if there was a topology change that the client was late to observe, the new node (targeted by the proxy) can still use the public signature and not fail the write.
On the write acknowledgment, the new node can attach the topological change evidence and update the client.

\subsubsection{Column Families vs. Key-Value semantics}
For clarity of presentation, the algorithms described so far reflect only key-value semantics.
Yet, our work also supports \Cassandra's column family semantics.
In the latter, a client has to sign each column separately, producing a number of signatures that is equivalent to the number of non-key columns.
This is needed in order to be able to reconcile partial columns writes correctly according to \Cassandra's semantics.
For example, consider a scheme with two non-key columns A and B.
One node can hold an updated version of A and a stale version of B while another node might hold the opposite state.
A correct read should return one row containing the latest columns versions for both A and B.

Nodes acknowledgments can still include only a single signature covering all columns.
This is because the purpose of signatures here is to acknowledge the operation.

\subsubsection{Comparing The Variants}
In Tables \ref{tbl:comparing_table_opt}, \ref{tbl:comparing_table_res} and \ref{tbl:comparing_table_worst}, we summarize the different algorithms proposed in this section.
We focus on the number of signing and verification operations of the digital signatures as these are the most time consuming.
We divide our analysis into three cases: (1) best case and no failures, (2) a benign mismatch on the read flow that requires resolving, and (3) worst case with $f$ Byzantine nodes.

\newcommand{\cell}[1]  	{\shortstack{#1}}

\begin{table*}
	%\footnotesize
	\tiny
	\captionsetup{font=footnotesize}
	\centering
	
	\begin{tabular}{|l|c|c|c|c|} \hline
		\textbf{Proxy Verifies?}		&\textbf{Op} &\textbf{MAC Tags} & \textbf{Signatures}  & \textbf{Verifications} \\ \hline
		Yes          & Write & None 			  & \cell{Client: $C$(p) \\ Nodes: $3f+1$(p) }	  & \cell{Nodes: $(3f+1)\cdot C$(p)\\Proxy: $2f+1$(p)\\Client: $2f+1$(p)} \\ \hline
		No 	& Write & None 			  & \cell{Client: $C$(p) \\ Nodes: $3f+1$(p) }	  & \cell{Nodes: $(3f+1)\cdot C$(p)\\Client: $2f+1$(p)} 				  \\ \hline
   		No   & Write & Nodes to client & \cell{Client: $C$(p) \\ Nodes: $3f+1$(s) }	  & \cell{Nodes: $(3f+1)\cdot C$(p)\\Client: $2f+1$(s)	} 				  \\ \hline
   		No   & Write & Both ways		  & \cell{Client: $C$(p) \& $3f+1$(s)	\\ Nodes: $3f+1$(s) }	  & \cell{Nodes: $3f+1$(s)\\Client: $2f+1$(s)	} 		  \\ \hline
   		
  		Yes          & Read & None 			 & \cell{Nodes: $2f+1$(p) }	 & \cell{Proxy: $2f+1$(p)\\Client: $2f+1$(p)} \\ \hline
  		No 	& Read & None			 & \cell{Nodes: $2f+1$(p) }	 & \cell{Client: $2f+1$(p)} 				  \\ \hline
  		No   & Read & Nodes to client & \cell{Nodes: $2f+1$(s) }	 & \cell{Client: $2f+1$(s)	} 				  \\ \hline  	
	  	  	
	\end{tabular}
	\caption[Comparing the variants of our solution in the read and write flows with the most optimist assumptions.]{Comparing the variants of our solution in the read and write flows with the most optimist assumptions. $C$ is the number of columns, (p) indicated public key signatures and (s) MAC tags. In the variants where the proxy does not verify, we refer both for the proxy resolves and client resolves modes. We assume that on a read, the proxy uses the optimization in the first phase and contacts only a Byzantine quorum and not all replicas. For example, the forth row presents a proxy that does not verify acknowledgments and MAC tags are used from client to nodes and from nodes to client. In this variant, the client signs the $C$ columns using public key signatures and adds $3f+1$ MAC tag, one for each node. All nodes ($3f+1$) have to store it and they verify only their MAC tags. All nodes issue verifiable acknowledgments ($3f+1$) and the client verifies only a Byzantine quorum ($2f+1$).}
	\label{tbl:comparing_table_opt}
\end{table*}

\begin{table*}[t!]
	%\footnotesize
	\tiny
	\captionsetup{font=footnotesize}
	\centering
	\begin{tabular}{|l|c|c|c|c|} \hline
		\textbf{Proxy Verifies?}		&  \cell{ \textbf{Mismatch} \\ \textbf{Resolving}} 	& \cell{ \textbf{MAC tags}} & \textbf{Signatures}  & \textbf{Verifications} \\ \hline	
		Yes        & Proxy   &  No 			 & \cell{Nodes: $5f+1+M$(p) }	 & \cell{Nodes: $M \cdot C$(p) \\ Proxy: $4f+1+C+M$(p)\\Client: $2f+1+M$(p)} \\ \hline
		
		% 2f+1 Nodes requested and answer to proxy - Mismatch - Proxy Requests all 3f+1 and verifies 2f+1. Resolves C columns and  updates M replicas
		% Nodes 2f+1 sign,          3f sign,                 M*C ver, M sign
		% Proxy           2f+1 ver,           2f ver, C ver,                  M ver
		% Client 2f+1 ver + 2f+1 resolving ver
		
		No & Proxy 	&  No			 & \cell{Nodes: $5f+1+M$(p) }	 & \cell{Nodes: $M \cdot C$(p) \\ Proxy: $C$(p)\\Client: $2f+1+M$(p)} \\ \hline	
		% Nodes 2f+1 sign,  3f+1 sign,         M*C ver, M sign
		% Proxy                        C ver,
		% Client 2f+1 ver + 2f+1 resolving ver

		No & Proxy   &  Yes			 & \cell{Nodes: $5f+1+M$(s) }	 & \cell{Nodes: $M \cdot C$(p) \\ Proxy: $C$(p)\\Client: $2f+1+M$(s)} \\ \hline	
		No & Client  &  No			 & \cell{Nodes: $5f+1+M$(p) }	 & \cell{Nodes: $M \cdot C$(p) \\Client: $2f+1+C+M$(p)} \\ \hline	
		No & Client  &  Yes			 & \cell{Nodes: $5f+1+M$(s) }    & \cell{Nodes: $M \cdot C$(p) \\Client: $2f+1+M$(s) \& $C$(p)} \\ \hline
		
	\end{tabular}
	\caption[Comparing the variants in the read flow in case of a benign mismatch that requires resolving.]{Comparing the variants in the read flow in case of a benign mismatch that requires resolving. $C$ is the number of columns, $M$ is the number of outdated replicas in the used quorum, (p) indicated public key signatures and (s) MAC tags. We assume that the proxy uses the optimization in the first phase and contacts only a Byzantine quorum. For example, the first row presents a proxy that verifies the acknowledgments and resolves conflicts when mismatch values are observed. MAC tags are not in use. On a read request, a Byzantine quorum of nodes ($2f+1$) have to retrieve the row and sign it. The proxy verifies their signatures ($2f+1$) and detects a conflict. Then, the proxy requests all relevant nodes (except for the one that returned data in the first phase) for the full data ($3f$ nodes sign and the proxy verifies only $2f$). The proxy resolves the mismatch (verifies $C$ columns) and sends the resolved row to the $M$ outdated nodes (write-back). These nodes verify the row ($C$) and sign the acknowledgments that are later verified by the proxy. The proxy supply the client with the original $2f+1$ answers and the resolved row signed also by $M$ nodes that approved the write-back.}
	\label{tbl:comparing_table_res}
\end{table*}

\begin{table*}[t!]
	\tiny
	\captionsetup{font=footnotesize}
	\centering
	
	\begin{tabular}{|l|c|c|c|c|c|c|} \hline
		\cell{\textbf{Proxy}  \\ \textbf{Verifies?}} 	&
		\textbf{Op} &
		\cell{\textbf{Mismatch}  \\ \textbf{Resolving}} 	&
		\cell{\textbf{MAC Tags}} &
		\textbf{Signatures}  &
		\textbf{Verifications} &
		\cell{\textbf{Client-Proxy} \\ \textbf{Requests}}
		\\ \hline
		
		Yes & Write & - & None 			  & \cell{ $C$(p) }	  & \cell{ $(2f+1)\cdot (f+1)$(p)} & $f+1$ \\ \hline
		No 	& Write & - & None 			  & \cell{ $C$(p) }	  & \cell{ $(3f+1)\cdot (f+1)$(p)} & $(f+1)\cdot (f+1)$ \\ \hline
		No  & Write & - & Nodes to client & \cell{ $C$(p) }	  & \cell{ $(3f+1)\cdot (f+1)$(s)} & $(f+1)\cdot (f+1)$ \\ \hline
		No  & Write & - & Both ways		  & \cell{ $C$(p) }	  & \cell{ $(3f+1)\cdot (f+1)$(s)} &  $(f+1)\cdot (f+1)$ \\ \hline
		Yes & Read & Proxy  & None 			 & \cell{ None }	 & \cell{ $(2f+1+M) \cdot (f+1)$(p)} &  $(f+1)$ \\ \hline
		No  & Read & Proxy  & None			 & \cell{ None }	 & \cell{ $(2f+1+M) \cdot (f+1) \cdot (f+1)$(p)} &  $(f+1)\cdot (f+1)$ \\ \hline
	
		No  & Read & Client  & None			 & \cell{ None }  	 & \cell{ $(2f+1) \cdot (f+1)\cdot (f+1)$ \\ $+ C + (M+f)\cdot(f+1)$(p)} &  \cell{$(f+1)\cdot (f+1)$ \\ $+(M+f)\cdot (f+1)$} \\ \hline
		% 2f+1 answers, one is not got, ask again f+1 time, fail, try f+1 proxy nodes
		% C verify signatures
		% Write back, requst M responses: Verify M fetch more f, switch f+1 proxy nodes
		
		No  & Read & Proxy  & Nodes to client & \cell{ None }	 & \cell{ $(2f+1+M) \cdot (f+1) \cdot (f+1)$(s)} &  $(f+1)\cdot (f+1)$ \\ \hline
		No  & Read & Client  & Nodes to client & \cell{ None } 	  & \cell{ $(2f+1) \cdot (f+1)\cdot (f+1)$ \\ $+(M+f)\cdot(f+1)$(s) \& $C$(p)} &  \cell{$(f+1)\cdot (f+1)$ \\ $+(M+f)\cdot (f+1)$} \\ \hline
	\end{tabular}
	\caption[Comparing the variants in the read and write flows in the worst case and $f$ Byzantine nodes.]{Comparing the variants in the read and write flows in the worst case and $f$ Byzantine nodes. Due to the wide options of Byzantine attacks and the fact that every Byzantine node can waste other node's cycles, we compare the variants only from the point of view of a correct client. $C$ is the number of columns, $M$ is the number of outdated replicas in the used quorum, (p) indicated public key signatures and (s) MAC tags. For example, the second row presents a proxy that does not verify the acknowledgments in a write operation. MAC tags are not in use. On a write request, the client signs the $C$ columns and sends it to the proxy. The client receives from the proxy responses from a Byzantine quorum of nodes ($2f+1$) and detects that one is incorrect. The client requests the proxy $f$ more times for the missing signature and every time gets a false signature. Then, the client uses alternative proxies and the story repeats itself $f$ additional times. In the last round, the client successfully retrieves all $2f+1$ correct signatures due to our assumption on $f$.}
	\label{tbl:comparing_table_worst}
\end{table*}

\subsection{Handling Byzantine Clients}
In addressing the challenge of handling Byzantine clients, we keep in mind that some actions are indistinguishable from correct clients behaviors.
For example, erasing data or repeatedly overwriting the same value.
Yet, this requires the client to have ACL permissions.

In our work, we focus on preserving the consistency of the data from the point of view of a correct client.
A correct client should not observe inconsistent values resulting from a split-brain-write.
Further, a correct client should not read values that are older than values returned by previous reads.

More precisely, we guarantee the following semantics, similar to plain \Cassandra:
(1) The order between two values with the same timestamp is their lexicographical order (breaking ties according to their value).
(2) A write of multiple values with the same timestamps is logically treated as multiple separate writes with the same timestamp.
(3) Deleting values is equivalent to overwriting these values with a \emph{tombstone}.
(4) A read performed by a correct client must return any value that is not older (in terms of timestamp order) than values returned by prior reads.
(5) A read performed after a correct write must return a value that is not older (in terms of timestamp order) than that value.

As mentioned before, in \Cassandra, if the proxy that handles a read observes multiple versions from different nodes, it first resolves the mismatch and writes the resolved value back to the nodes.
The resolved version will be a row with the most updated columns according to their timestamps.
If the proxy observes two values with the same timestamp, it will use the lexicographical order of the values as a tie breaker.

For performing split-brain-writes, Byzantine clients may collude with Byzantine proxies and sign multiple values with the same timestamp.
Proxies can send these different values with the same timestamps to different nodes, setting the system in an inconsistent state.
Even though we consider a split-brain-write as a Byzantine behavior, this kind of write could occur spontaneously in plain \Cassandra{} by two correct clients that write in parallel since in \Cassandra{} clients provide the write's timestamp, typically by reading their local clock.

Consider a Byzantine client that colludes with a proxy.
If they try to perform a split-brain-write, due to the resolve mechanism, all reads that witness both values will return only the latest value in lexicographical order.
On a client read, no quorum will agree on one version.
Consequently, the proxy will resolve the conflict and update a quorum of servers with that version, leaving the system consistent for that value after the first read.

If the Byzantine client and colluding proxy will try to update only part of the nodes with a certain $v$, a read operation may return two kinds of values:
(1) If the read quorum will witness $v$, it will be resolved and propagated to at least a quorum of nodes meaning that $v$ will be written correctly to the system.
As a result of this resolve, every following read will return $v$ (or a newer value).
(2) If a read will not witness $v$, the most recent version of a correct write will be returned.
Hence, the hardened system protects against such attempts.

Finally, if a Byzantine client is detected by the system administrator and removed, its ACL and certificate can be revoked immediately.
This way any potentially future signed writes saved by a colluder will be voided and the future influence of that client will cease.

%In order to handle Byzantine clients and ensure that correct nodes hold the same value for the same key and timestamp, we exploit the work of Liskov \& Rodrigues \cite{liskov2005byzantine}. On a write request, the proxy contacts the nodes and asks for a signed commitment for storing the proposed value with the current timestamp (binding a key, value and a timestamp). Then it transmits the set of commitments to the nodes. A node will store a value on receiving and verifying a set of at least $2f+1$ commitments.

%In case a timestamp is already committed for a given key, the first phase will fail. In this case, the nodes that were asked to provide a commitment will propose a new timestamp and the proxy will ask them again with the latest timestmap received.

%A client will complete a read query on receiving from a proxy the value along with a set of $2f+1$ responses each containing a set of at least $2f+1$ commitments for the same value summery with the same timestamp.

%The client also has to send a \e{nonce} that the nodes will sign in order to prevent retransmission.

%In Fiqure~\ref{fig:read-flow}, line 22, we use the timestamp as a nonce.

\subsection{Deleting Values}
In \Cassandra, deleting a value is done by replacing it with a \e{tombstone}.
This tombstone is served to any system node that requests that value to indicate that it is deleted.
Once in a while, every node runs a garbage collector that removes all tombstones that are older than a configurable time (10 days by default).

Even in a benign environment, some failures might result in deleted values reappearing.
One case is when a failed node recovers after missing a delete operation and passing the garbage collection interval in all other nodes.
In a Byzantine setting, a Byzantine node can ignore all delete operations and later (after the garbage collection interval) propagate the deleted values to correct nodes.

To overcome this, we define the following measures:
(1) Every delete should be signed by a client as in the write operation previously defined.
This signature will be stored in the tombstone.
A client will complete a delete only after obtaining a Byzantine quorum of signed acknowledgments.
(2) In the period of every garbage collection interval, a node will run at least once the anti-entropy tool against a Byzantine quorum of nodes, fetching all missed tombstones.
(3) A node will accept writes of values that are not older than the configured time for garbage collection interval as previously defined.
Since the node runs the anti-entropy tool periodically, even if a deleted value is being fetched, the tombstone will overwrite it.
(4) A node that receives a store value that is older than the configured time for the garbage collector will handle this case as follows.
It will issue a read for the value and accept it only if a Byzantine quorum approves that the value is live.
When a new node joins the system, reading every value from a Byzantine quorum might be very expensive.
In this case, we can improve the performance by batching these requests.

\subsection{Membership View}

The membership implementation of \Cassandra{} is not Byzantine proof as faulty nodes can temper other's views by sending false data~\cite{CassandraGossip}.
In addition, Byzantine seed nodes can partition the system into multiple subsystems that do not know about each other.
This is by exposing different sets of nodes to different nodes.

To overcome this, in a Byzantine environment, each node installation should be signed by the trusted system administrator with a logical timestamp.
The logical timestamp is used so a node will make sure it is using the updated configuration.
Each node should be configured to contact at least $f+1$ seed nodes in order to get at least one correct view.
This solution requires also the system administrator to pre-configure manually the first $f+1$ nodes view as they cannot trust the rest of the nodes.
We would like to emphasize that Byzantine seeds cannot forge false nodes existence.
Rather, they can only hide some nodes by not publishing them.

Here, we adopt the work on BFT push-pull gossip by Aniello et al.~\cite{CassandraGossip}.
Their solution solves the dissemination issues by using signatures on the gossiped data.
This way, a node's local view cannot be mislead to think that a node is responsive or suspected.

%We are also interested in exploring other dissemination solutions like using a random graph dissemination technique such as Araneola~\cite{araneola}. The last offers a quickly converging dissemination structure with a strong  for failures but not for Byzantine failures.

\subsection{Synchronized Clock} \label{clock}

In plain \Cassandra, as well as in our solution, each write includes a wall-clock timestamp that implies an order on the writes.
Using this method, strong consistency cannot be promised unless local clocks are perfectly synchronized.
For example, consider two clients that suffer from a clock skew of $\Delta$.
If both clients write to the same object in a period that is shorter than $\Delta$, the later write might be attached with a smaller timestamp.
As a result, the older write wins.

In a benign environment, when ensuring a very low clock skew, for most applications, these writes can be considered as parallel writes so any ordering of them is correct.
For time synchronization, \Cassandra~requires the administrator to provide an external solution such as NTP.
In our work, we follow this guideline using the latest version of NTP that can tolerate Byzantine faults when ensuring the usage of SSL and authentication measures~\cite{baldoni2008peerTime, mills2010network}.
%using equivalent BFT NTP solutions~\cite{}.
We configure this service so that all servers could use it as is and clients would be able only to query it, without affecting the time.

%Let us comment that the BFT synchronization clock protocols we are aware off~\cite{}, all share our basic assumption of a system consisting of at least $3f+1$ servers, masking up to $f$ Byzantine serves.
Alternatively, one could use external clocks such as GPS clocks, atomic clocks or equivalents~\cite{fetzer1997integrating}, assuming Byzantine nodes can neither control them nor the interaction with them.
Finally, \Cassandra{} nodes can ignore writes with timestamps that are too far into the future to be the result of a normal clock's skew.

\subsection{Other Network Attacks}

\Cassandra{} might be targeted with known overlay networks attacks, such as \e{Sybil attacks}~\cite{douceur2002sybil} and \e{Eclipse attacks}~\cite{singh2006eclipse}.
In a Sybil attack, attackers create multiple false entities.
In \Cassandra, they may create multiple node ids that lead to the same node, thereby fooling a client into storing its data only on a single Byzantine replica.
As suggested in~\cite{douceur2002sybil}, here we rely on a trusted system administrator to be the sole entity for approving new entities that can be verified using PKI.

In an Eclipse attack, attackers try to divert requests towards malicious entities.
In our solution, we authenticate each part of the communication using SSL.
In \Cassandra, a proxy might try to target only Byzantine replicas.
To overcome this, clients request verifiable acknowledgments and count the number of correct repliers.
If a proxy fails to provide these, alternative proxies are contacted until enough correct nodes have been contacted.
Additionally, Section~\ref{irrelevant} explains how we handle a proxy that diverts requests to irrelevant nodes.

Yet, we currently do not provide any protection for data theft even when a single node has been breached.
This can be overcome by encrypting the data at the client application side.

%%%%%%%%%%%%%%%%%%%%%%%%% Section 6 : Performance %%%%%%%%%%%%%%%%%%%%%%%
\section{Performance}
\label{sec:perf}
\begin{figure*}[t]
	\begin{subfigure}{\textwidth}
			
	\begin{subfigure}{\textwidth}
		\centering
		\includegraphics[width=0.8\textwidth]{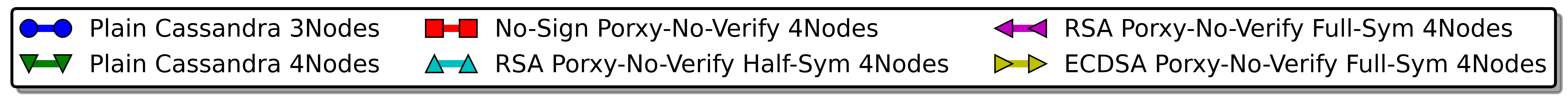}
	\end{subfigure}
	\\
	\begin{subfigure}{.331\textwidth}
		\centering
		\includegraphics[width=\textwidth]{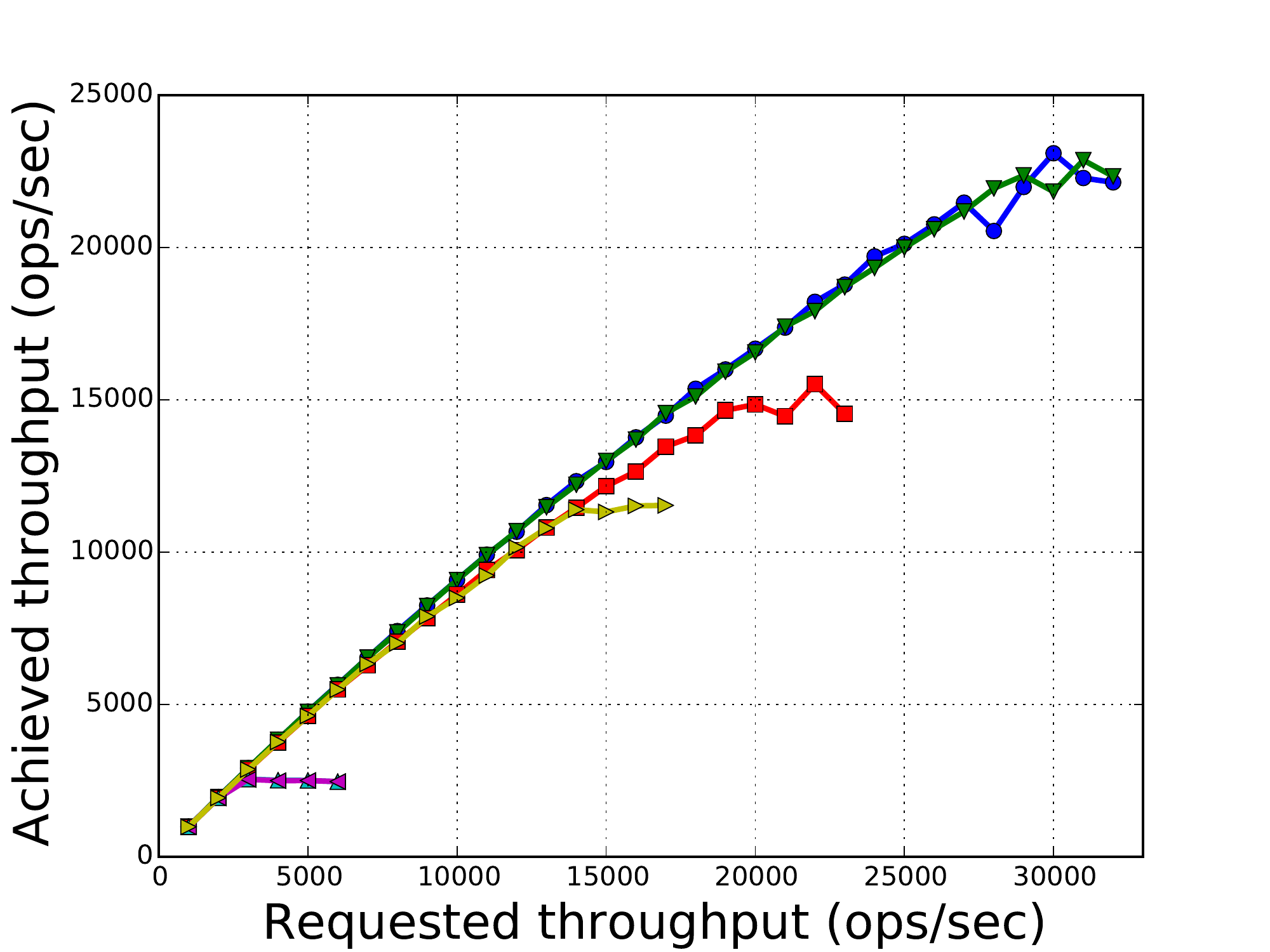}
	\end{subfigure}%
	\begin{subfigure}{.331\textwidth}
		\centering
		\includegraphics[width=\textwidth]{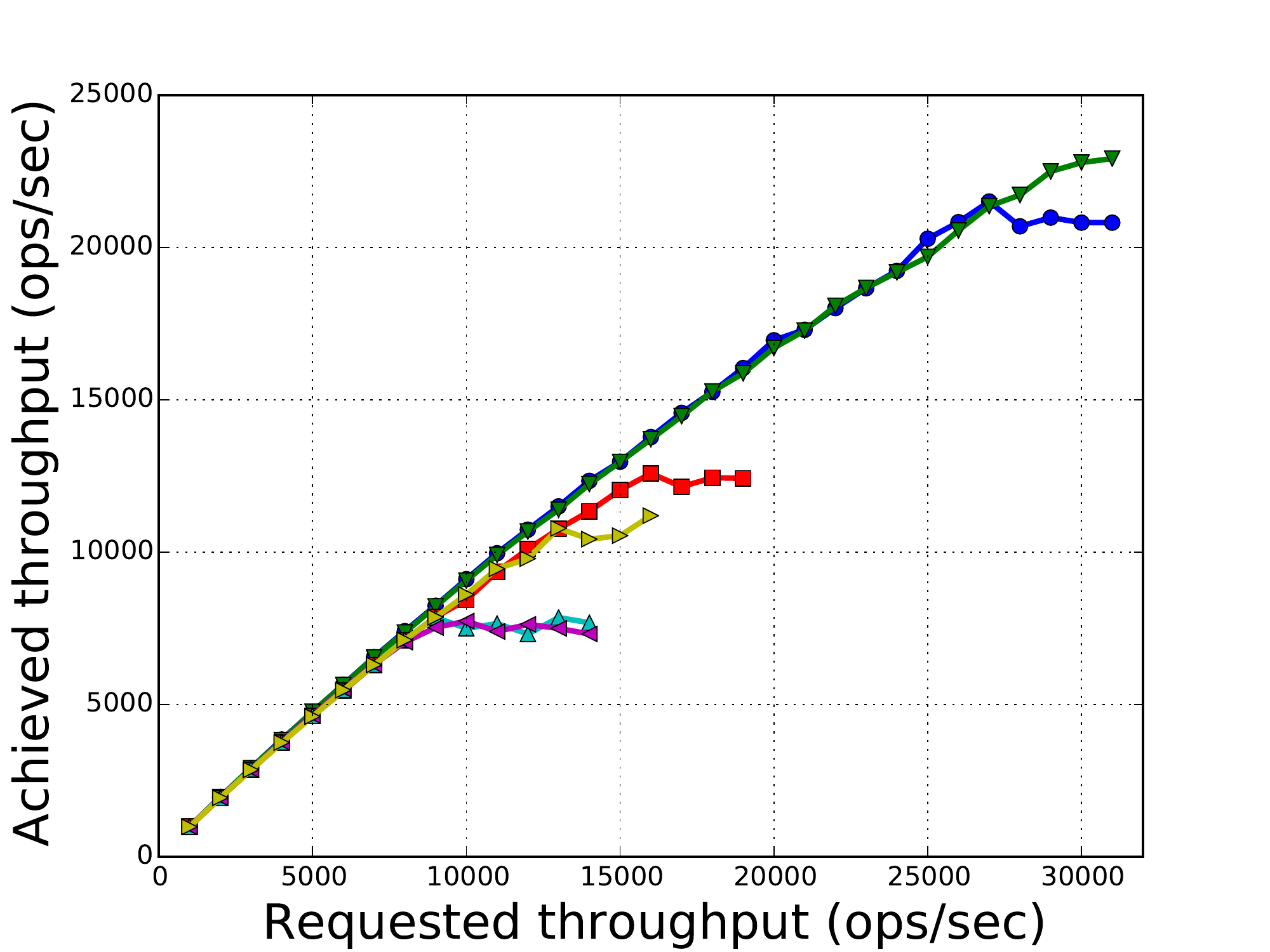}
	\end{subfigure}
	\begin{subfigure}{.331\textwidth}
		\centering
		\includegraphics[width=\textwidth]{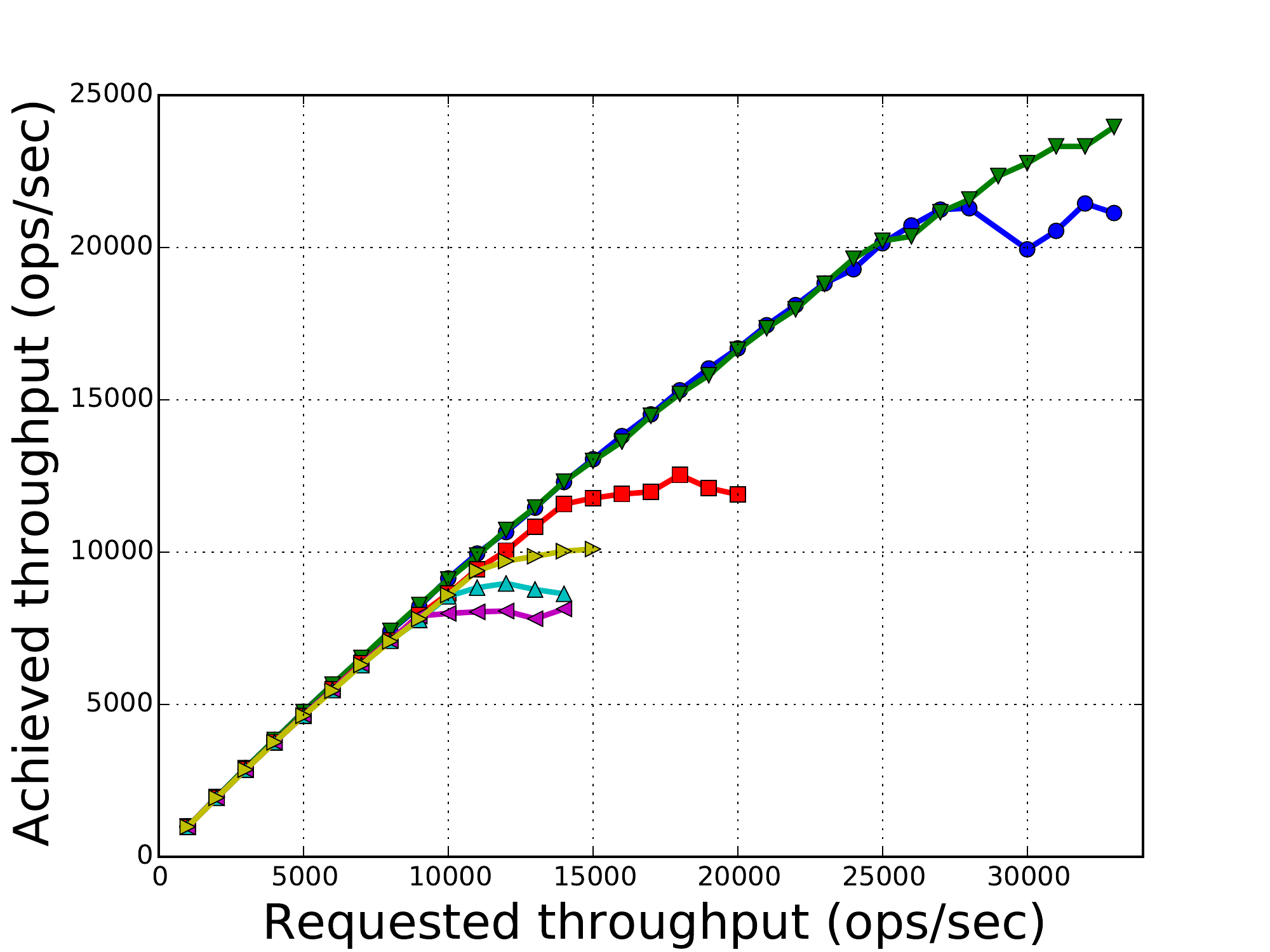}
	\end{subfigure} \\
	\begin{subfigure}{.331\textwidth}
		\centering
		\includegraphics[width=\textwidth]{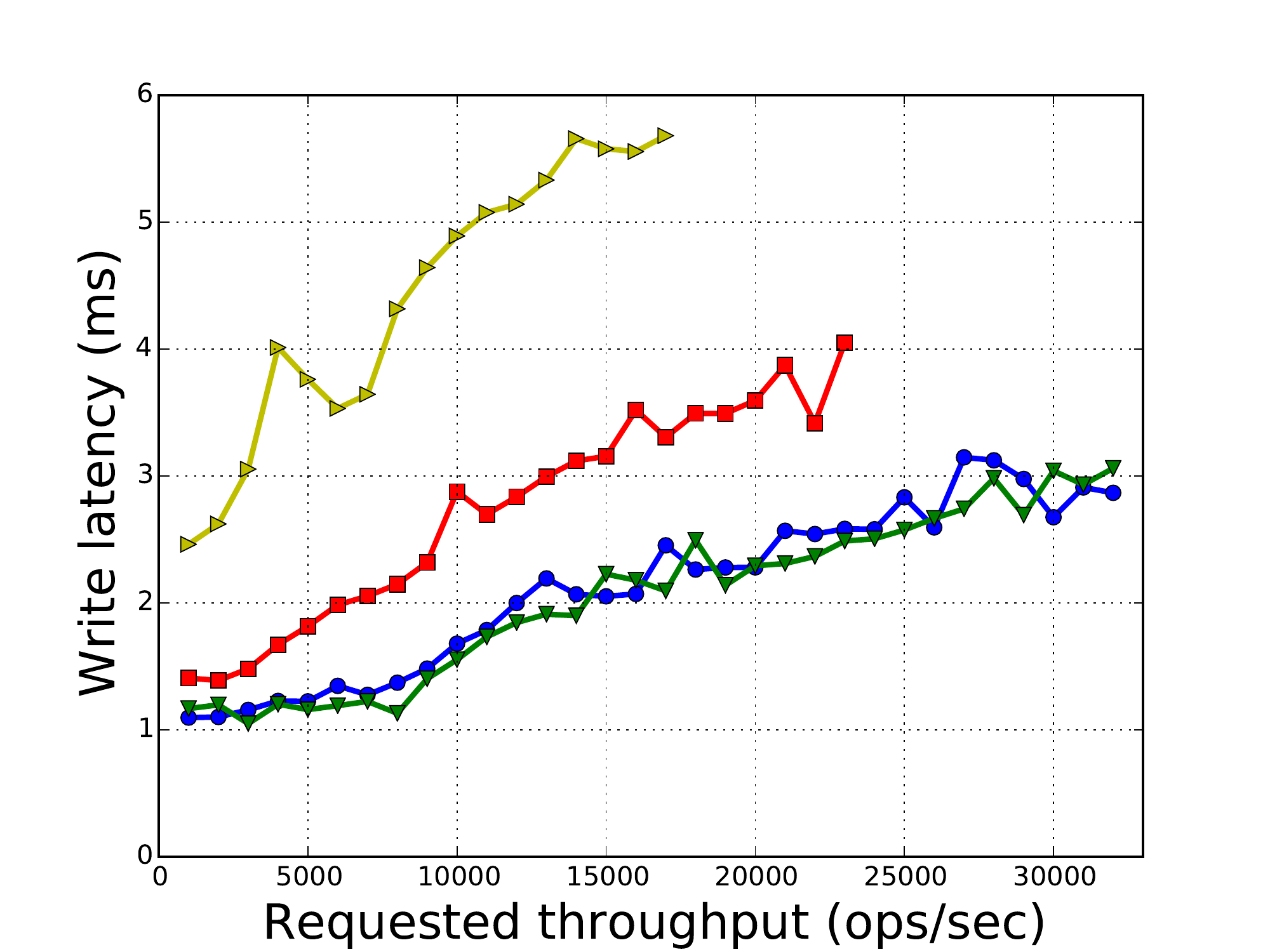}
	\end{subfigure}%
	\begin{subfigure}{.331\textwidth}
		\centering
		\includegraphics[width=\textwidth]{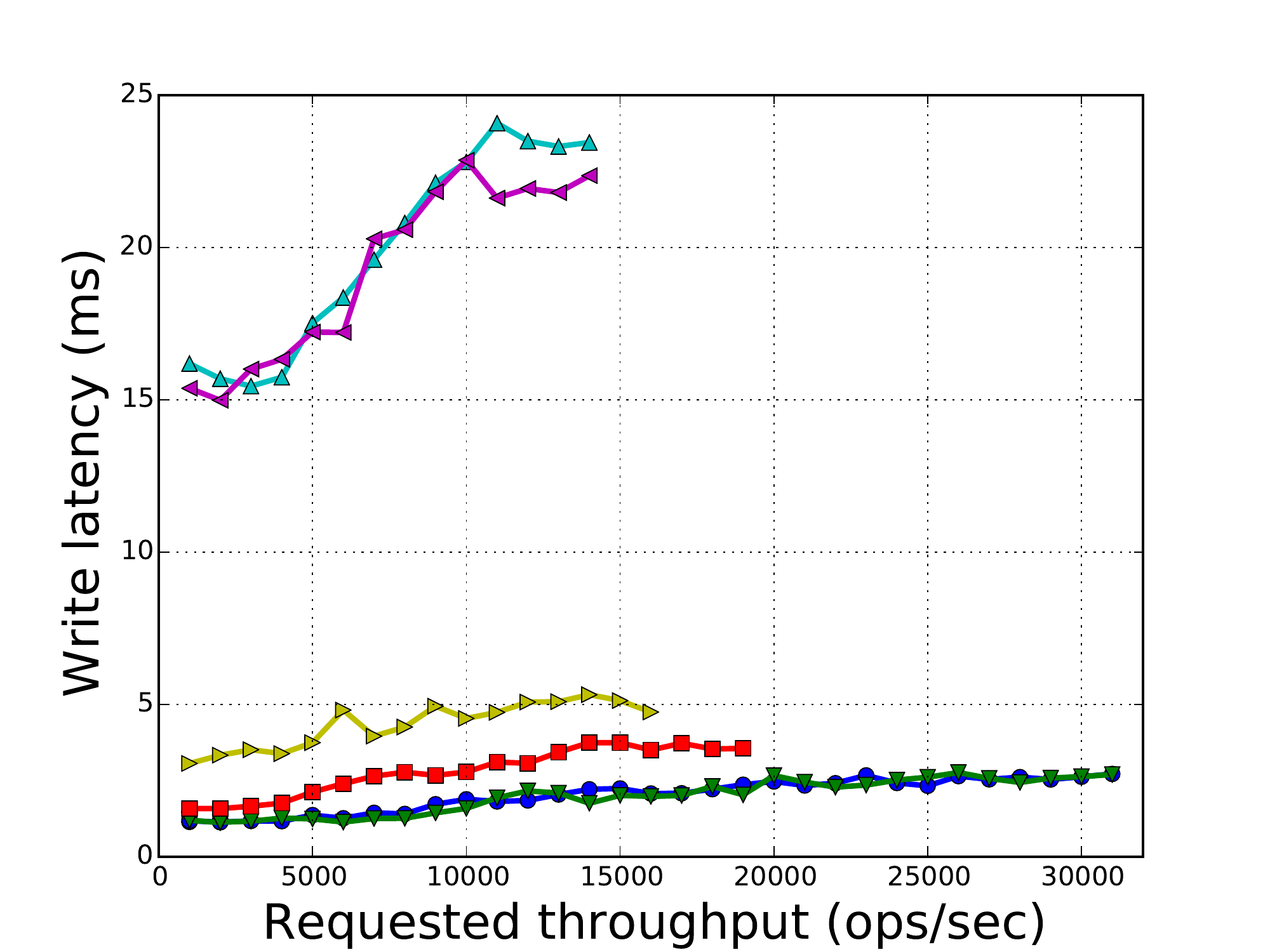}
	\end{subfigure} \\
	\begin{subfigure}{.331\textwidth}
		\centering
		\includegraphics[width=\textwidth]{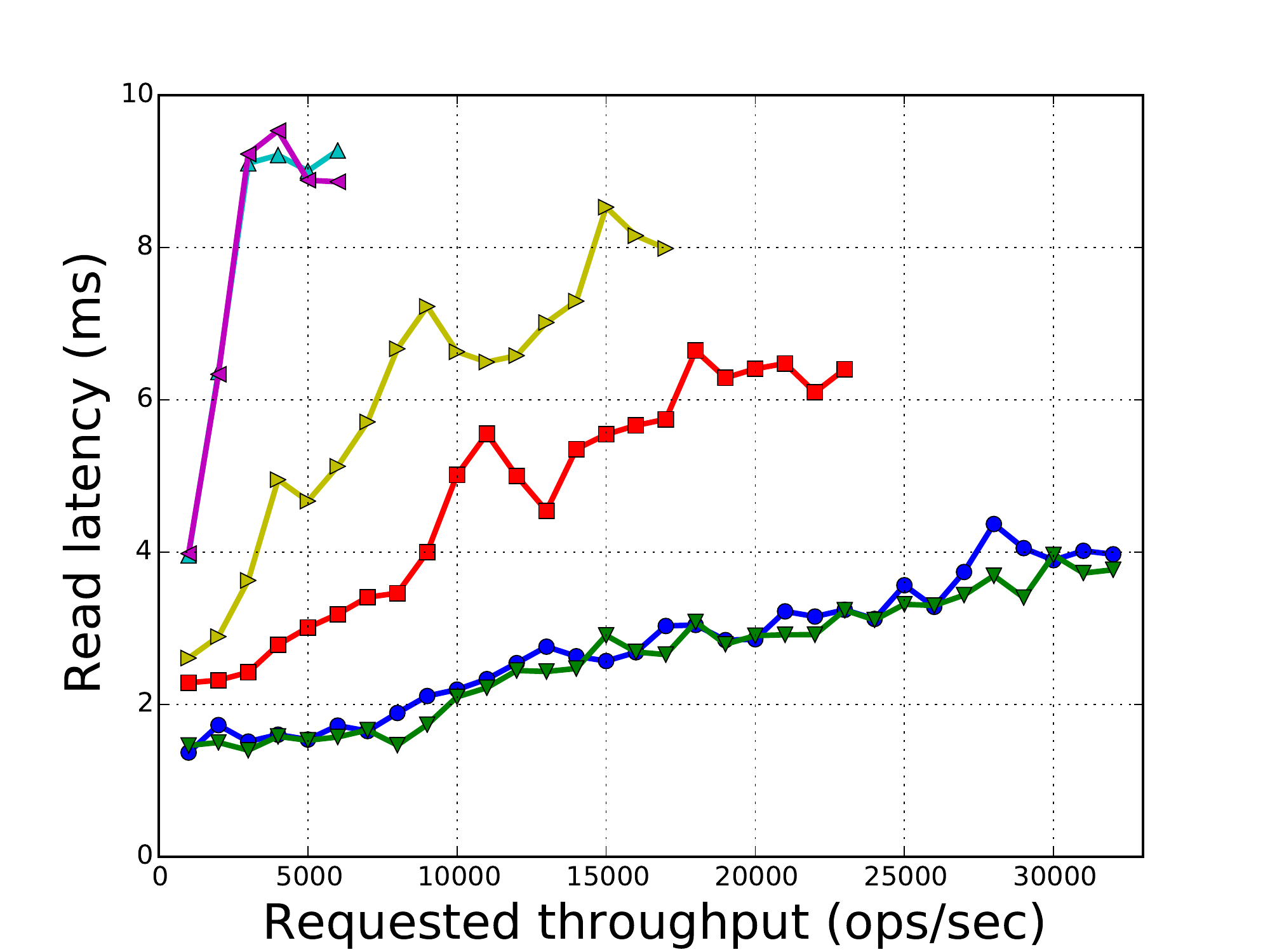}
		\caption{Workload A}
	\end{subfigure}%
	\begin{subfigure}{.331\textwidth}
		\centering
		\includegraphics[width=\textwidth]{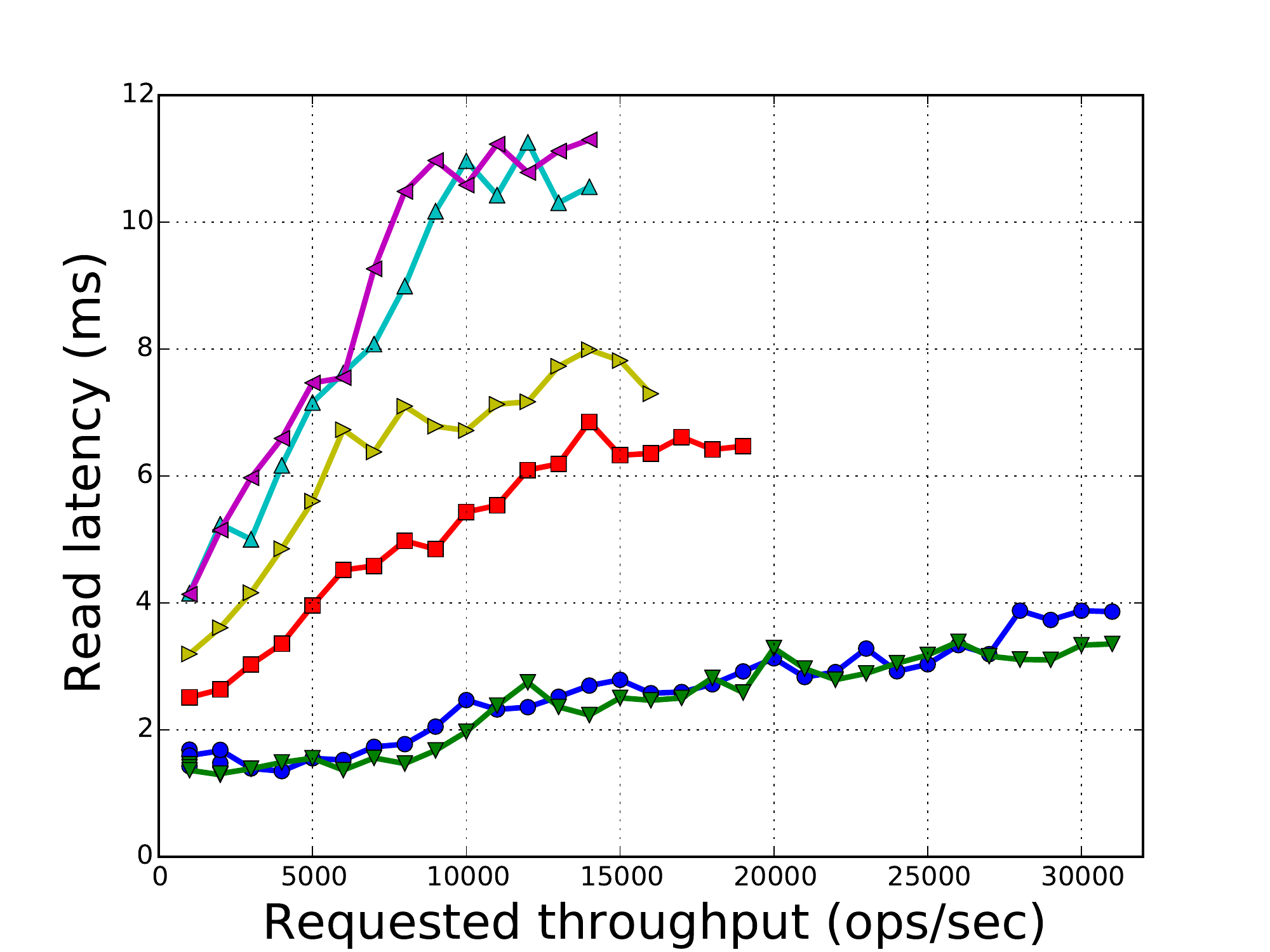}
		\caption{Workload B}
	\end{subfigure}
	\begin{subfigure}{.331\textwidth}
		\centering
		\includegraphics[width=\textwidth]{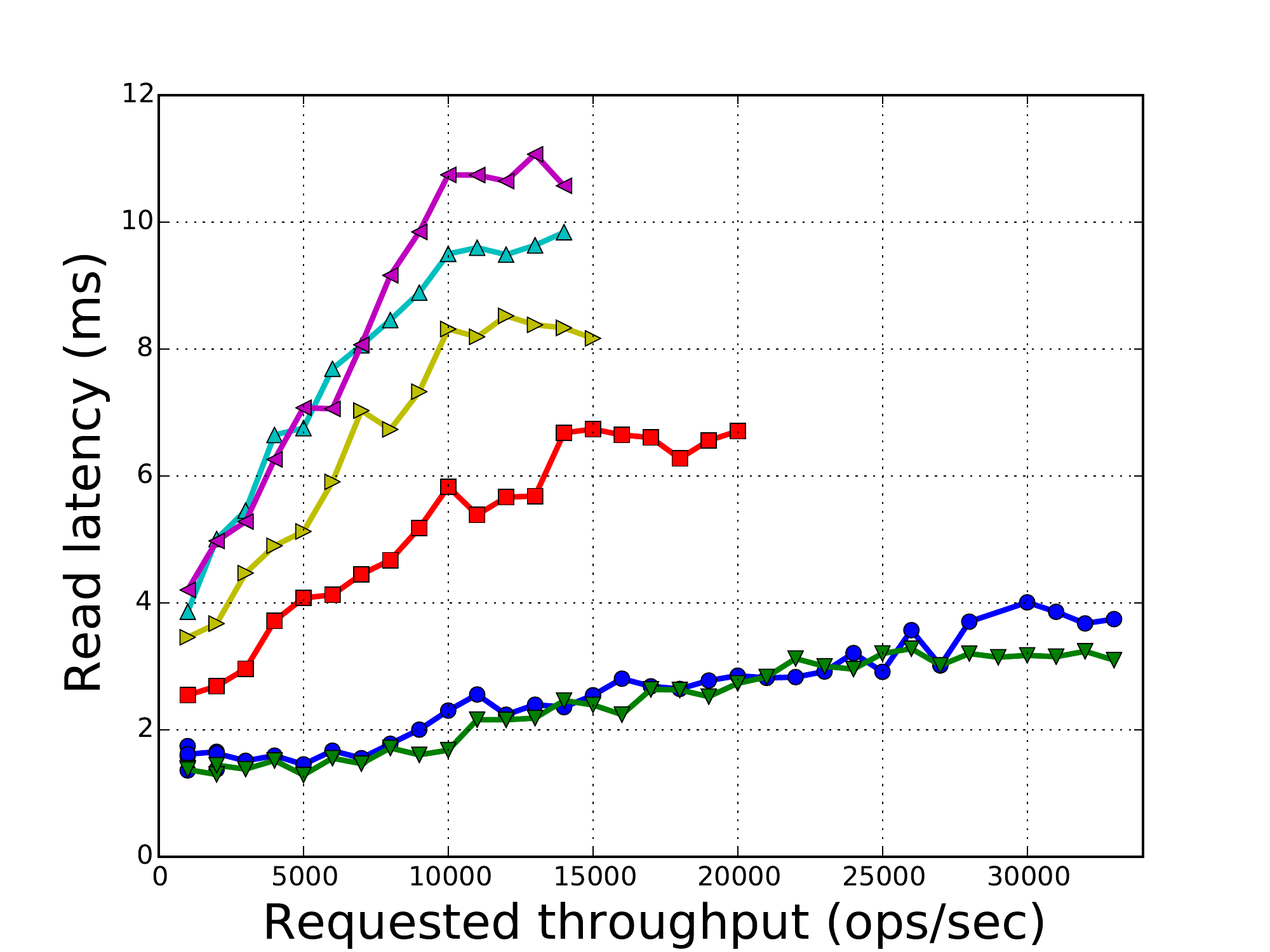}	
		\caption{Workload C}
	\end{subfigure}

	\end{subfigure}
	\caption[Comparing the best variants against plain \Cassandra~and the algorithm with \e{No-Sign} using workloads A, B and C.]{Comparing the best variants against plain \Cassandra{} and the algorithm with \e{No-Sign} using workloads A, B and C. In the write latency of (a), we left the RSA variants out as they rapidly grew to $\approx$65ms latency.} %(can be observed in Figure \ref{fig:two-a_to_c}).}
	\label{fig:one-a_to_c}
\end{figure*}

\begin{figure*}[t]
\begin{center}
	\begin{subfigure}{.331\textwidth}
		\centering
		\includegraphics[width=\linewidth]{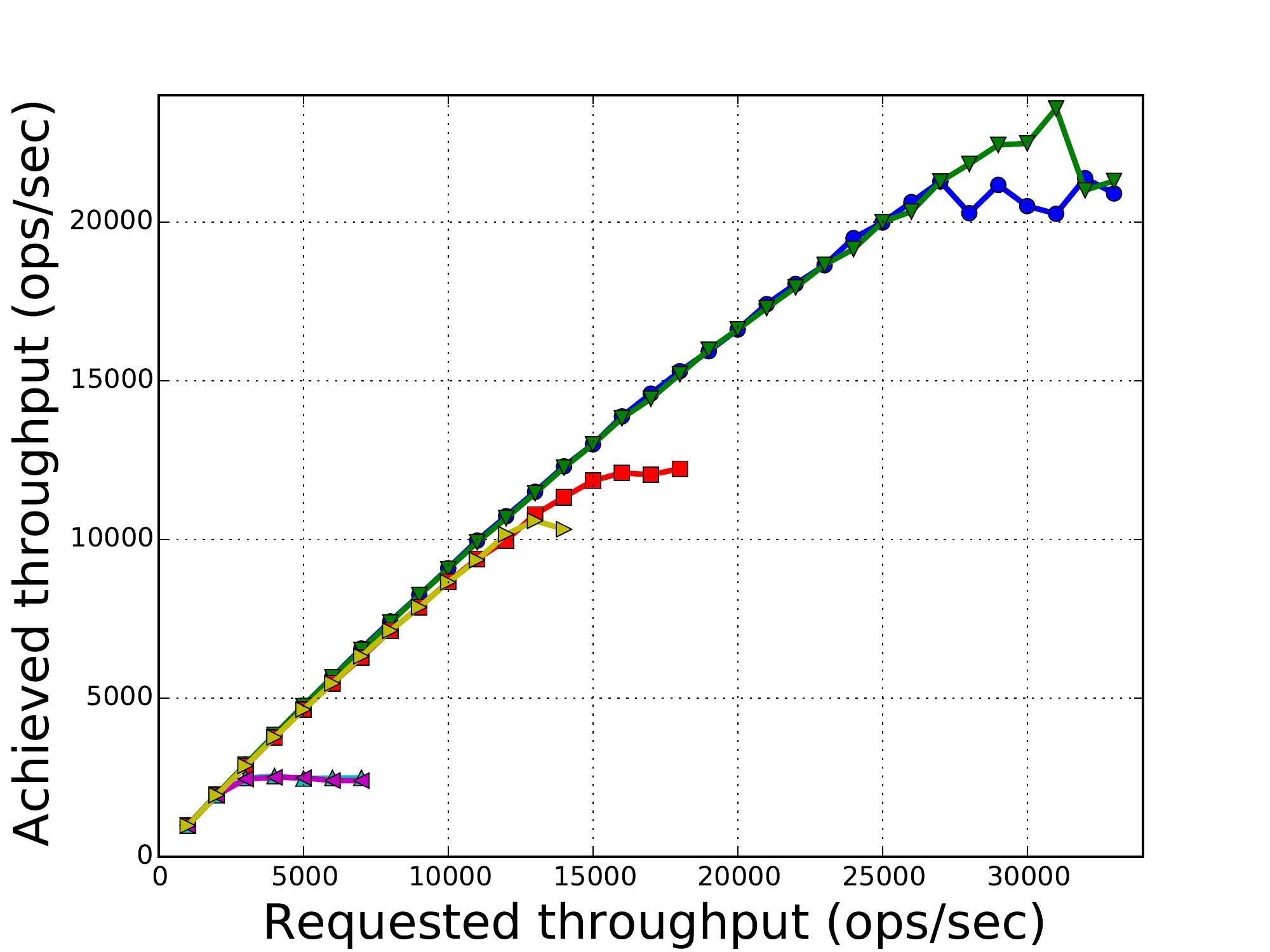}
	\end{subfigure}%
	\begin{subfigure}{.331\textwidth}
		\centering
		\includegraphics[width=\linewidth]{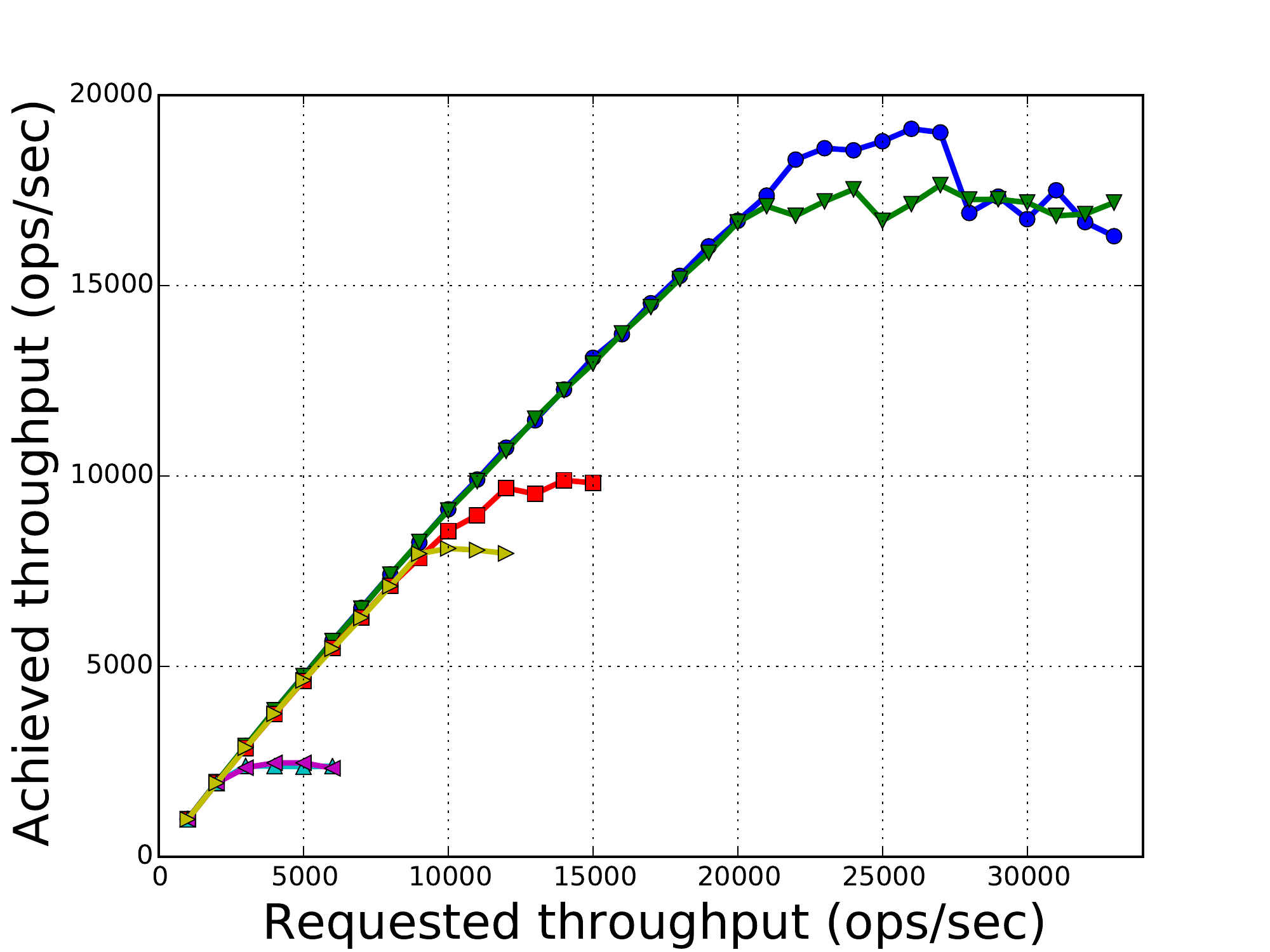}
	\end{subfigure} \\
	\begin{subfigure}{.331\textwidth}
		\centering
		\includegraphics[width=\linewidth]{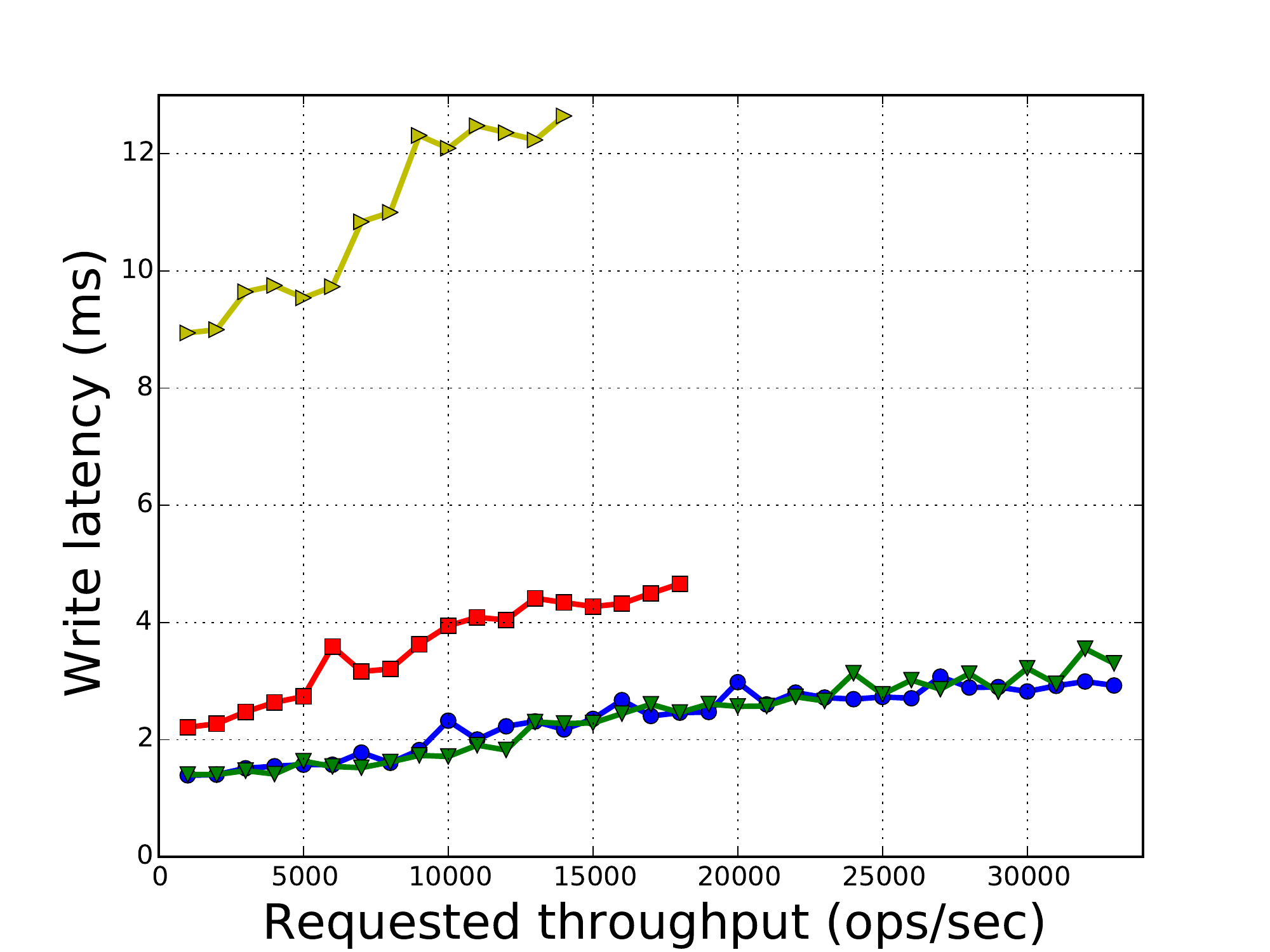}
	\end{subfigure}%
	\begin{subfigure}{.331\textwidth}
		\centering
		\includegraphics[width=\linewidth]{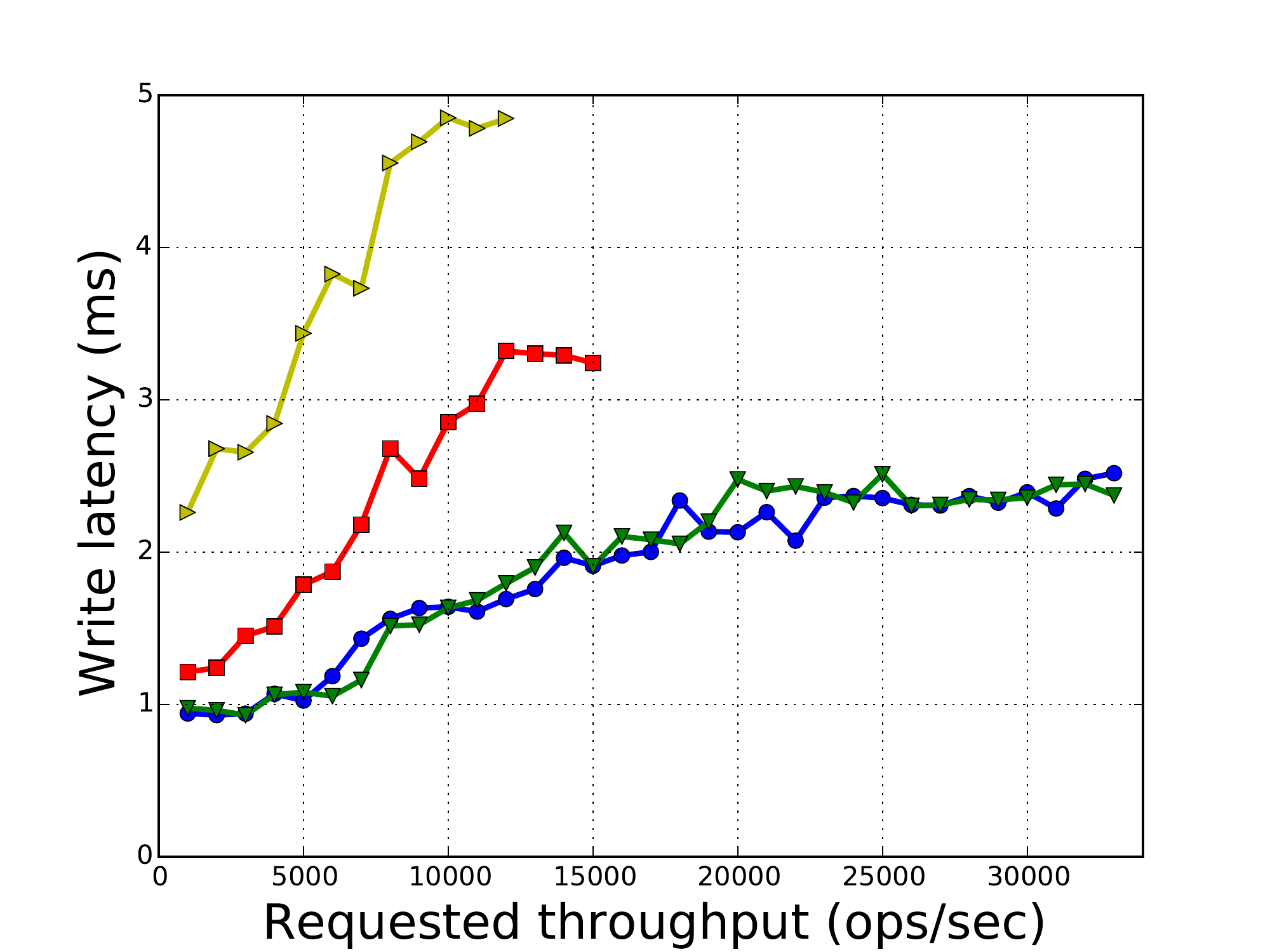}
	\end{subfigure} \\
	\begin{subfigure}{.331\textwidth}
		\centering
		\includegraphics[width=\linewidth]{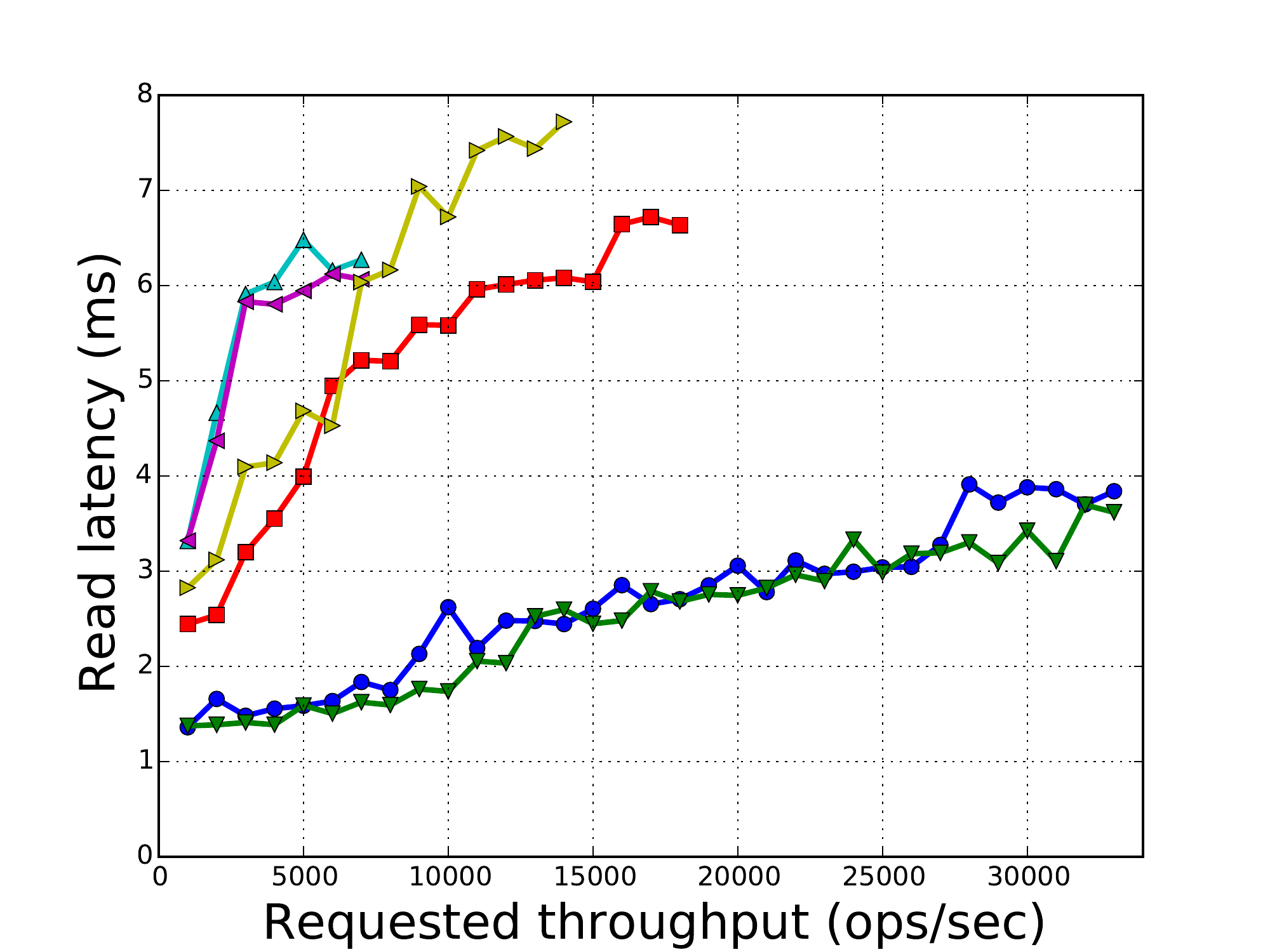}
		\caption{Workload D}
	\end{subfigure}%
	\begin{subfigure}{.331\textwidth}
		\centering
		\includegraphics[width=\linewidth]{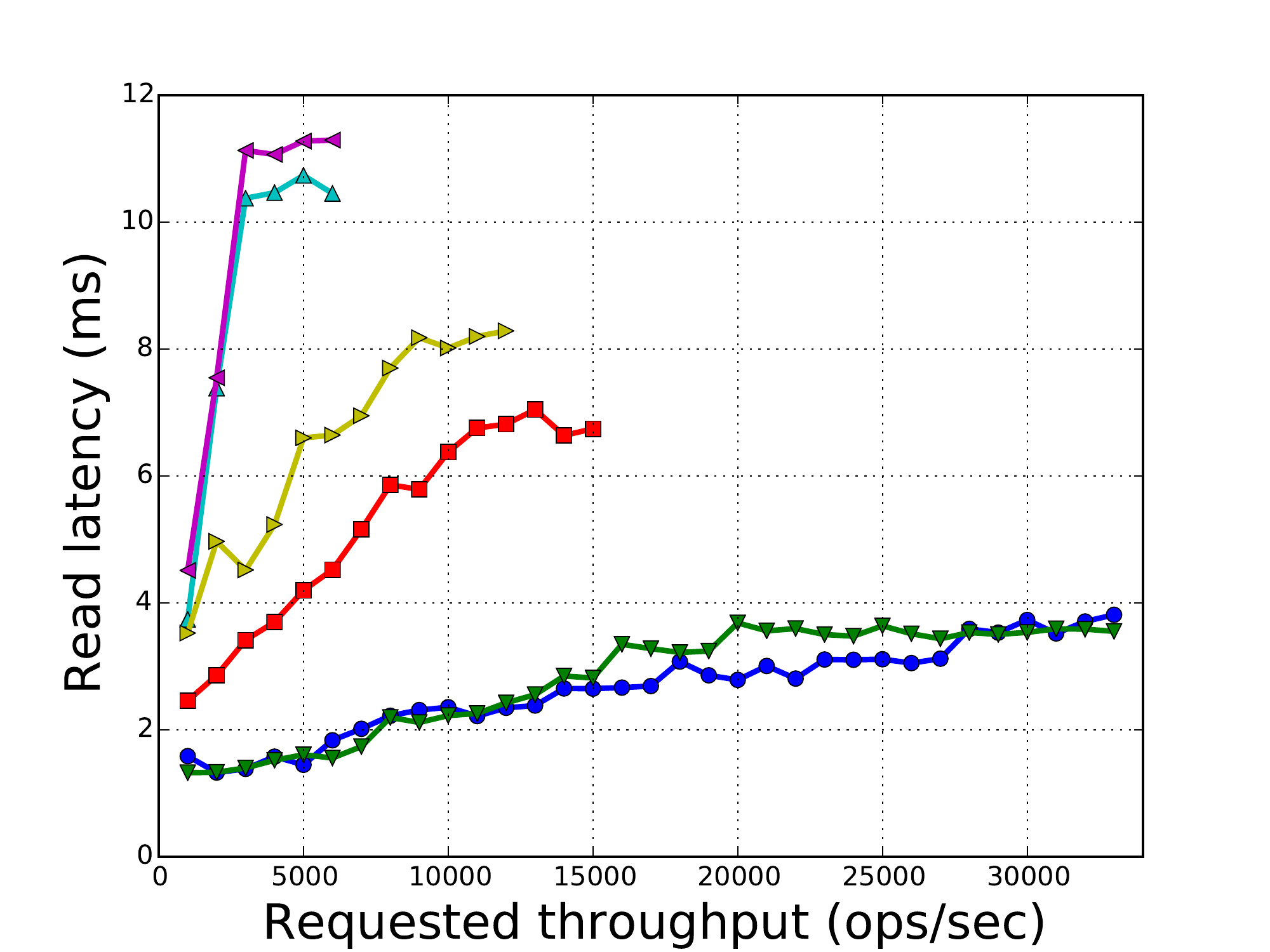}
		\caption{Workload F}
	\end{subfigure}%
\end{center}
	\caption[Same as Figure \ref{fig:one-a_to_c} while using workloads D and F.]{Same as Figure \ref{fig:one-a_to_c} while using workloads D and F. Here, the write latency graphs do not include the RSA variants as they rapidly reached the areas of 600 ms and 65 ms latency, respectively.} %(this can be observed in Figure \ref{fig:two-d_to_f}).}
	\label{fig:one-d_to_f}
\end{figure*}

\begin{figure*}[t]
	\begin{subfigure}{\textwidth}
		\centering
		\includegraphics[width=0.8\linewidth]{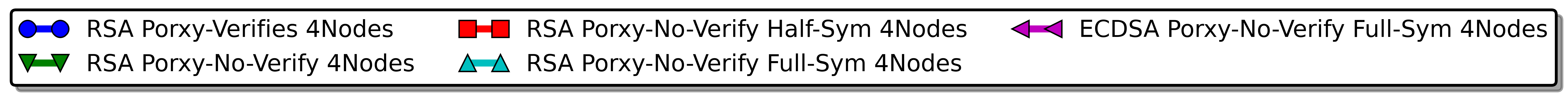}
	\end{subfigure}
	\\
	\begin{subfigure}{.331\textwidth}
		\centering
		\includegraphics[width=\linewidth]{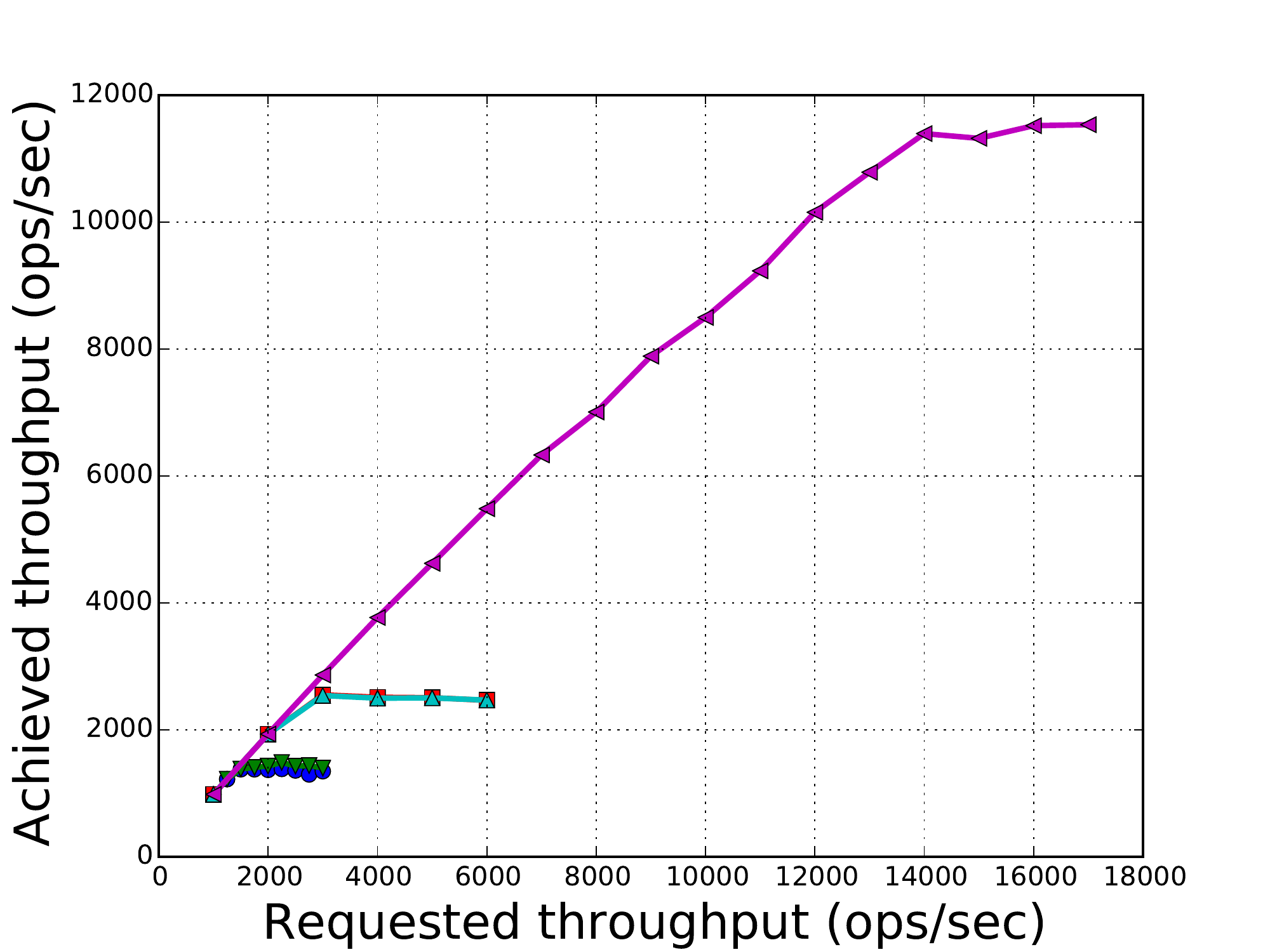}
	\end{subfigure}%
	\begin{subfigure}{.331\textwidth}
		\centering
		\includegraphics[width=\linewidth]{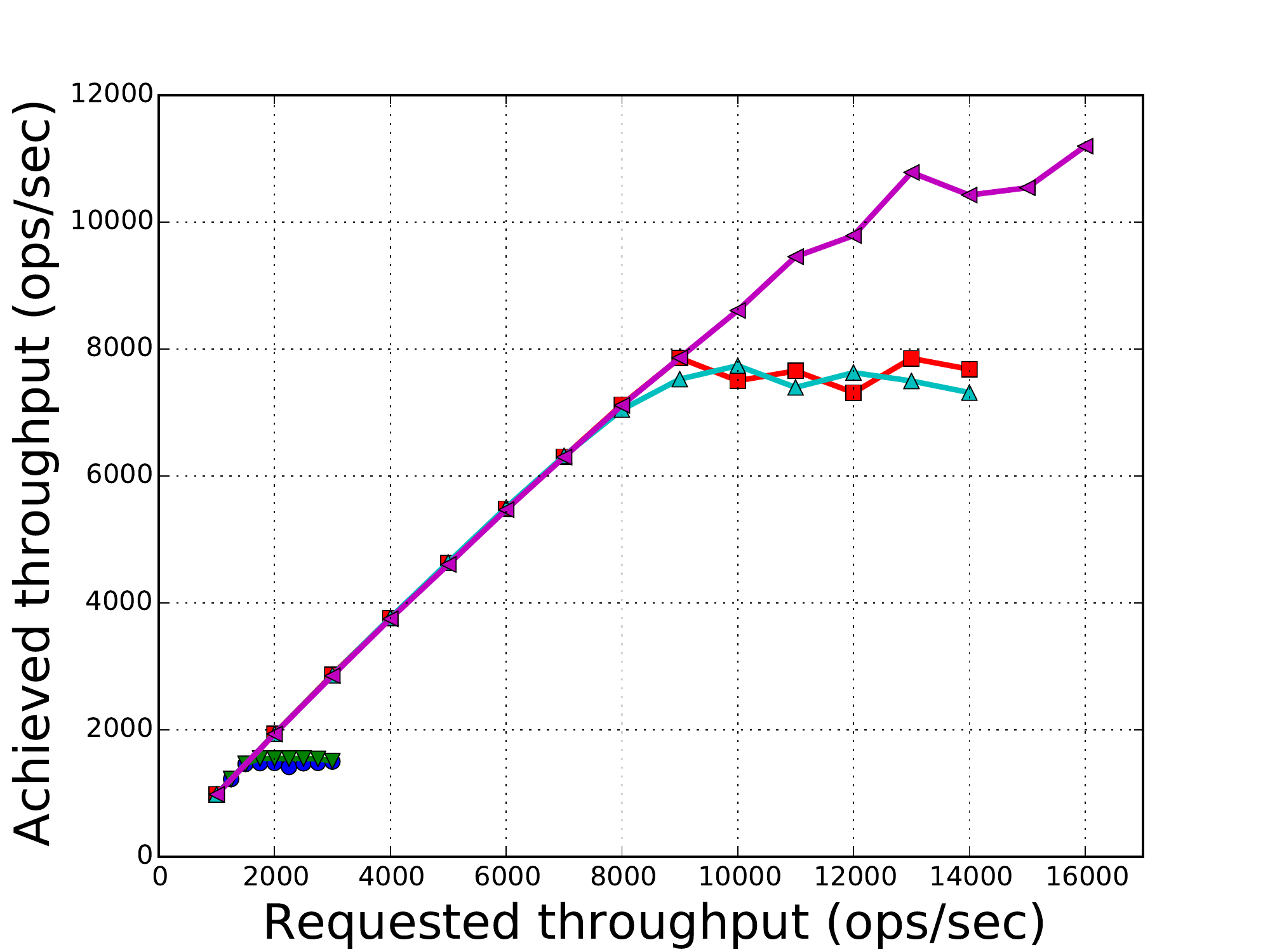}
	\end{subfigure}
	\begin{subfigure}{.331\textwidth}
		\centering
		\includegraphics[width=\linewidth]{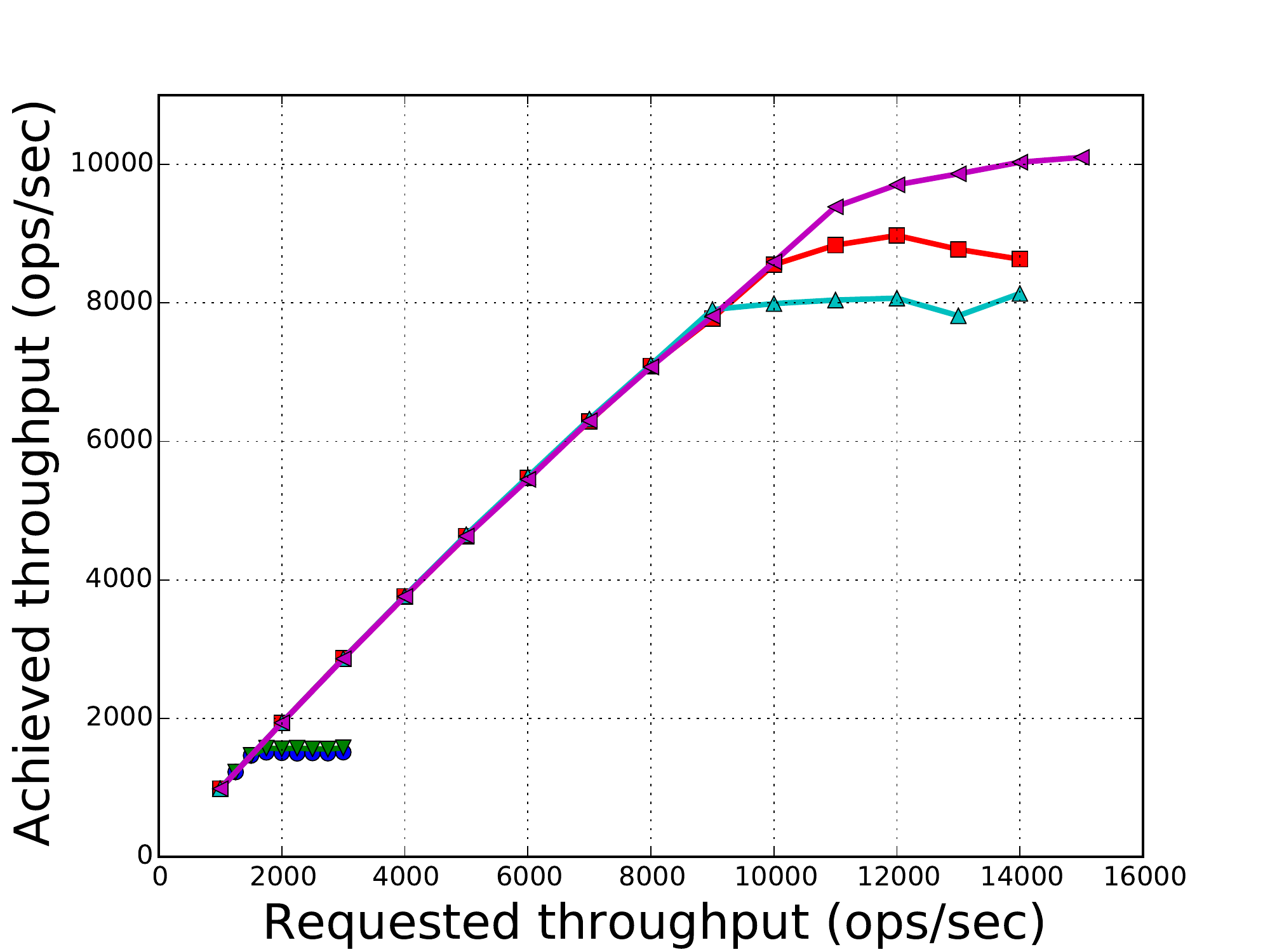}
	\end{subfigure} \\
	\begin{subfigure}{.331\textwidth}
		\centering
		\includegraphics[width=\linewidth]{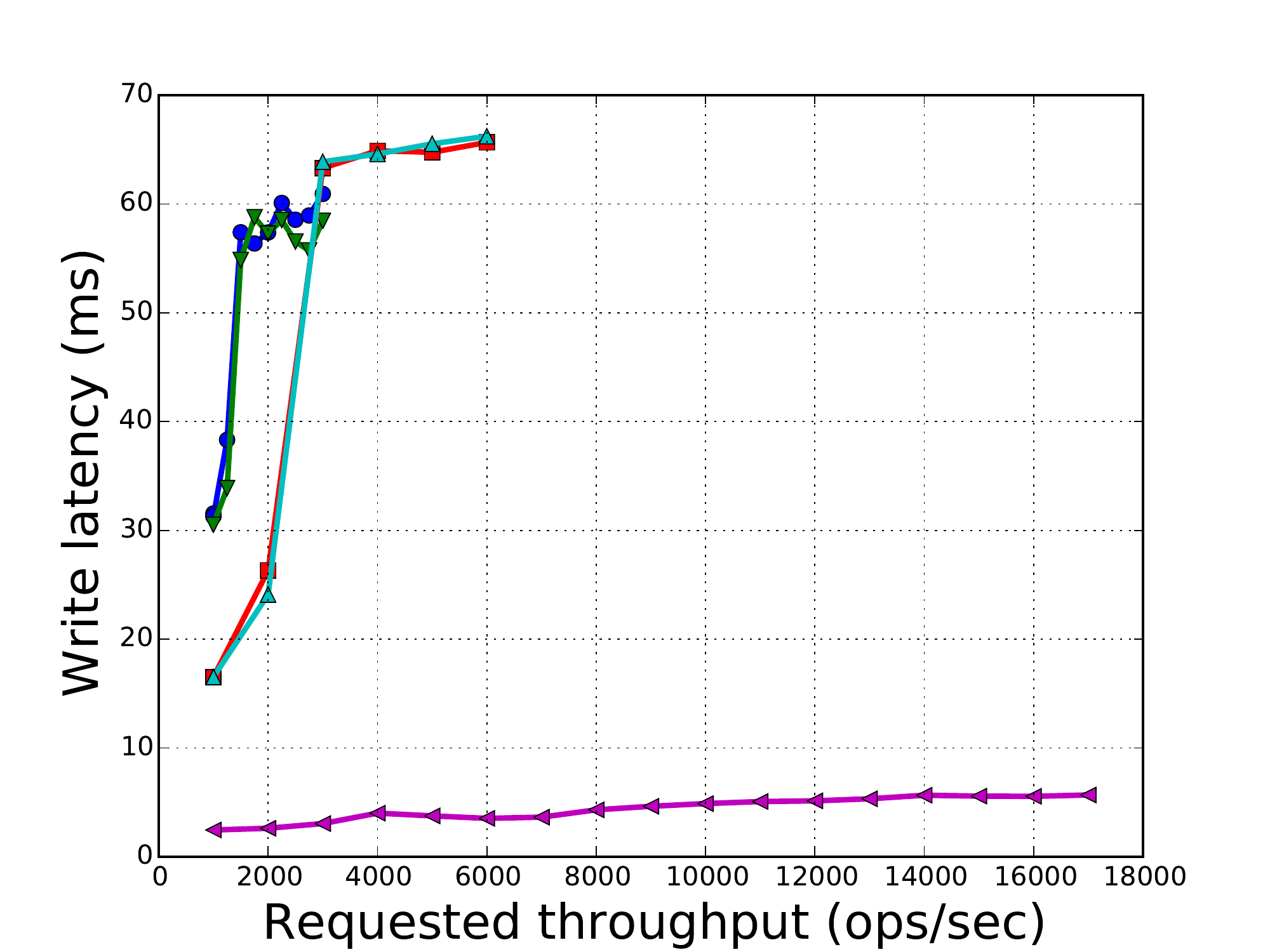}
	\end{subfigure}%
	\begin{subfigure}{.331\textwidth}
		\centering
		\includegraphics[width=\linewidth]{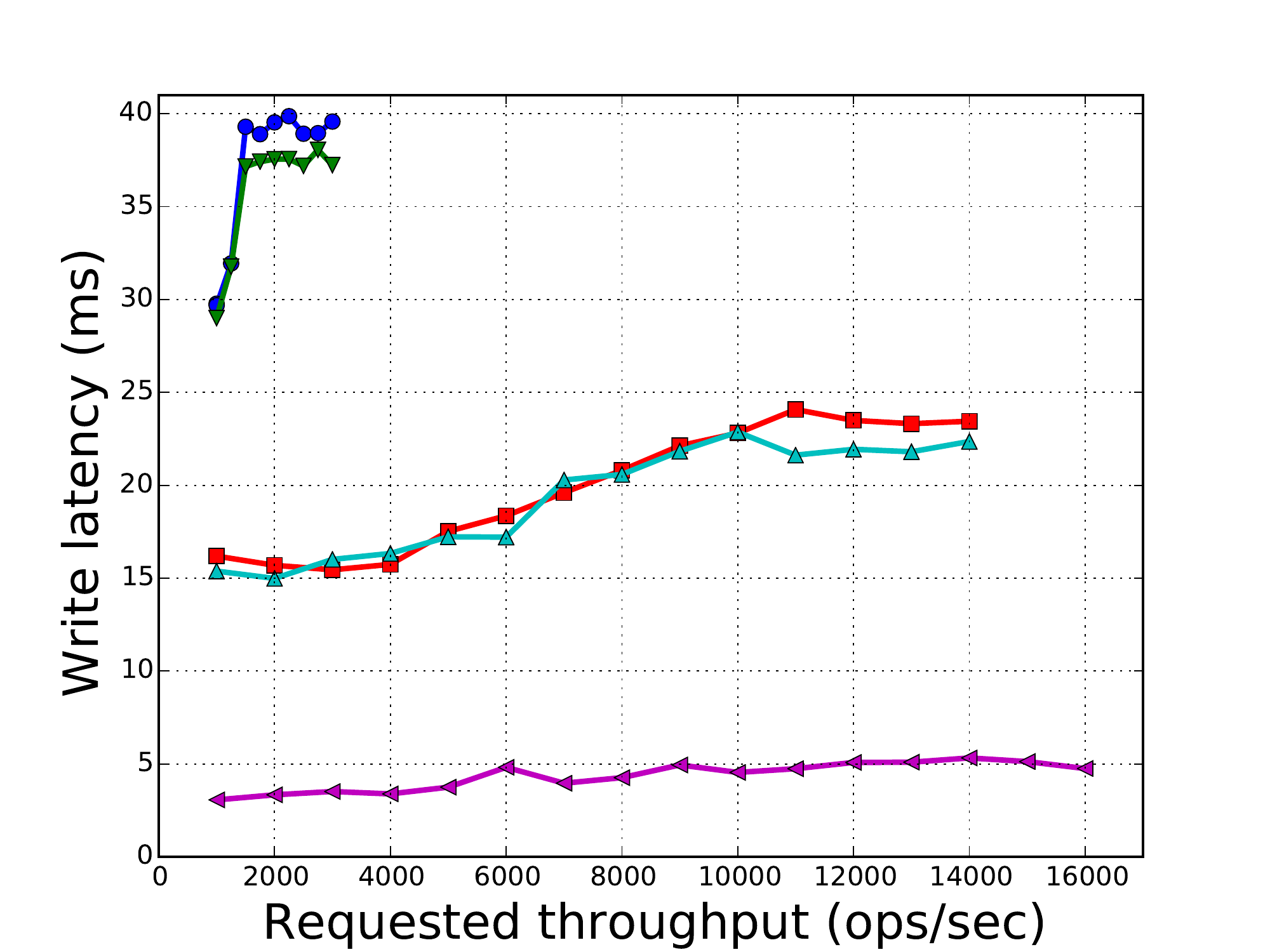}
	\end{subfigure} \\
	\begin{subfigure}{.331\textwidth}
		\centering
		\includegraphics[width=\linewidth]{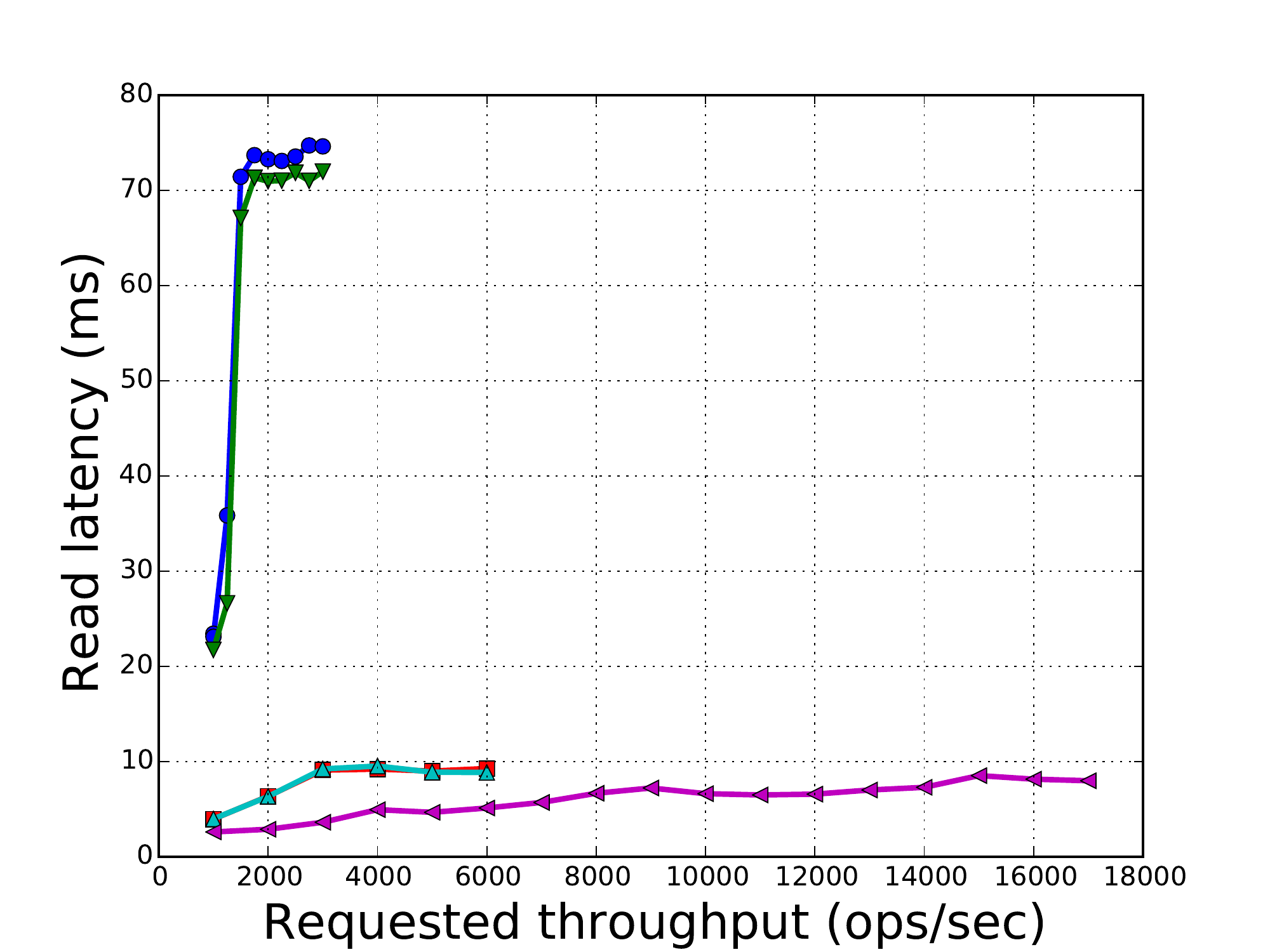}
		\caption{Workload A}
	\end{subfigure}%
	\begin{subfigure}{.331\textwidth}
		\centering
		\includegraphics[width=\linewidth]{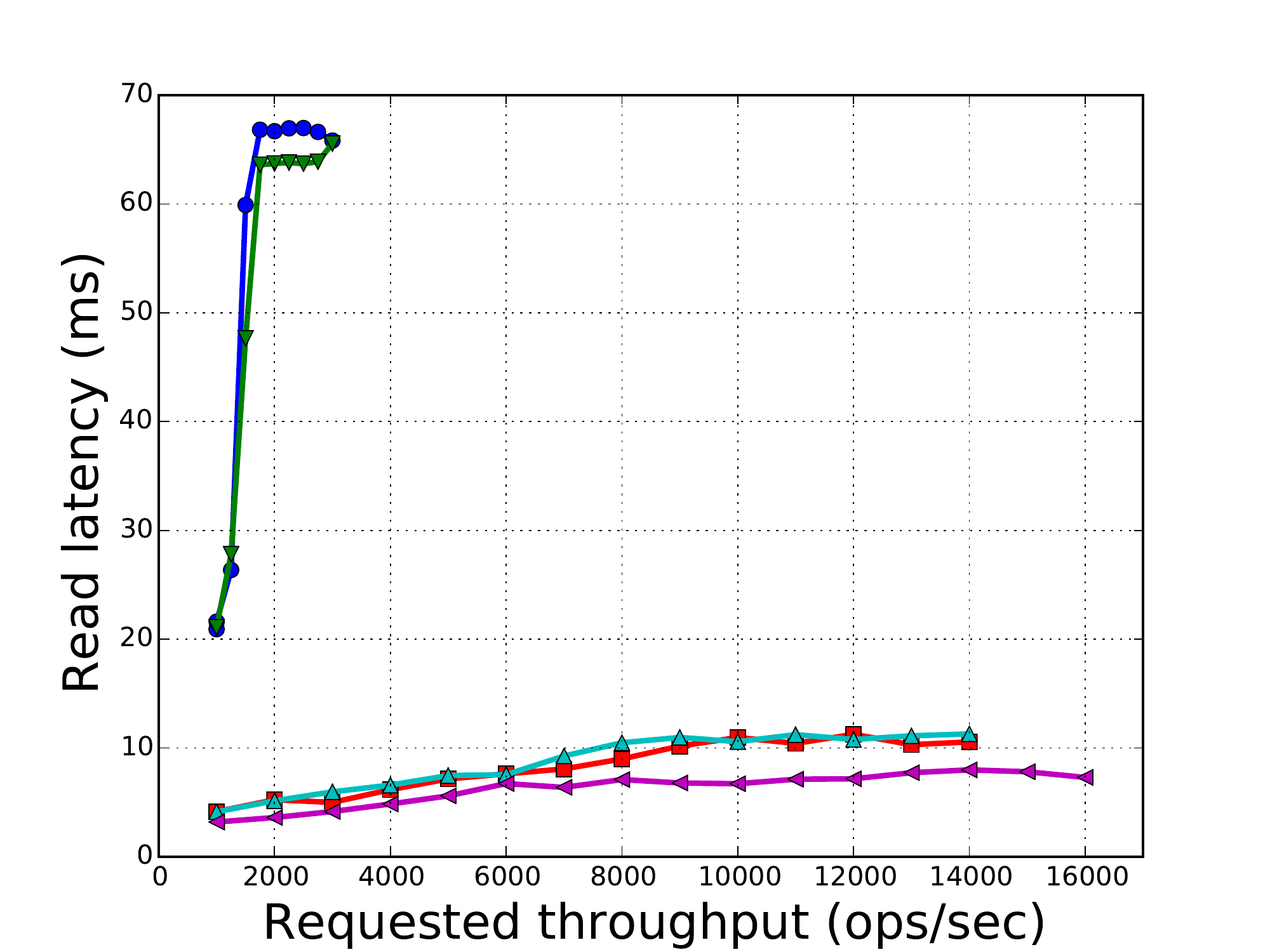}
		\caption{Workload B}
	\end{subfigure}
	\begin{subfigure}{.331\textwidth}
		\centering
		\includegraphics[width=\linewidth]{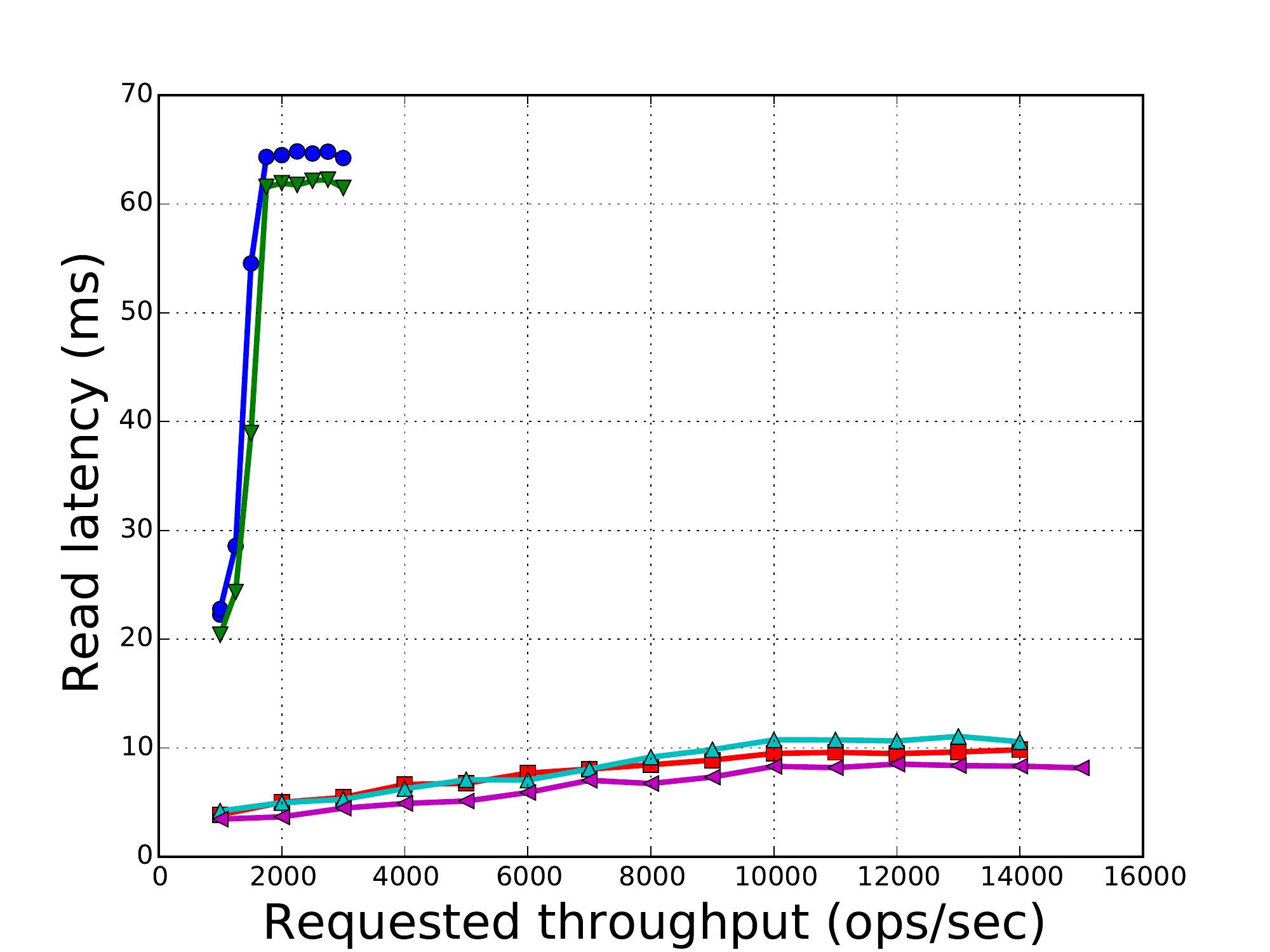}
		\caption{Workload C}
	\end{subfigure}
	\caption{Focusing on the hardened variants only - finer scale than Figure \ref{fig:one-a_to_c}.}
	\label{fig:two-a_to_c}
\end{figure*}

\begin{figure*}[t]
\begin{center}
	\begin{subfigure}{.331\textwidth}
		\centering
		\includegraphics[width=\linewidth]{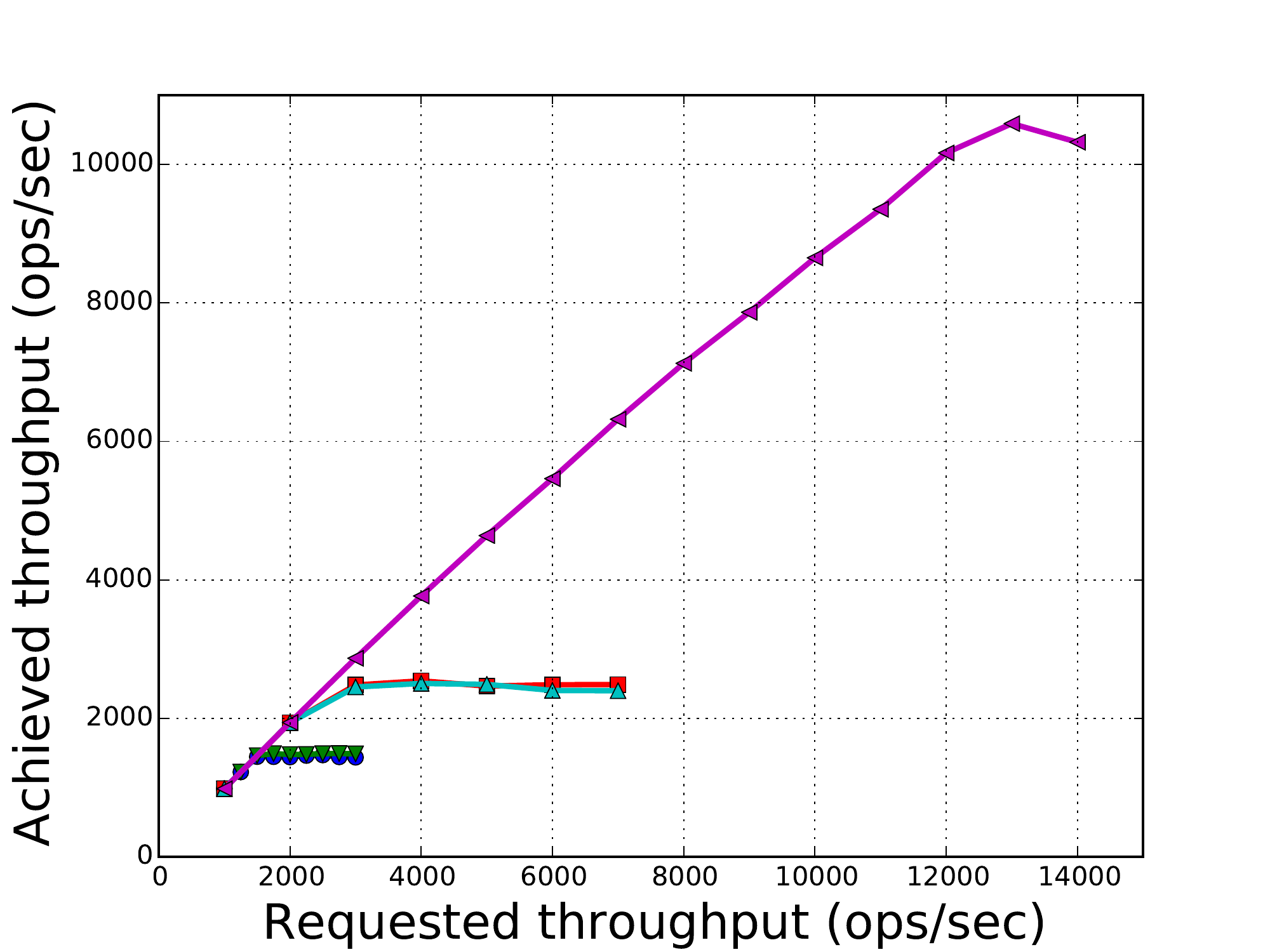}
	\end{subfigure}%
	\begin{subfigure}{.331\textwidth}
		\centering
		\includegraphics[width=\linewidth]{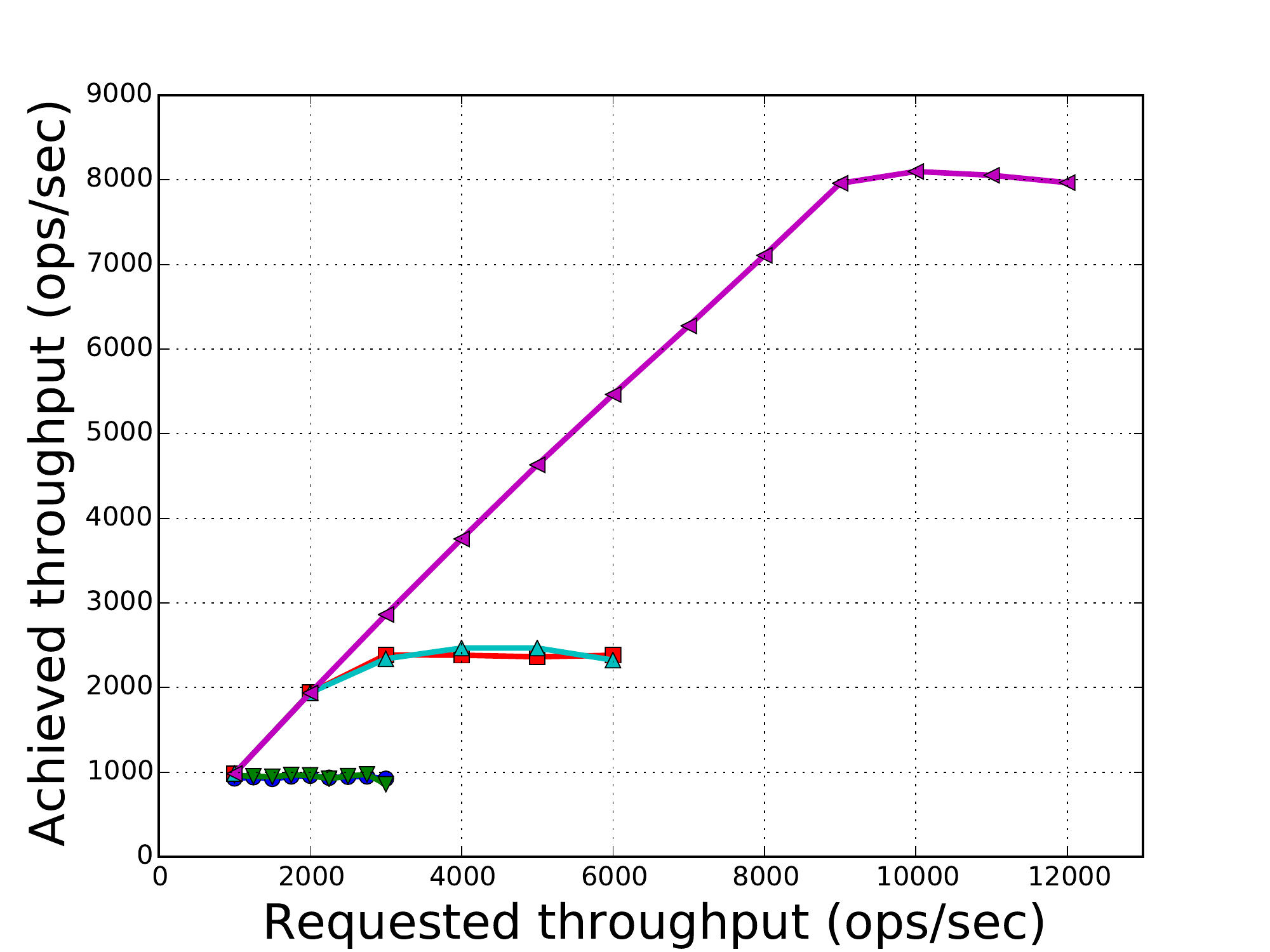}
	\end{subfigure} \\
	\begin{subfigure}{.331\textwidth}
		\centering
		\includegraphics[width=\linewidth]{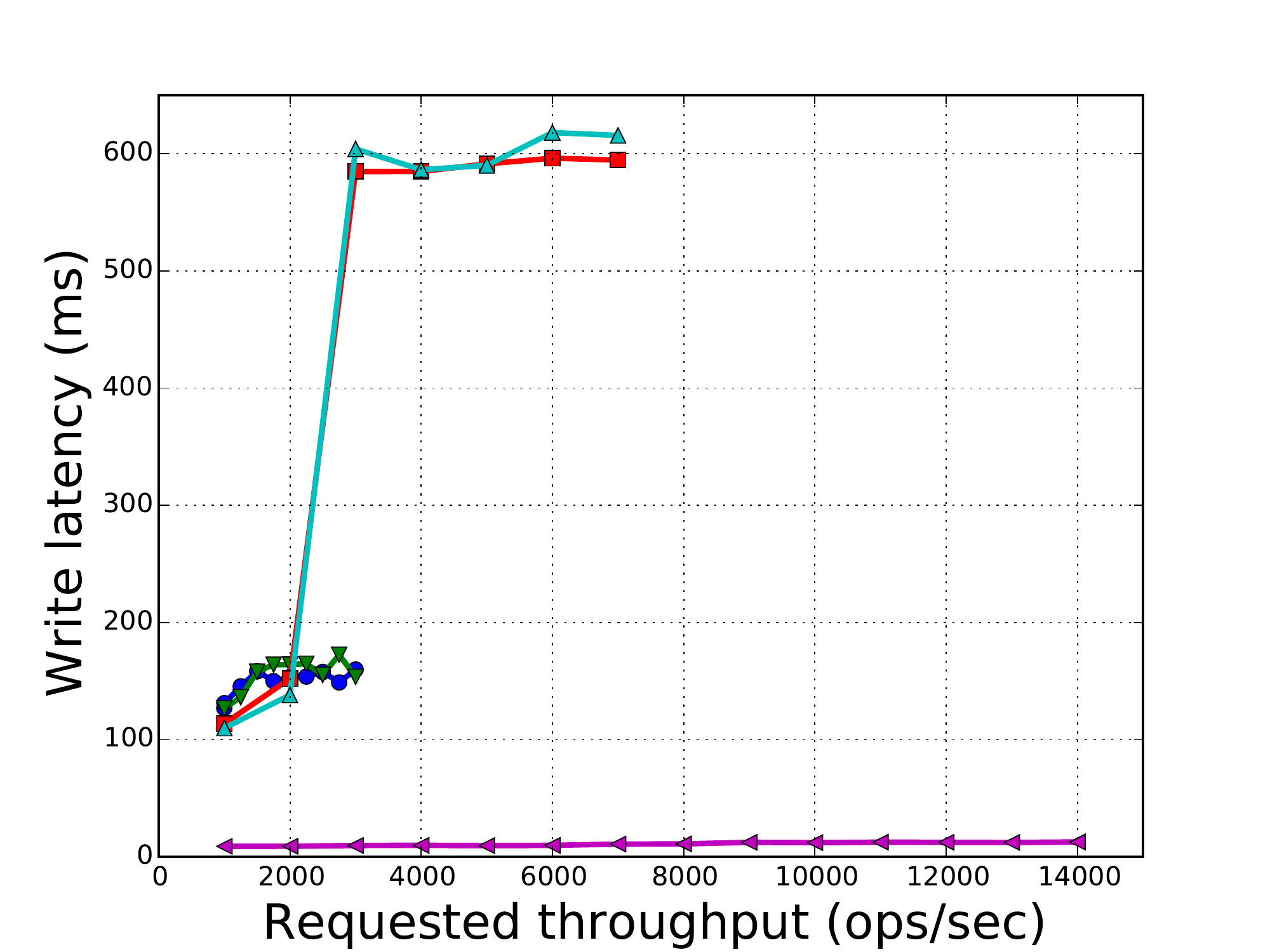}
	\end{subfigure}%
	\begin{subfigure}{.331\textwidth}
		\centering
		\includegraphics[width=\linewidth]{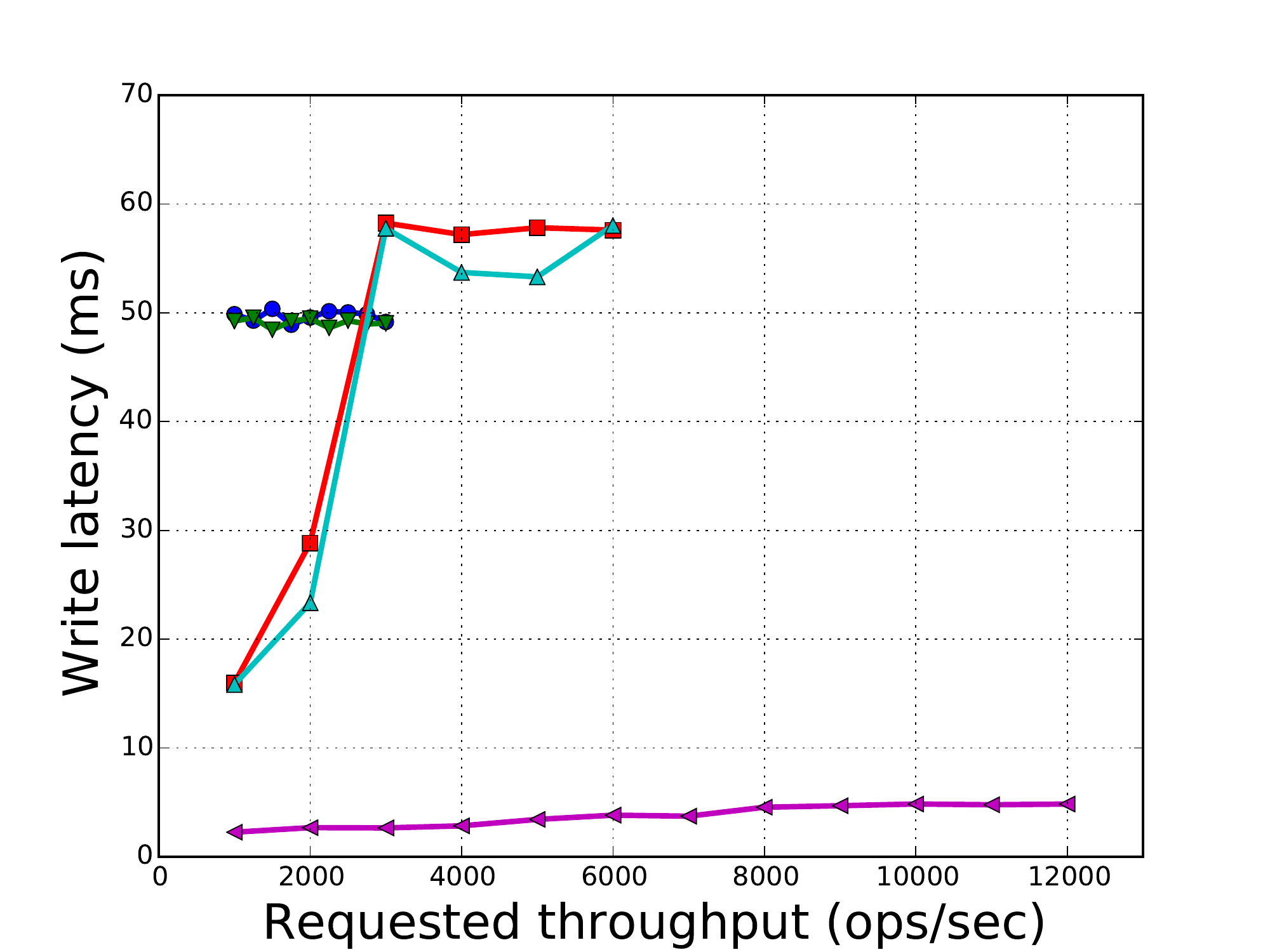}
	\end{subfigure} \\
	\begin{subfigure}{.331\textwidth}
		\centering
		\includegraphics[width=\linewidth]{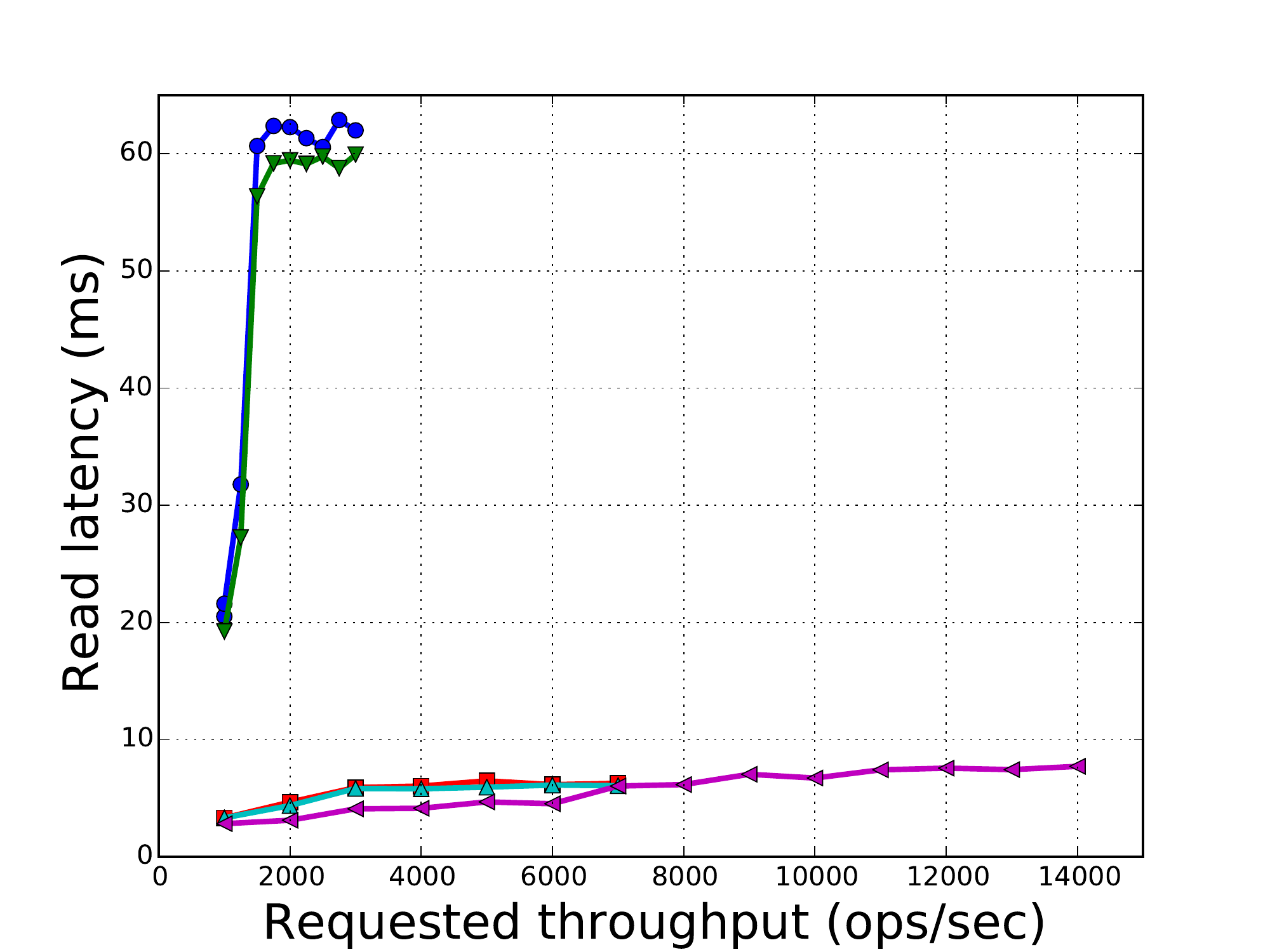}
		\caption{Workload D}
	\end{subfigure}%
	\begin{subfigure}{.331\textwidth}
		\centering
		\includegraphics[width=\linewidth]{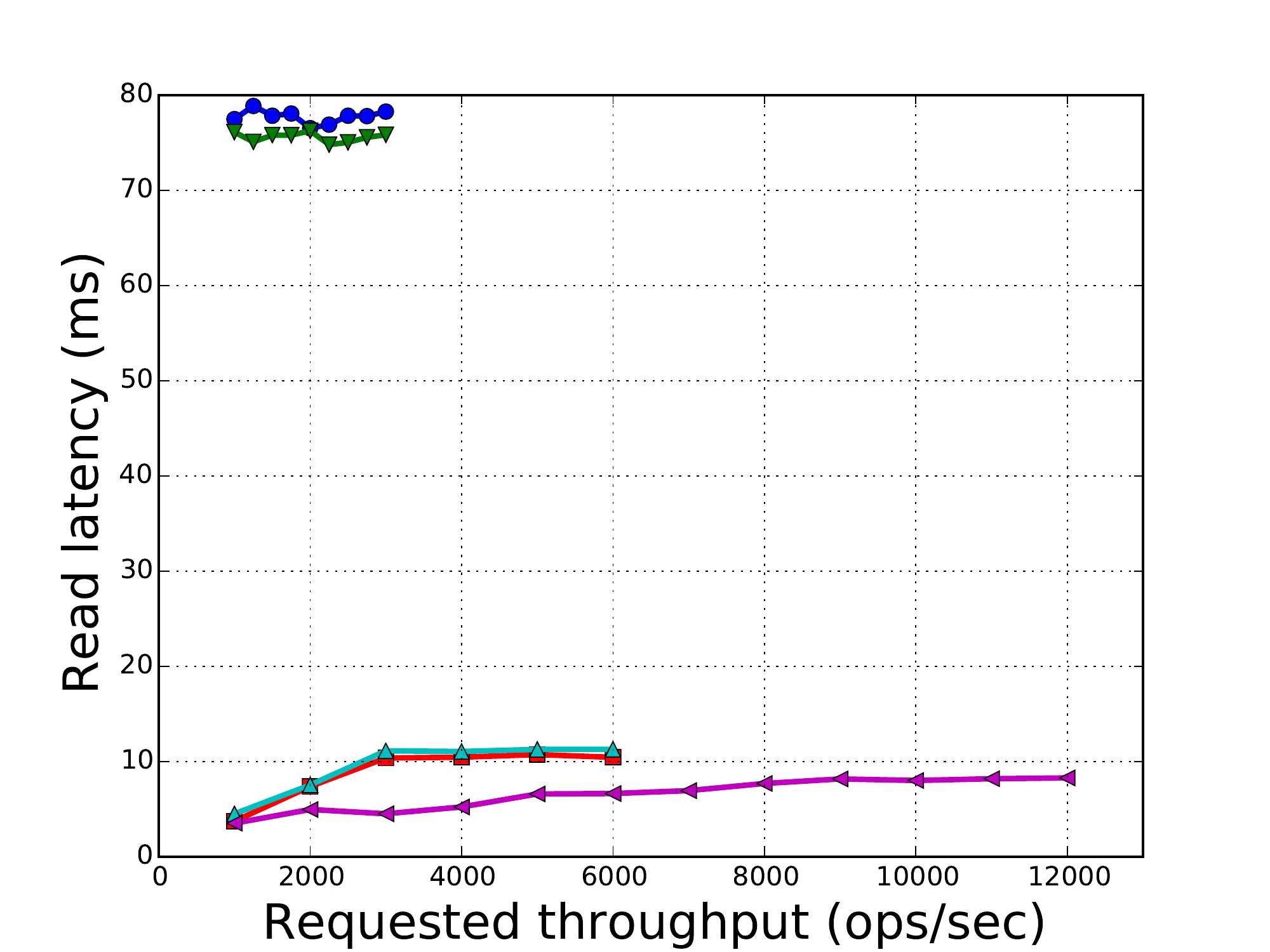}
		\caption{Workload F}
	\end{subfigure}%
\end{center}
	\caption{Same as Figure \ref{fig:two-a_to_c}~(hardened variants only), but with workloads D and F.}
	\label{fig:two-d_to_f}
\end{figure*}

\begin{figure*}[t]
	\begin{subfigure}{\textwidth}
		\centering
		\includegraphics[width=0.6\linewidth]{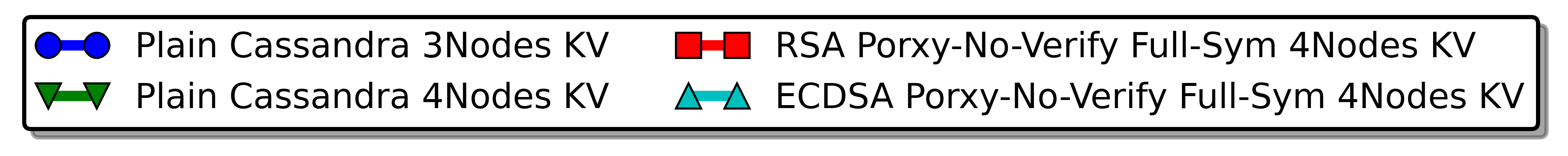}
	\end{subfigure}
	\\
	\begin{subfigure}{.331\textwidth}
		\centering
		\includegraphics[width=\linewidth]{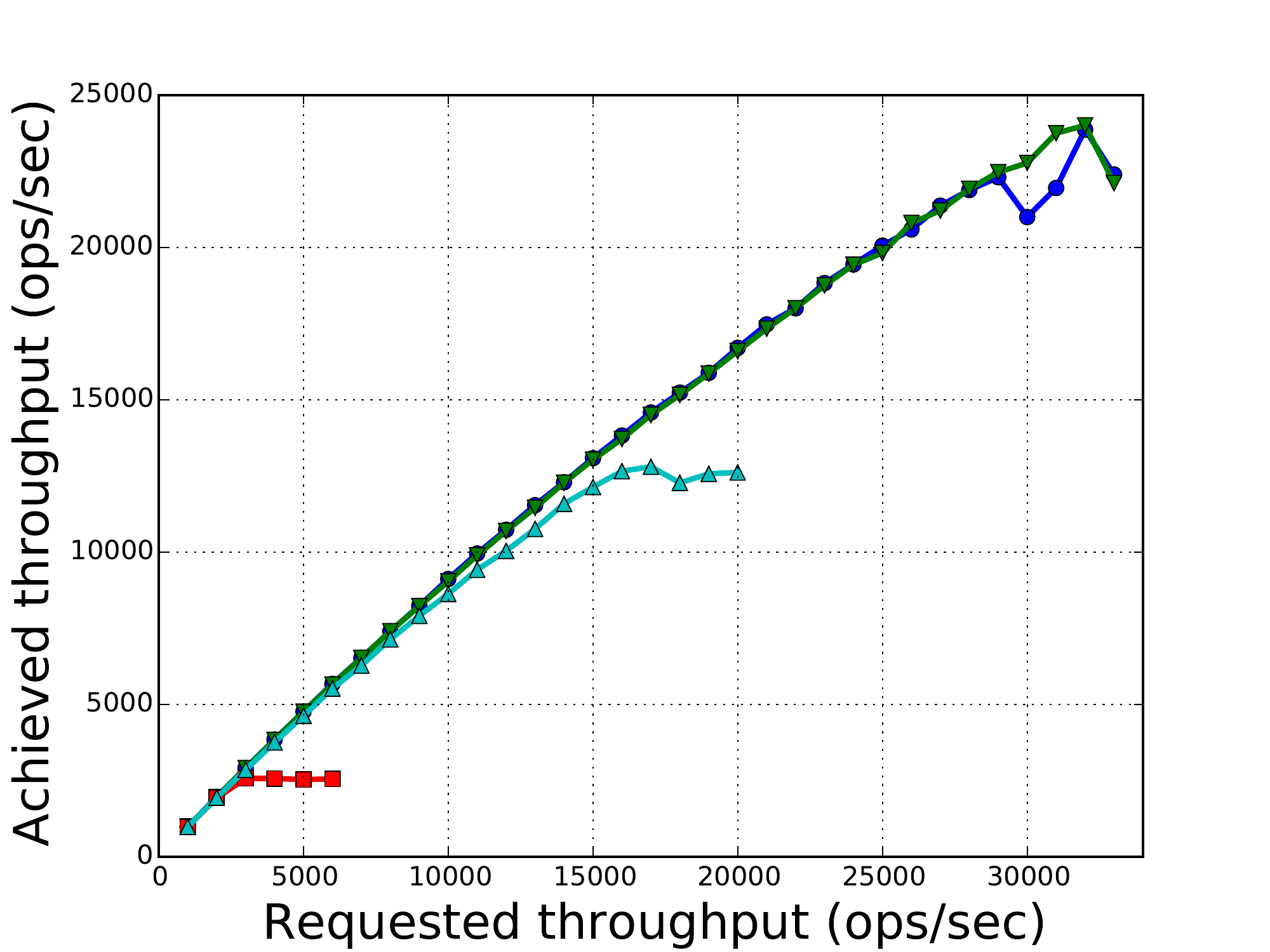}
	\end{subfigure}%
	\begin{subfigure}{.331\textwidth}
		\centering
		\includegraphics[width=\linewidth]{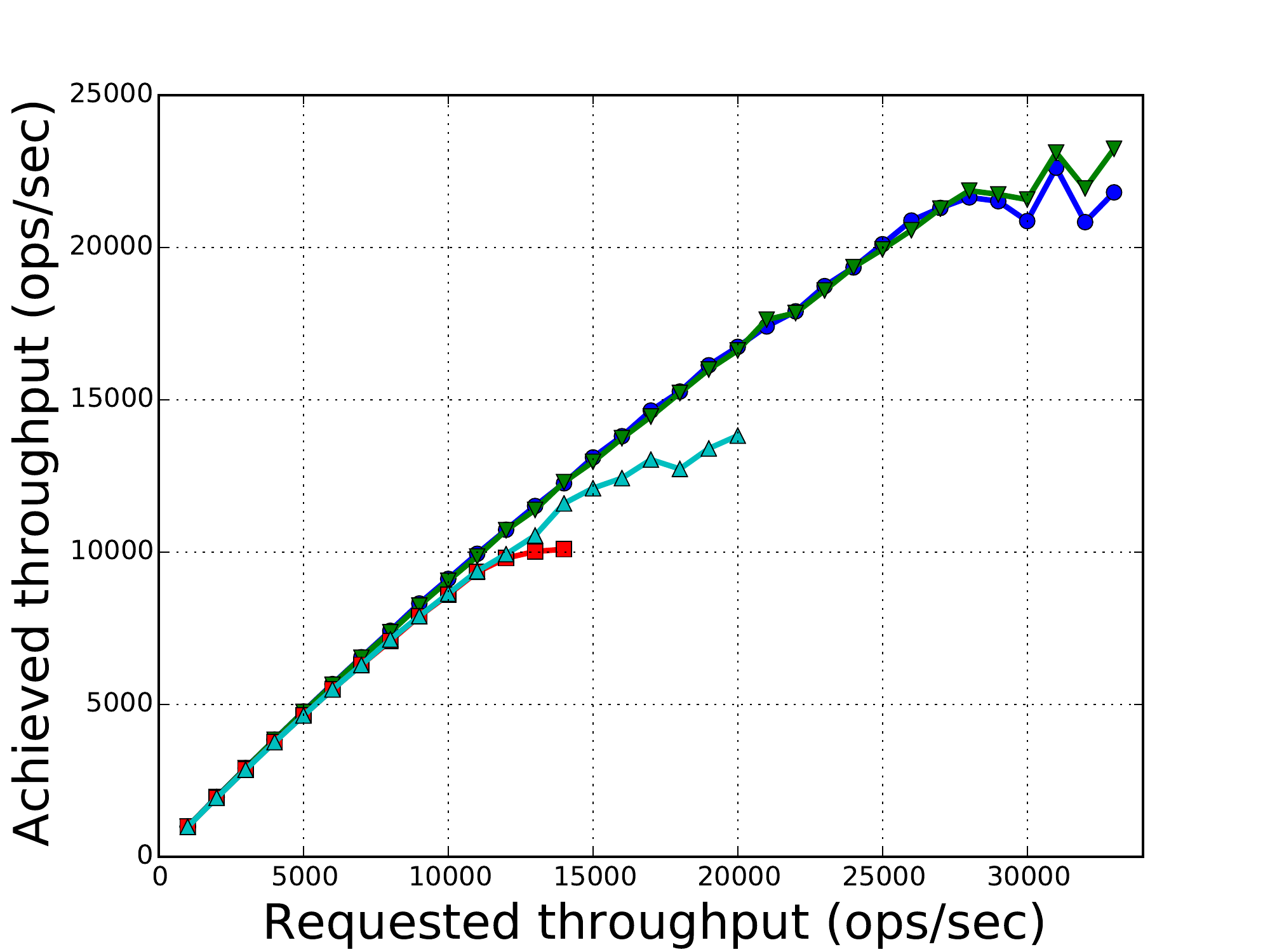}
	\end{subfigure}
	\begin{subfigure}{.331\textwidth}
		\centering
		\includegraphics[width=\linewidth]{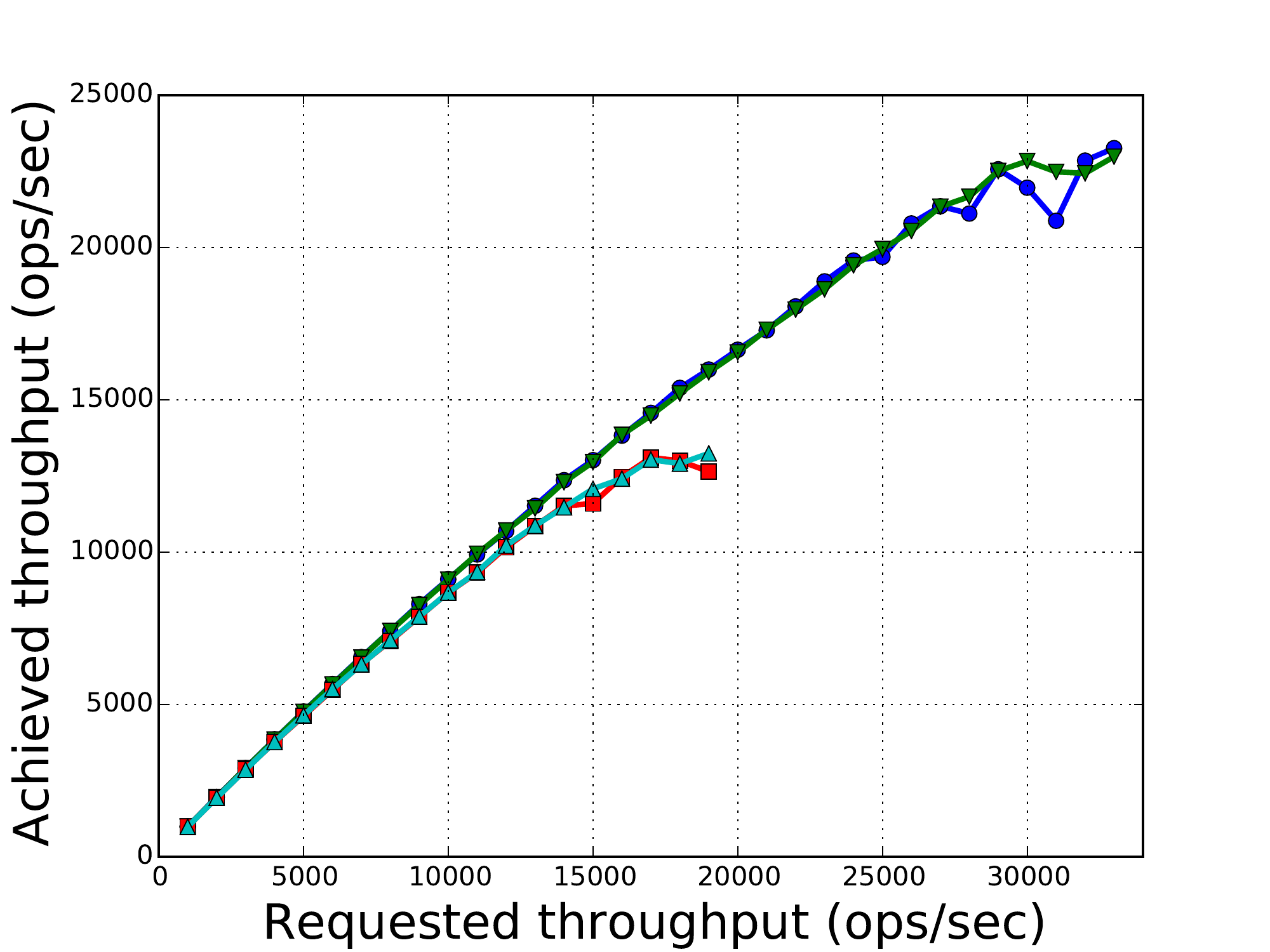}
	\end{subfigure} \\
	\begin{subfigure}{.331\textwidth}
		\centering
		\includegraphics[width=\linewidth]{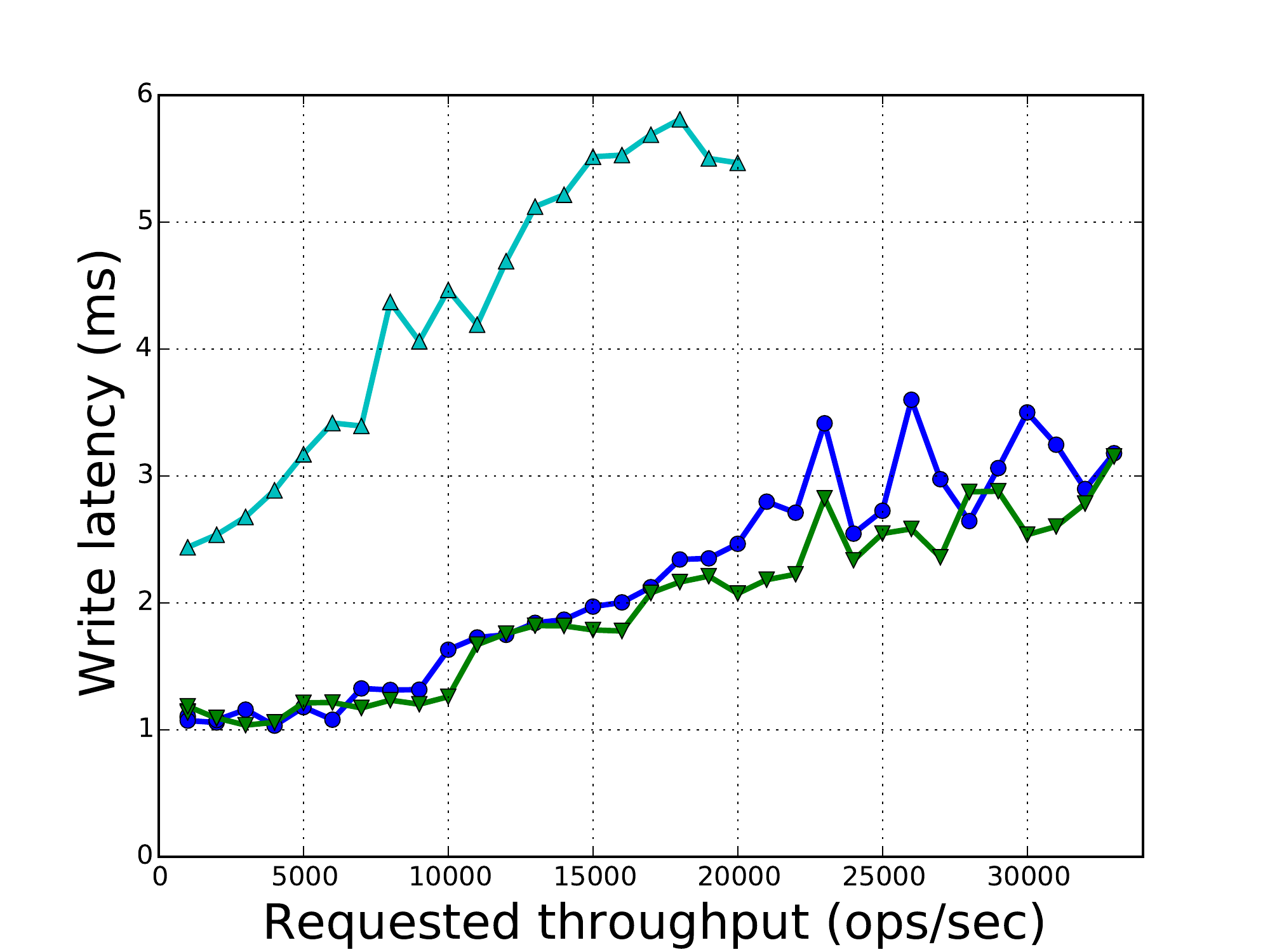}
	\end{subfigure}%
	\begin{subfigure}{.331\textwidth}
		\centering
		\includegraphics[width=\linewidth]{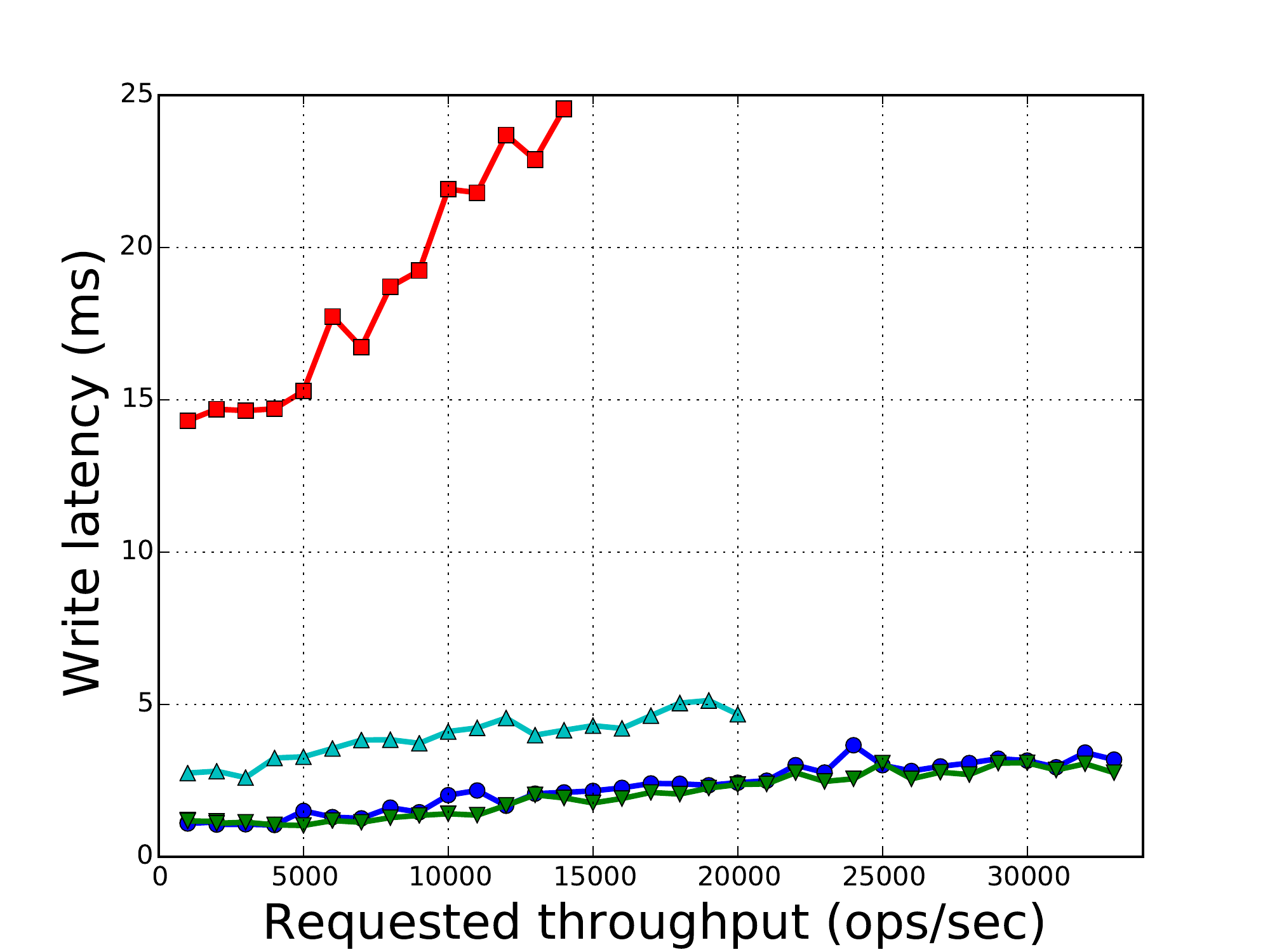}
	\end{subfigure} \\
	\begin{subfigure}{.331\textwidth}
		\centering
		\includegraphics[width=\linewidth]{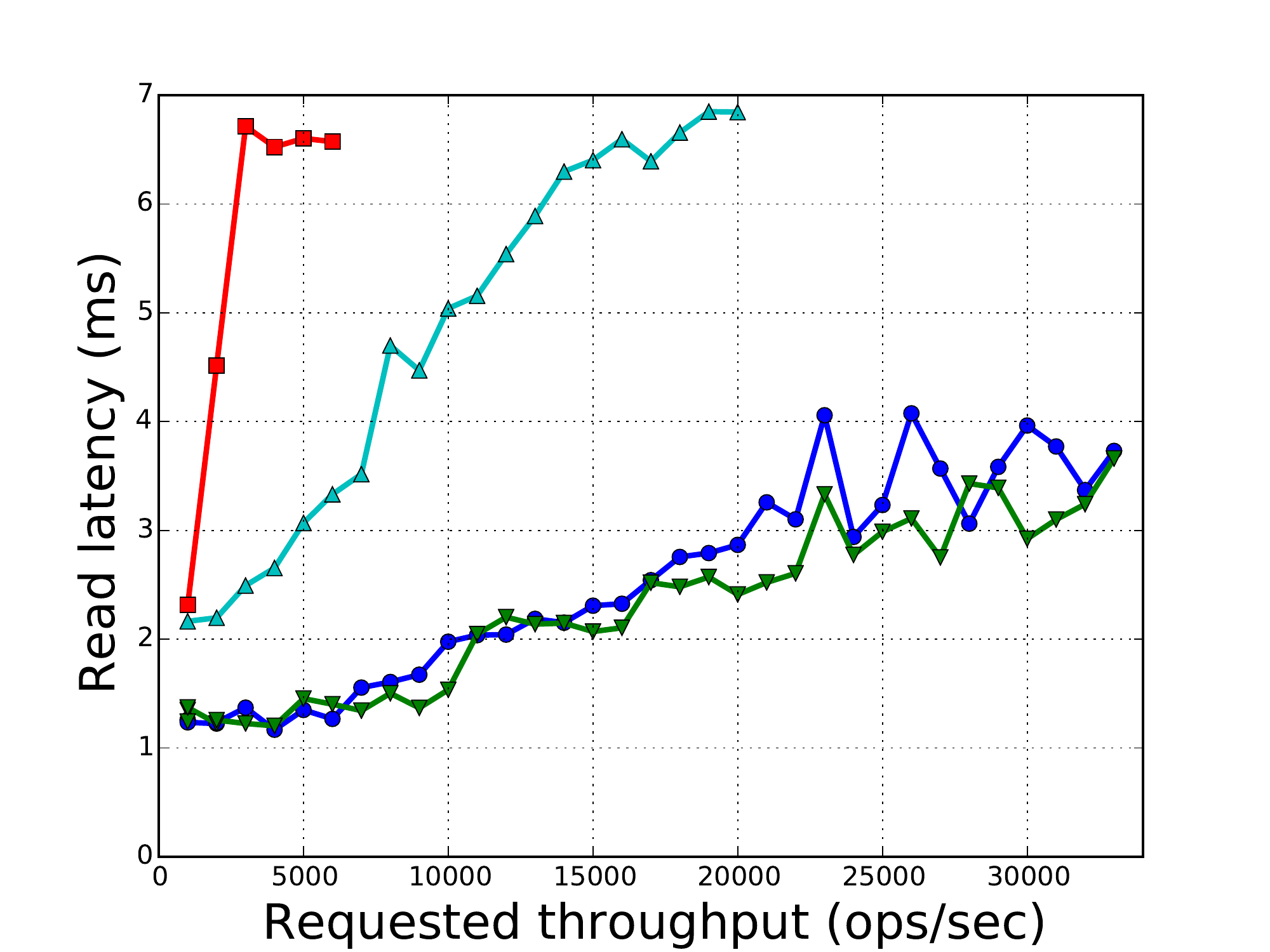}
		\caption{Workload A}
	\end{subfigure}%
	\begin{subfigure}{.331\textwidth}
		\centering
		\includegraphics[width=\linewidth]{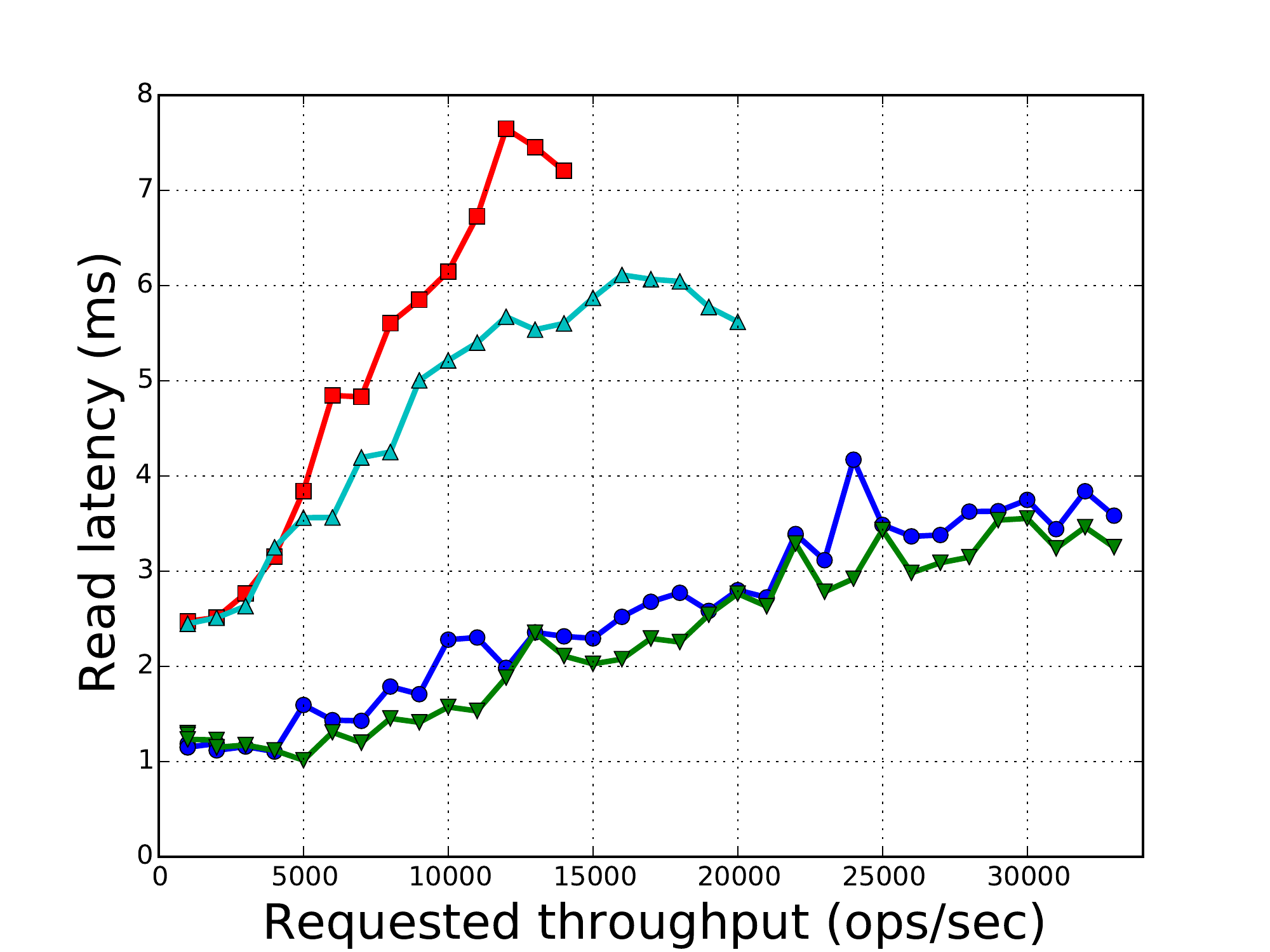}
		\caption{Workload B}
	\end{subfigure}
	\begin{subfigure}{.331\textwidth}
		\centering
		\includegraphics[width=\linewidth]{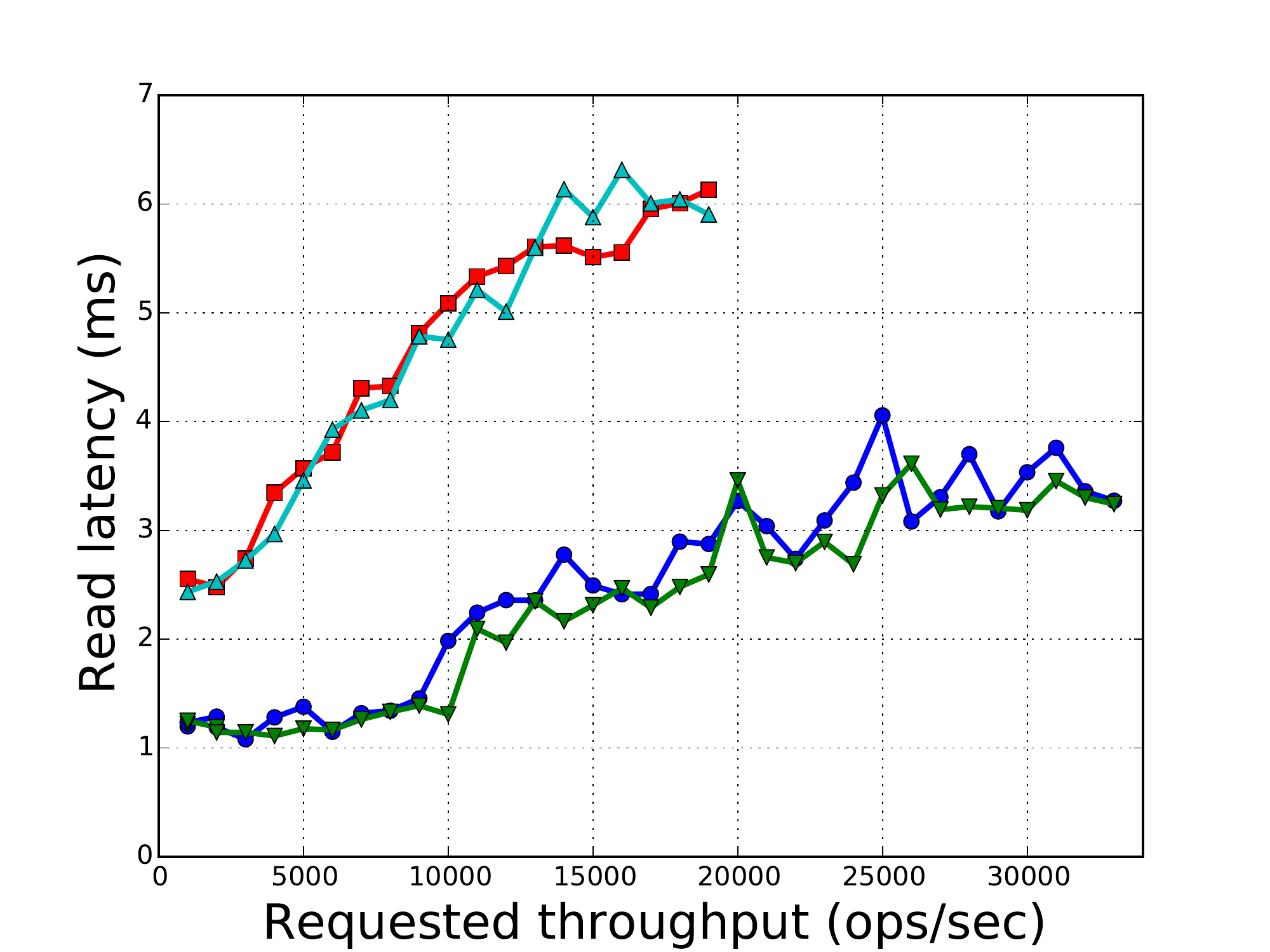}
		\caption{Workload C}
	\end{subfigure}
	\caption[Comparing the best variants using a key-value model.]{Comparing the best variants using a key-value model. In the write latency of sub-figure (a), we left the RSA variants out as they rapidly grew to around 65ms latency.}
	\label{fig:kv-a_to_c}
\end{figure*}

\begin{figure*}[t]
\begin{center}
	\begin{subfigure}{.331\textwidth}
		\centering
		\includegraphics[width=\linewidth]{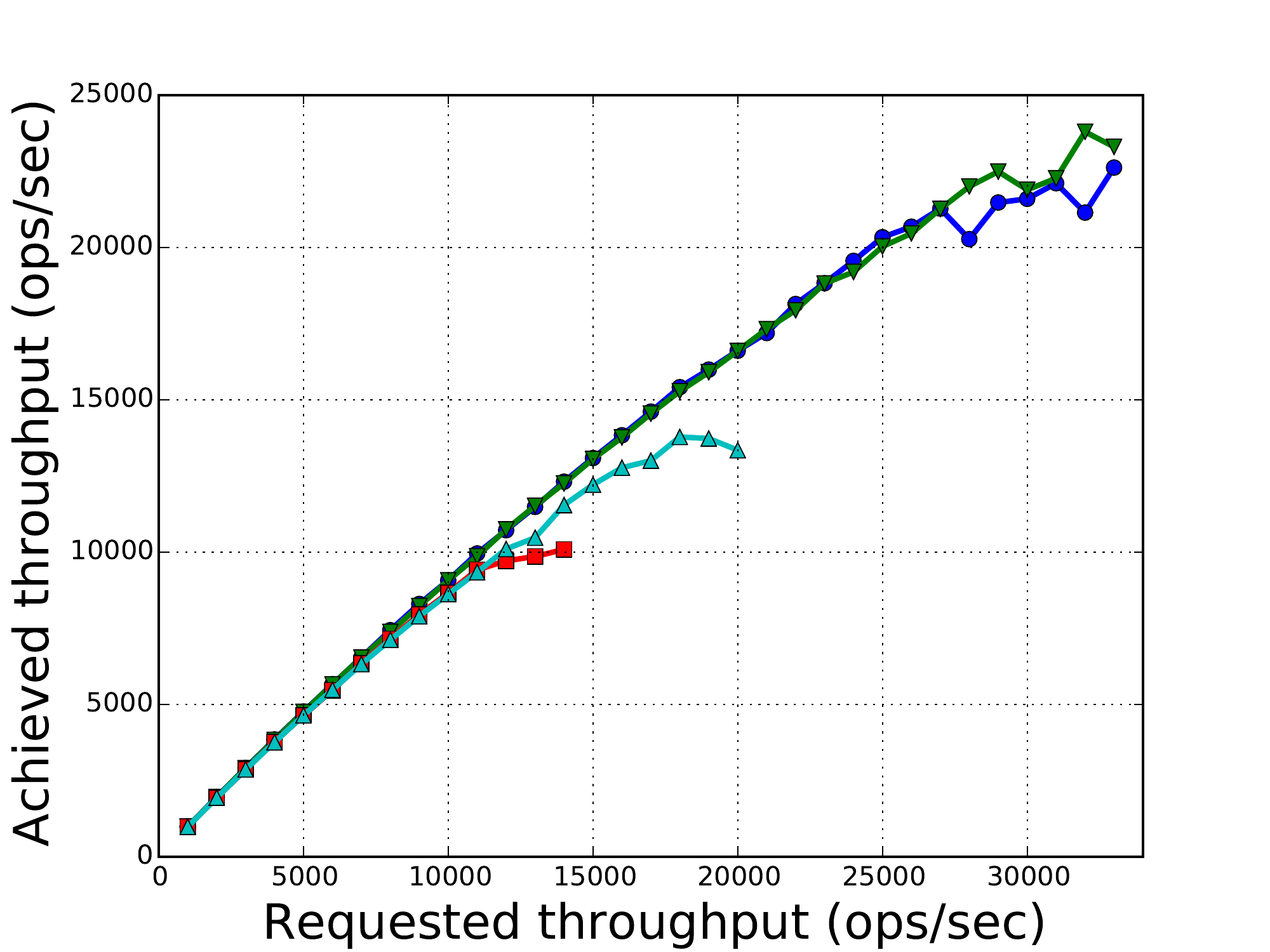}
	\end{subfigure}%
	\begin{subfigure}{.331\textwidth}
		\centering
		\includegraphics[width=\linewidth]{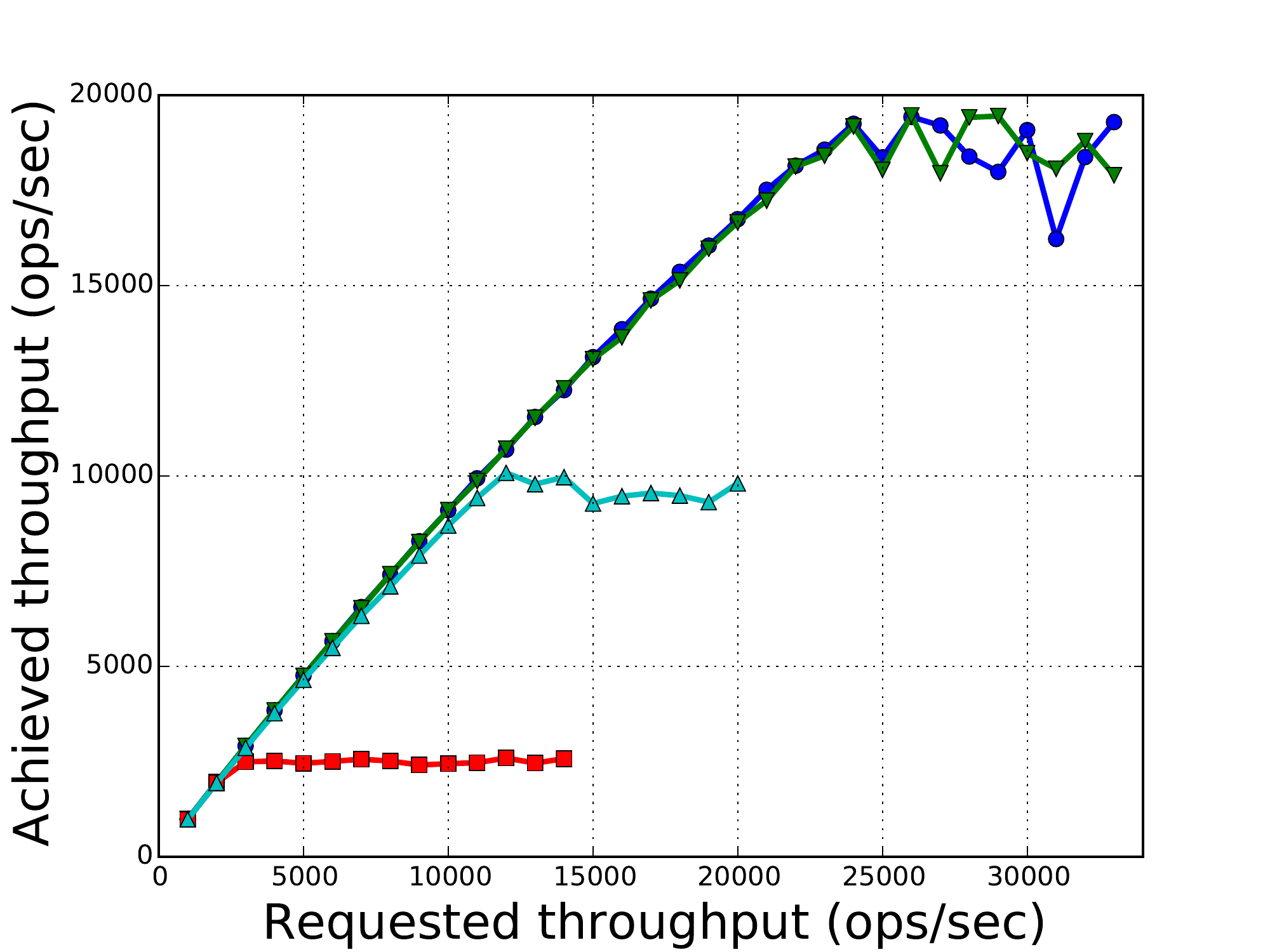}
		%\label{fig:sfig2}
	\end{subfigure} \\
	\begin{subfigure}{.331\textwidth}
		\centering
		\includegraphics[width=\linewidth]{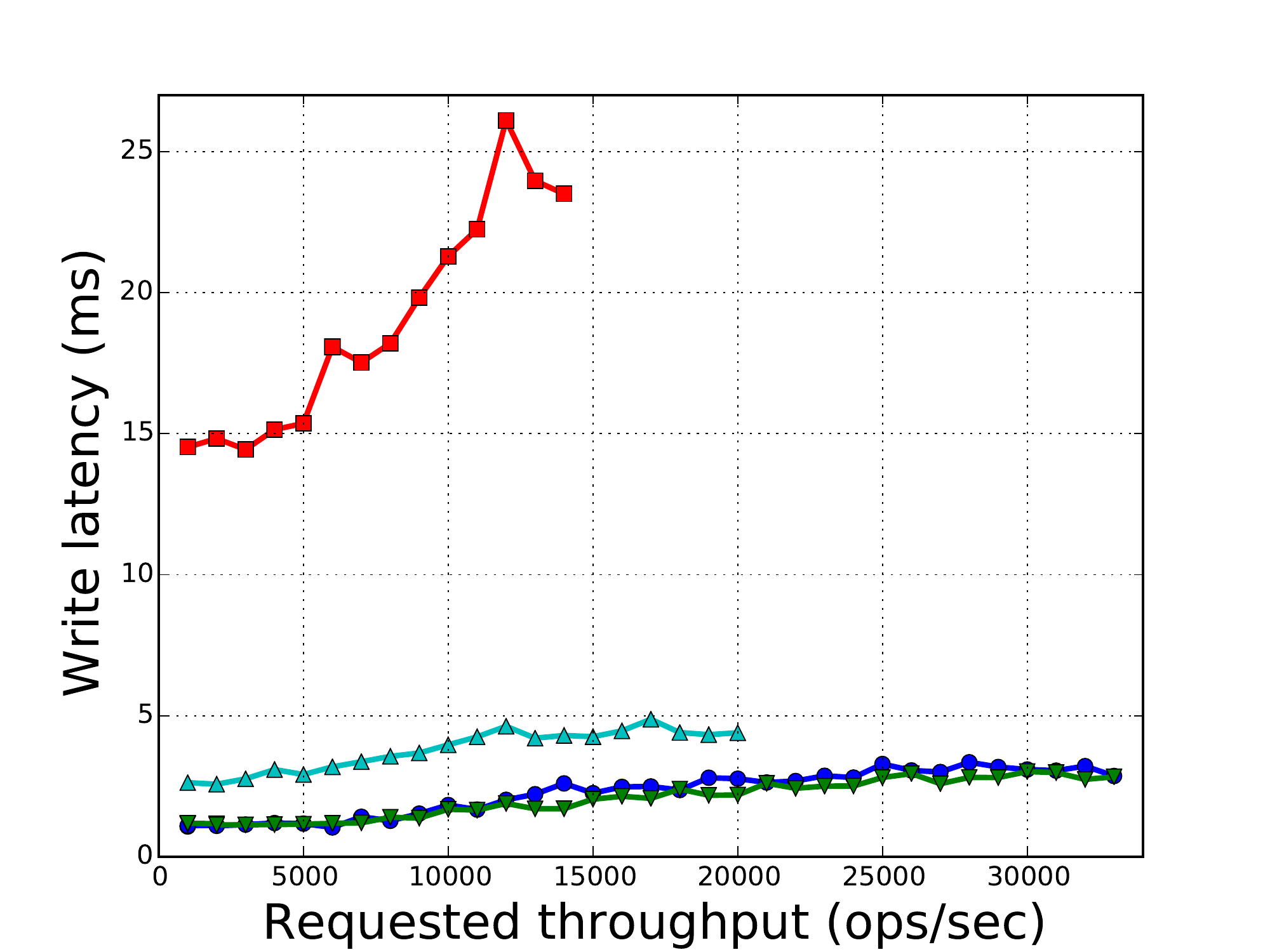}
	\end{subfigure}%
	\begin{subfigure}{.331\textwidth}
		\centering
		\includegraphics[width=\linewidth]{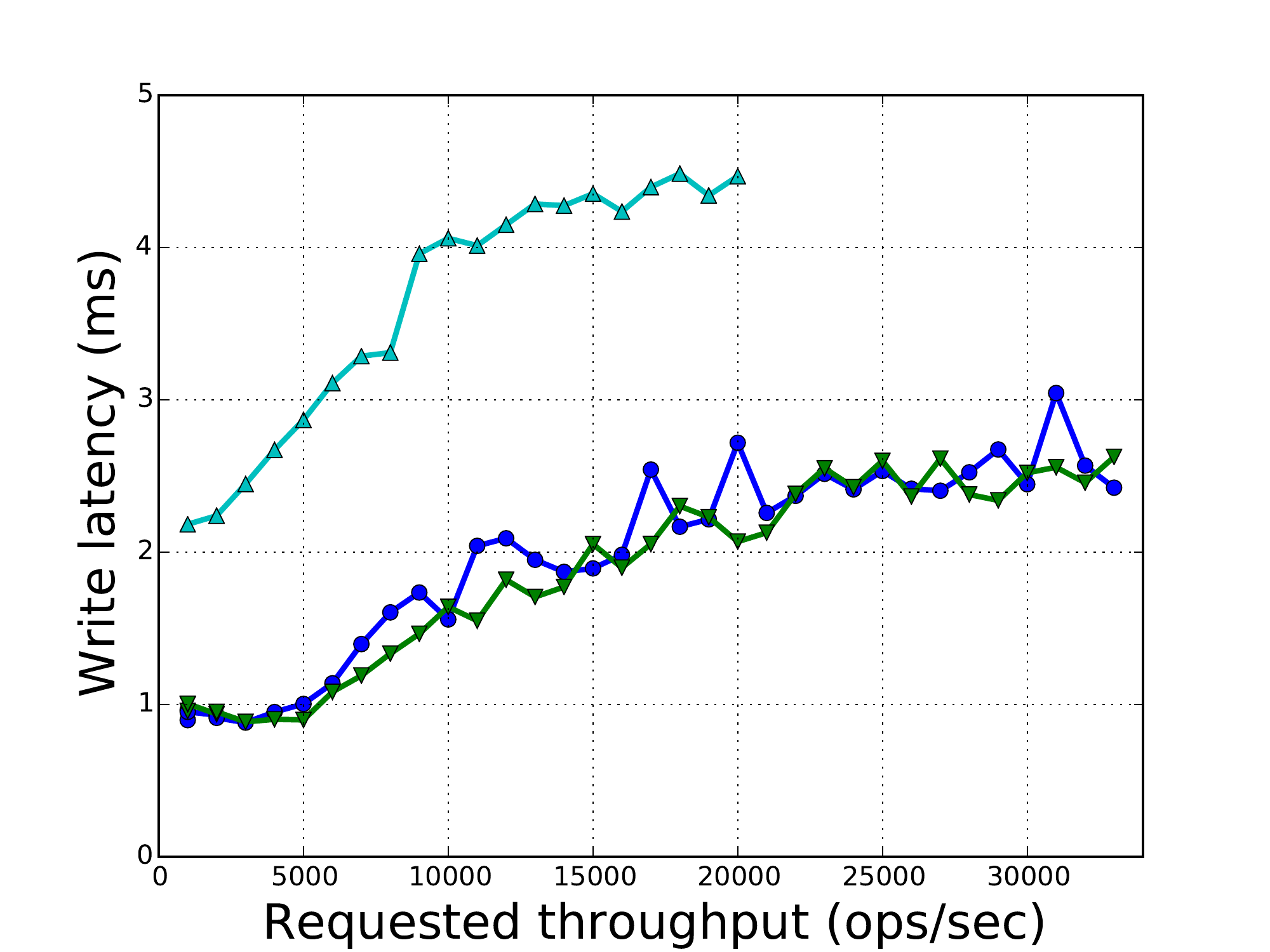}
	\end{subfigure} \\
	\begin{subfigure}{.331\textwidth}
		\centering
		\includegraphics[width=\linewidth]{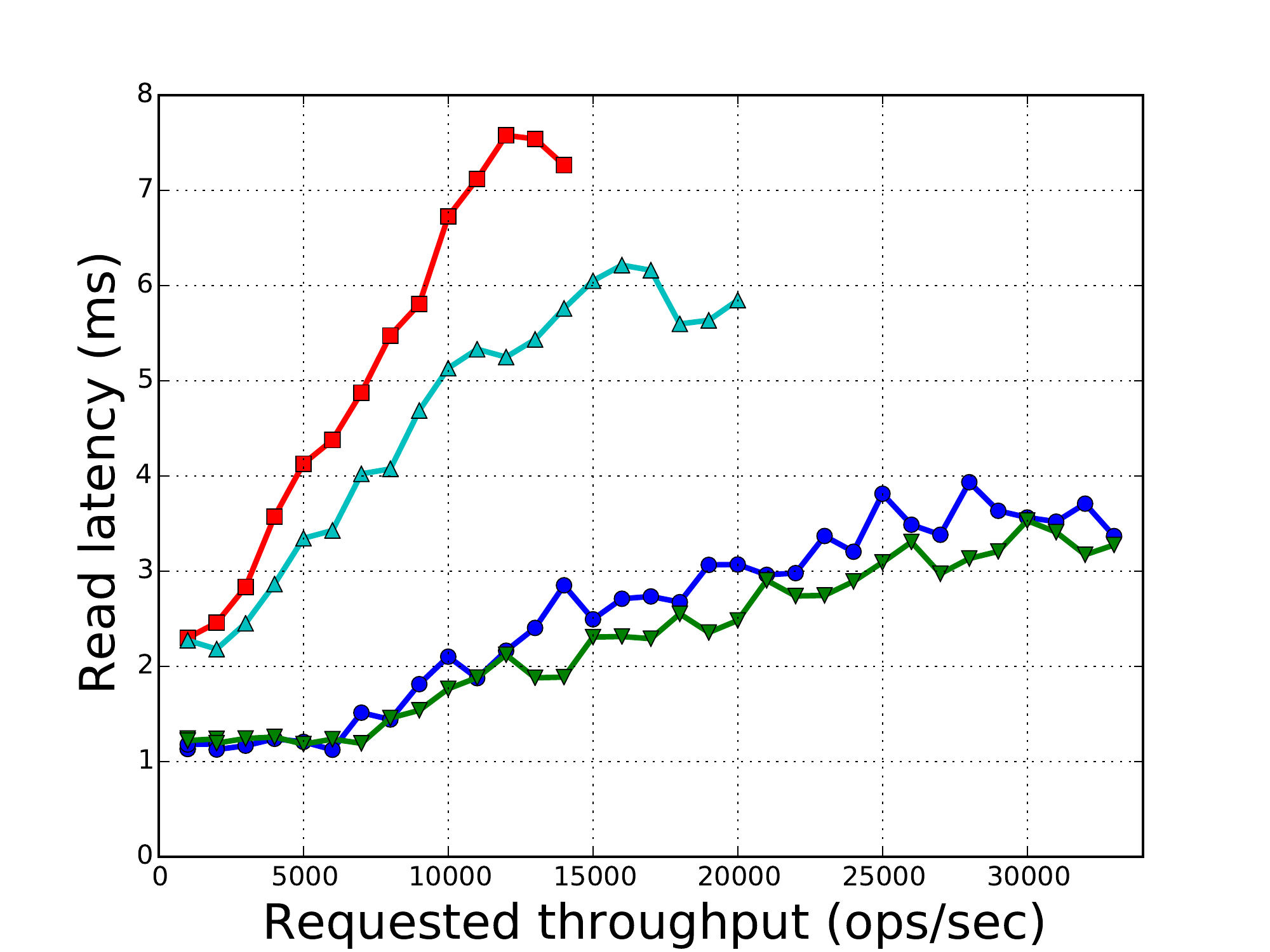}
		\caption{Workload D}
	\end{subfigure}%
	\begin{subfigure}{.331\textwidth}
		\centering
		\includegraphics[width=\linewidth]{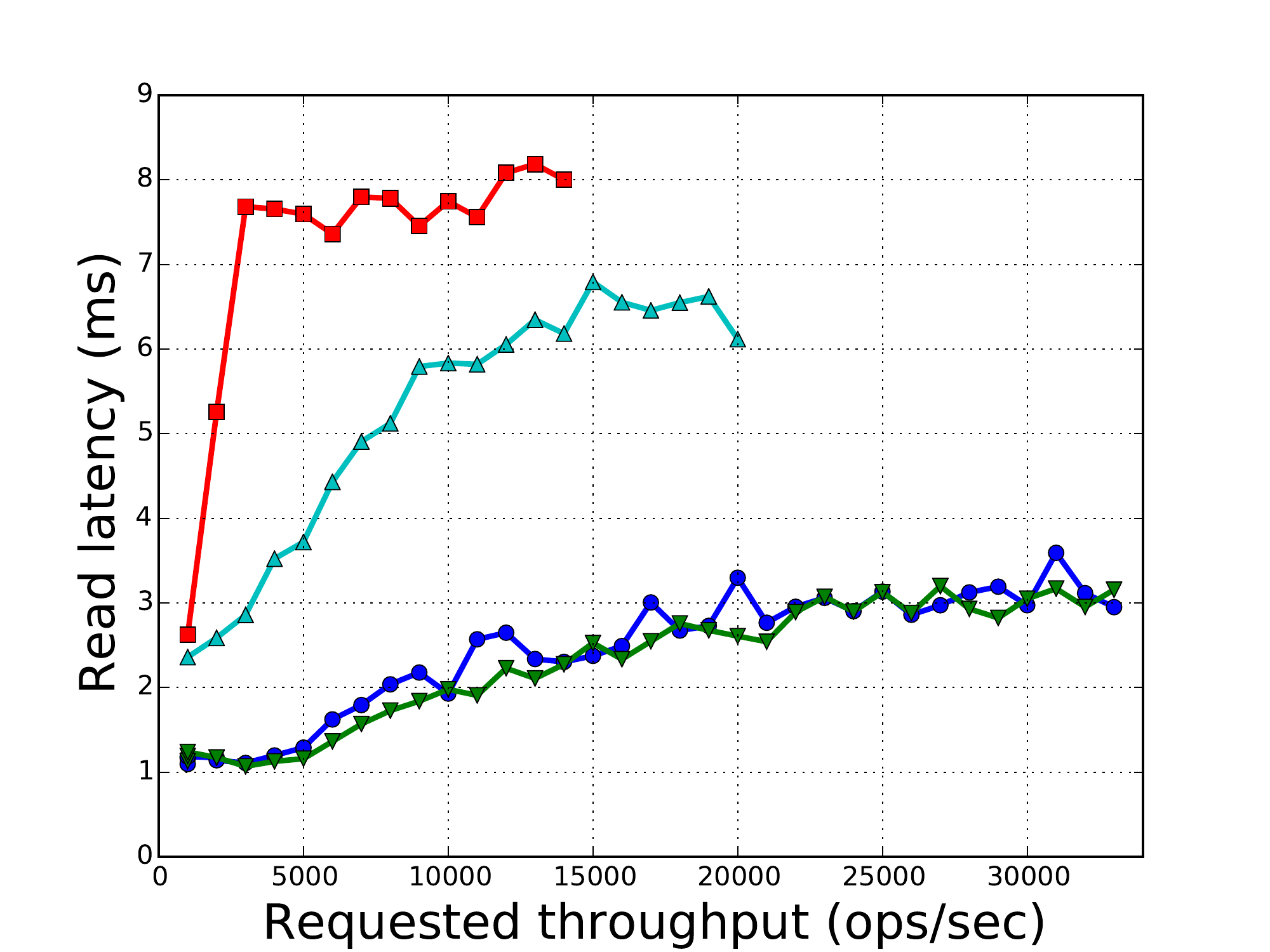}
		\caption{Workload F}
	\end{subfigure}%
\end{center}
	\caption{Same as Figure \ref{fig:kv-a_to_c} (a key-value model), but using workloads D and F.}
	\label{fig:kv-d_to_f}
\end{figure*}

\begin{figure*}[t]
		\begin{subfigure}{\textwidth}
			\centering
			\includegraphics[width=0.6\linewidth]{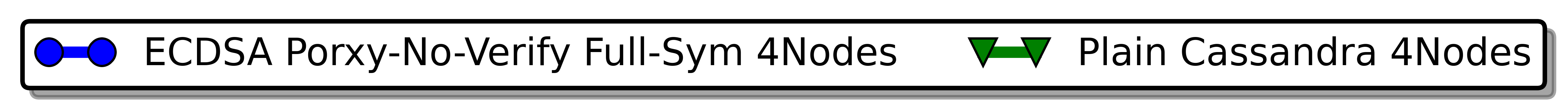}
		\end{subfigure}
	\begin{subfigure}{.331\textwidth}
		\centering
		\includegraphics[width=\linewidth]{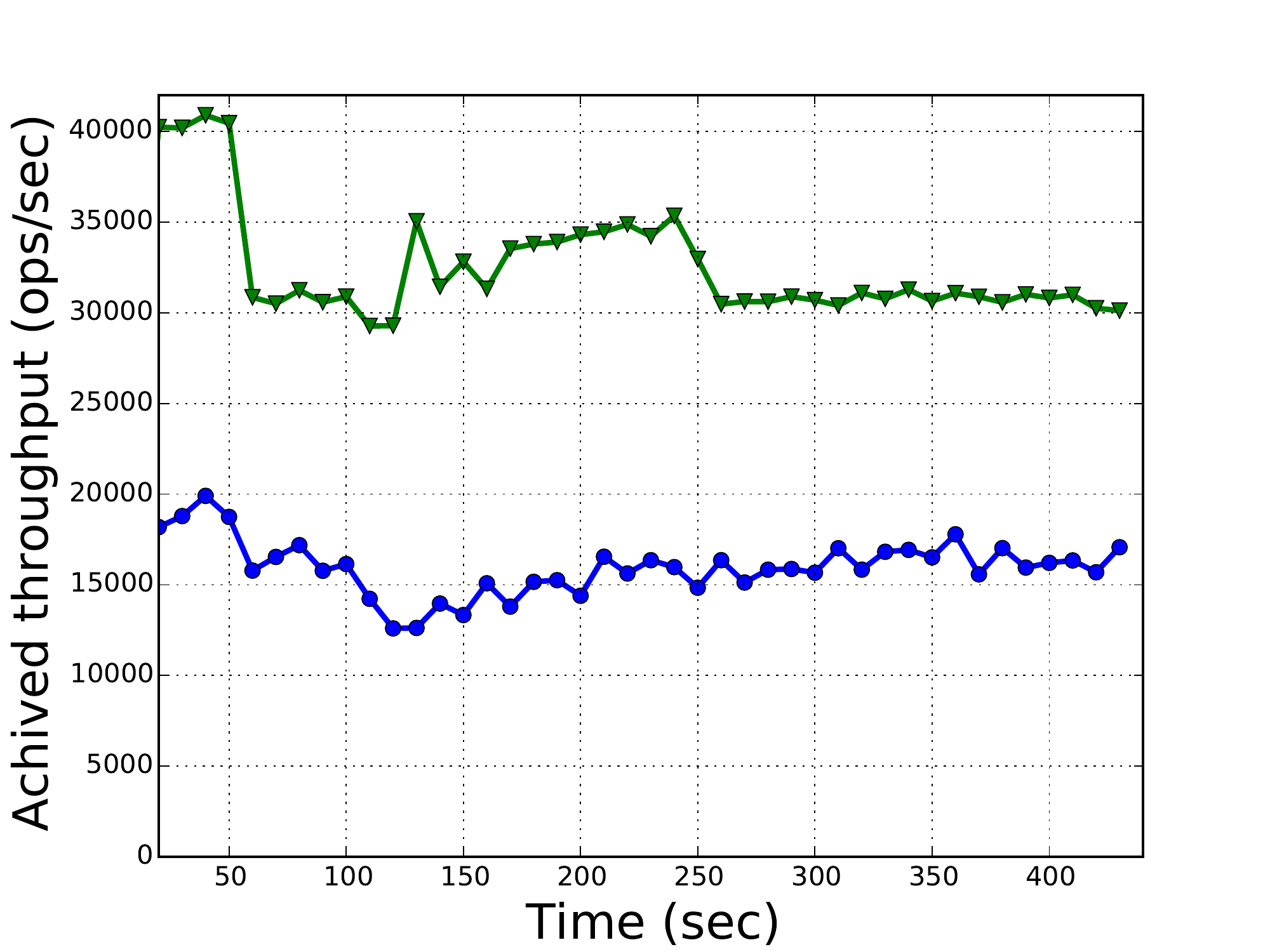}
	\end{subfigure}%
	\begin{subfigure}{.331\textwidth}
		\centering
		\includegraphics[width=\linewidth]{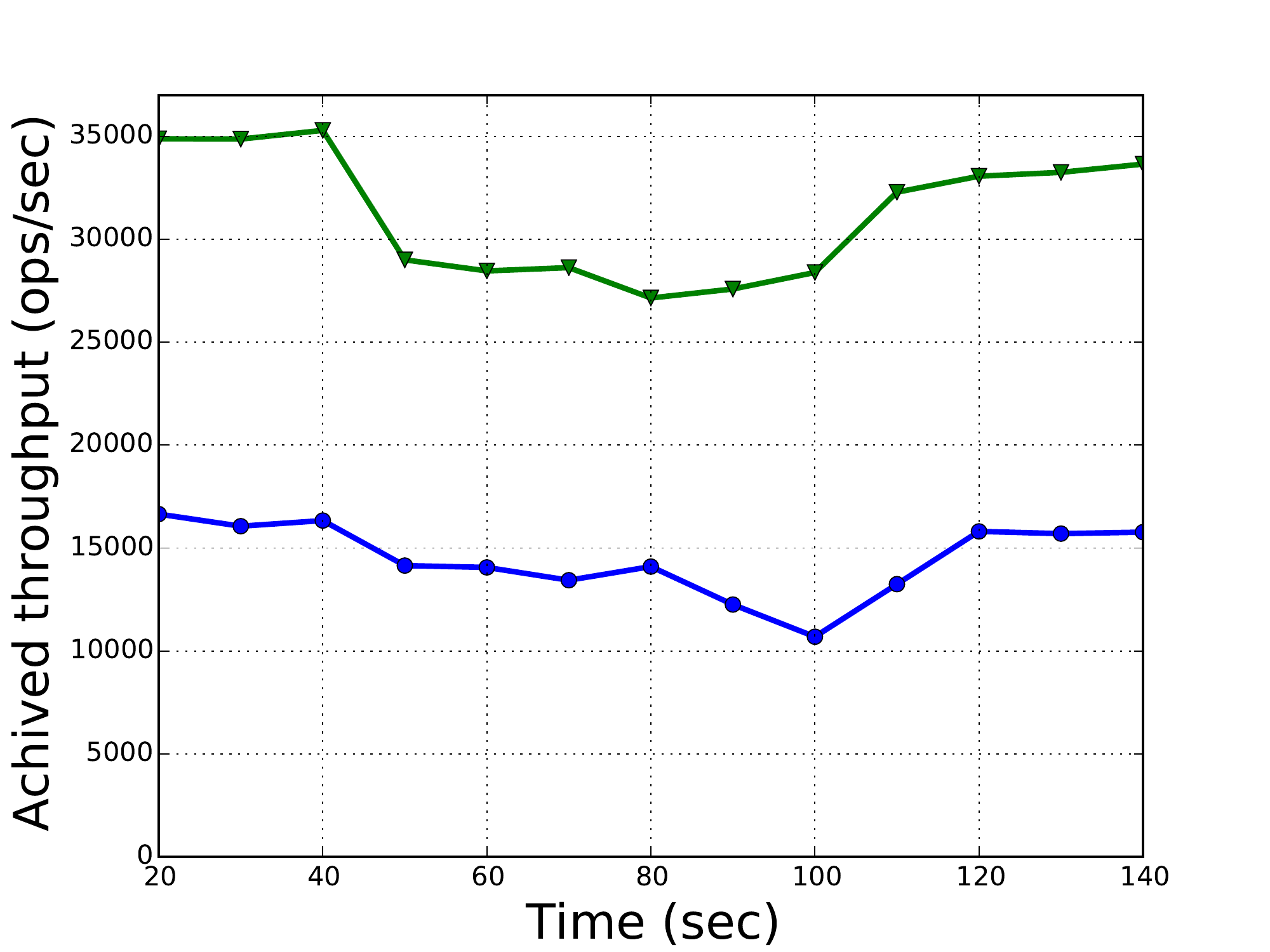}
	\end{subfigure}
	\begin{subfigure}{.331\textwidth}
		\centering
		\includegraphics[width=\linewidth]{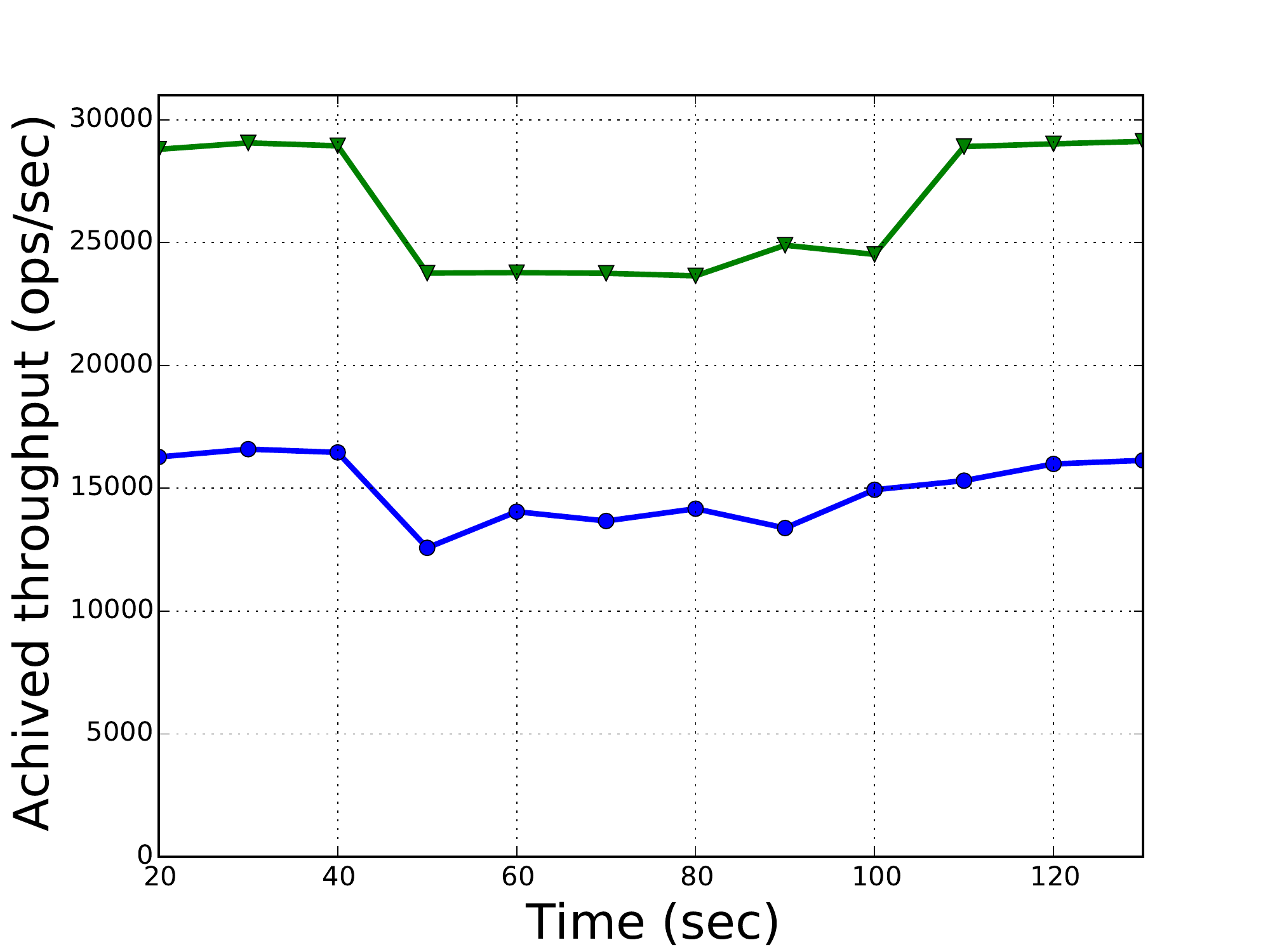}
	\end{subfigure} \\
	\begin{subfigure}{.331\textwidth}
		\centering
		\includegraphics[width=\linewidth]{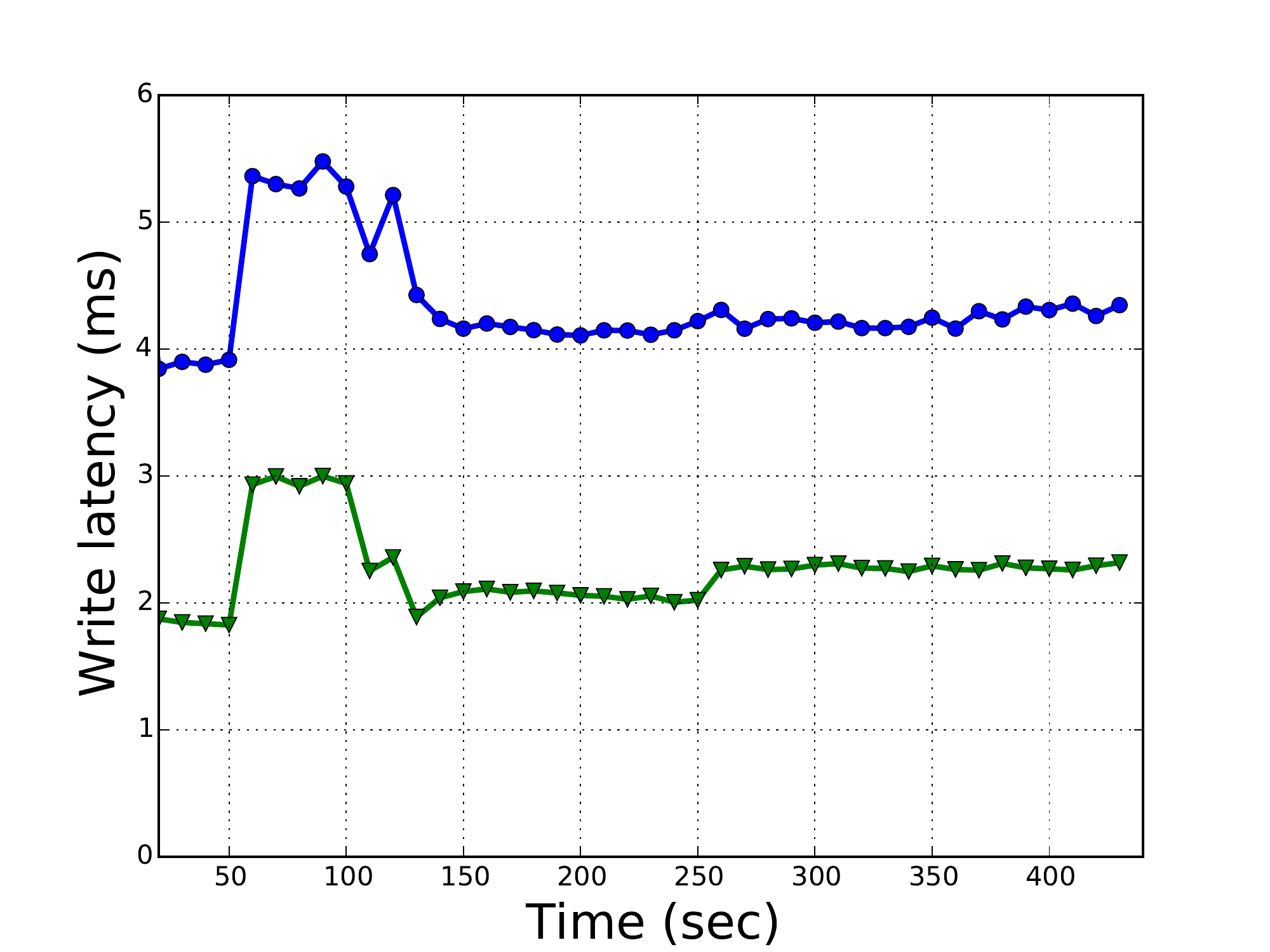}
	\end{subfigure}%
	\begin{subfigure}{.331\textwidth}
		\centering
		\includegraphics[width=\linewidth]{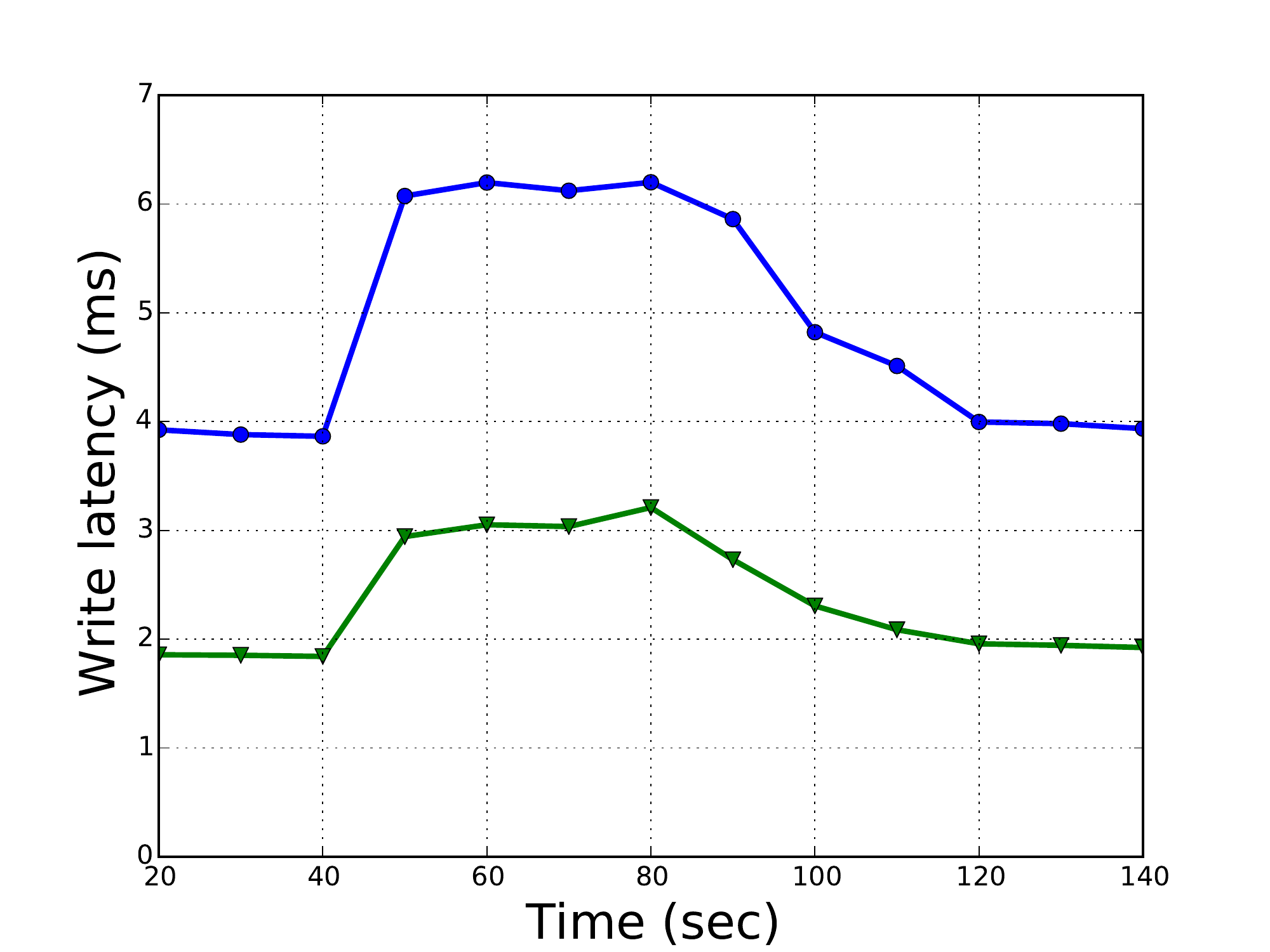}
	\end{subfigure} \\
	\begin{subfigure}{.331\textwidth}
		\centering
		\includegraphics[width=\linewidth]{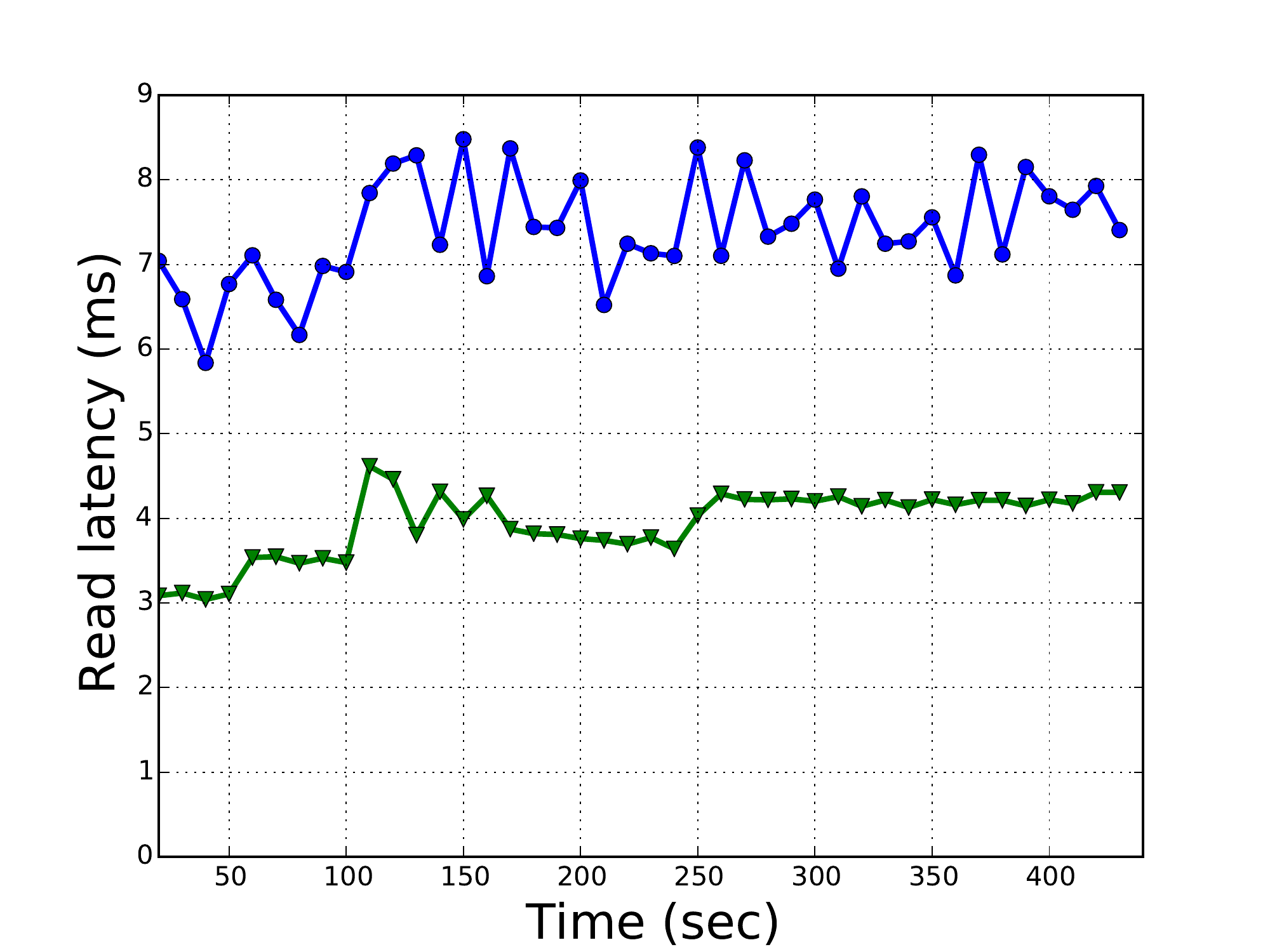}
		\caption{Workload A}
	\end{subfigure}%
	\begin{subfigure}{.331\textwidth}
		\centering
		\includegraphics[width=\linewidth]{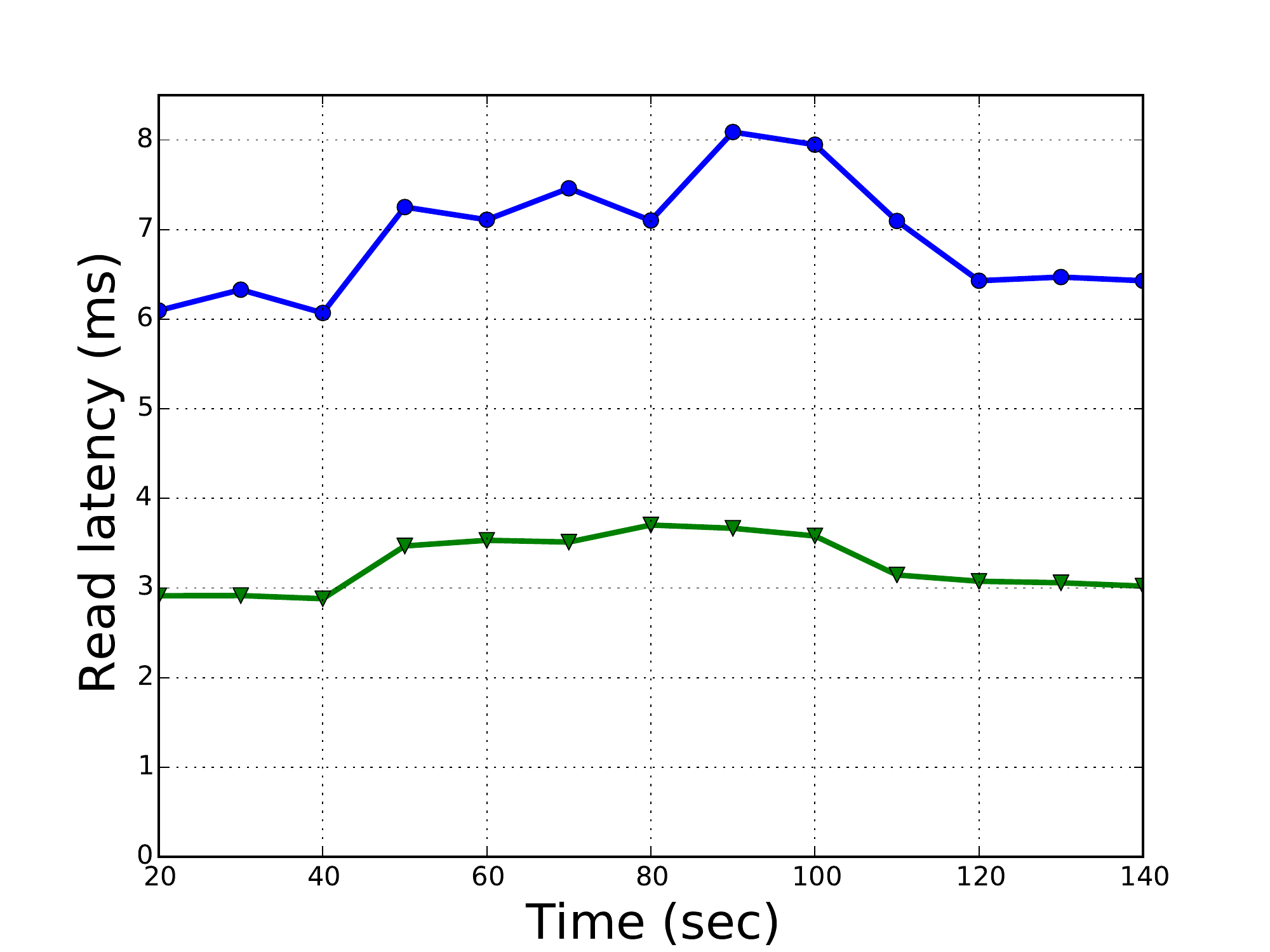}
		\caption{Workload B}
	\end{subfigure}
	\begin{subfigure}{.331\textwidth}
		\centering
		\includegraphics[width=\linewidth]{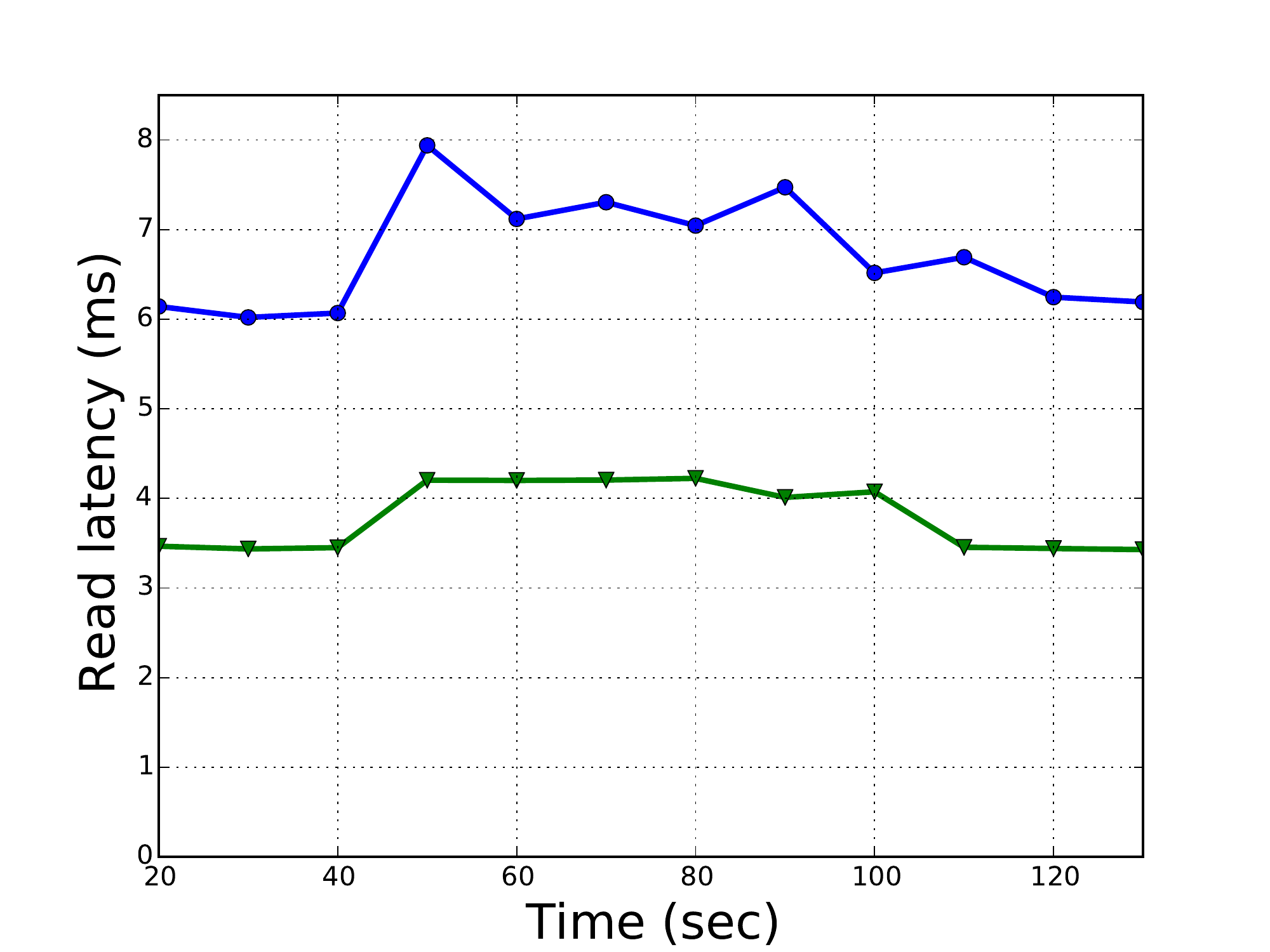}
		\caption{Workload C}
	\end{subfigure}
	\caption{Comparing the best solution against plain Cassandra in a benign failure of one node.}
	\label{fig:byz_test1}
\end{figure*}

\begin{figure*}[t]
	\begin{subfigure}{\textwidth}
		\centering
		\includegraphics[width=0.6\linewidth]{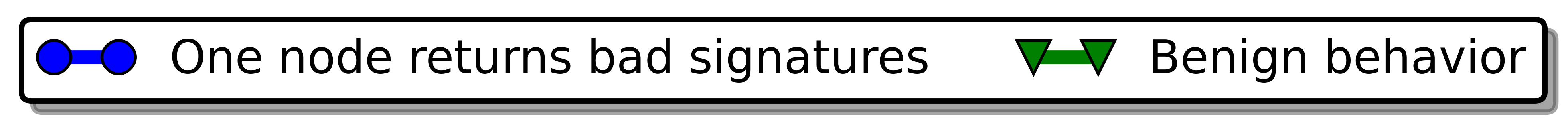}
	\end{subfigure}
	\begin{subfigure}{.331\textwidth}
		\centering
		\includegraphics[width=\linewidth]{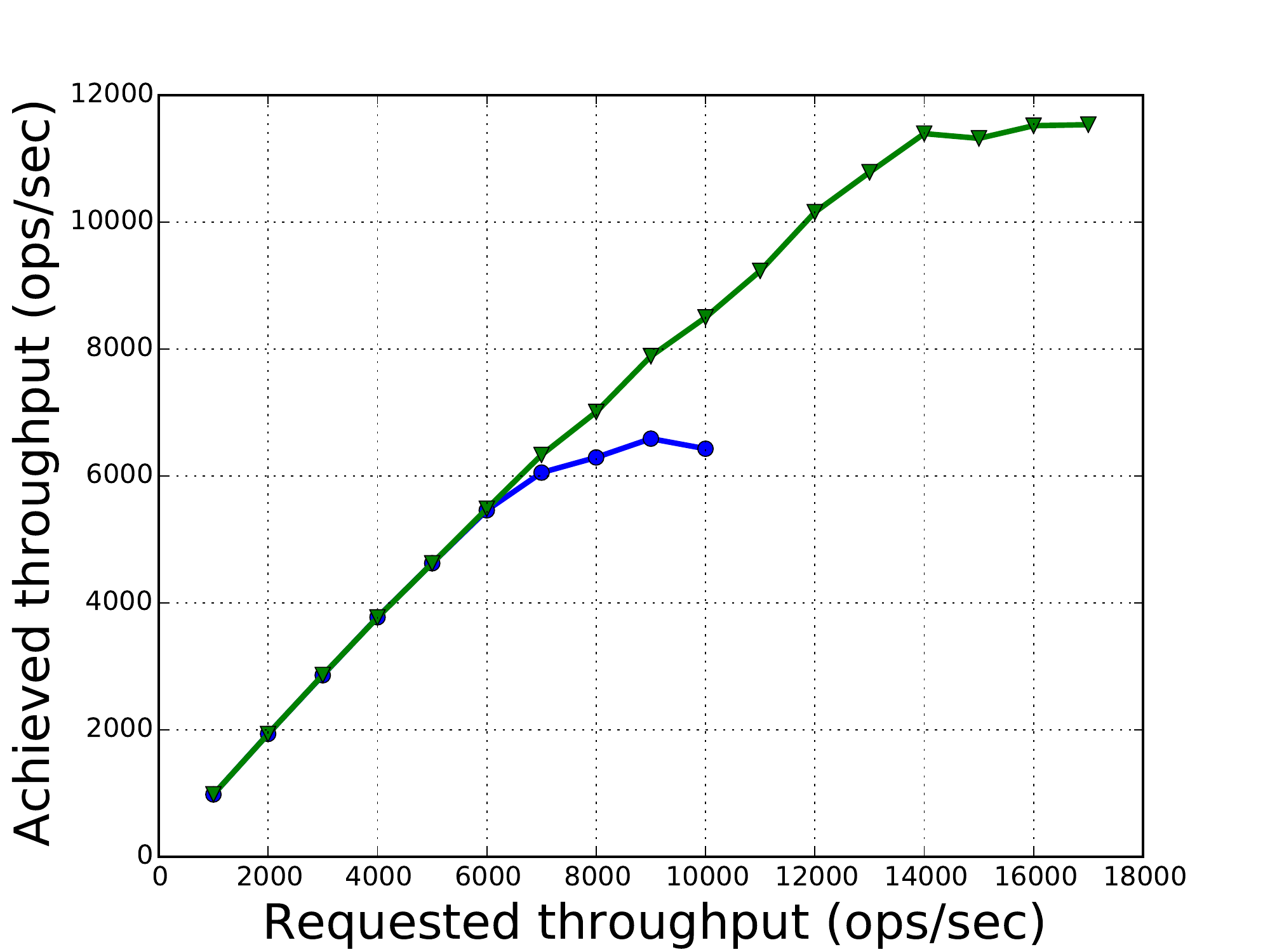}
	\end{subfigure}%
	\begin{subfigure}{.331\textwidth}
		\centering
		\includegraphics[width=\linewidth]{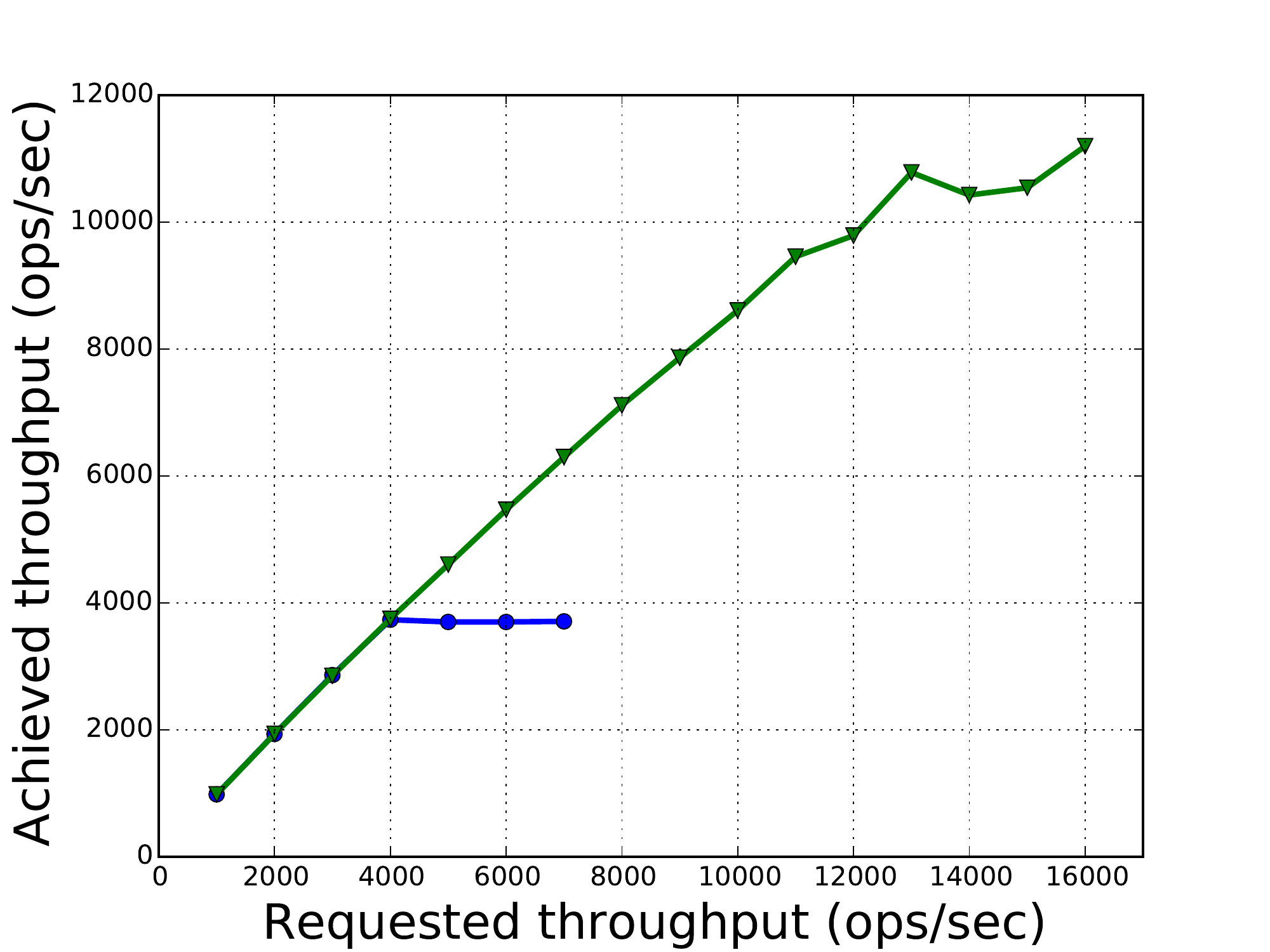}
	\end{subfigure}
	\begin{subfigure}{.331\textwidth}
		\centering
		\includegraphics[width=\linewidth]{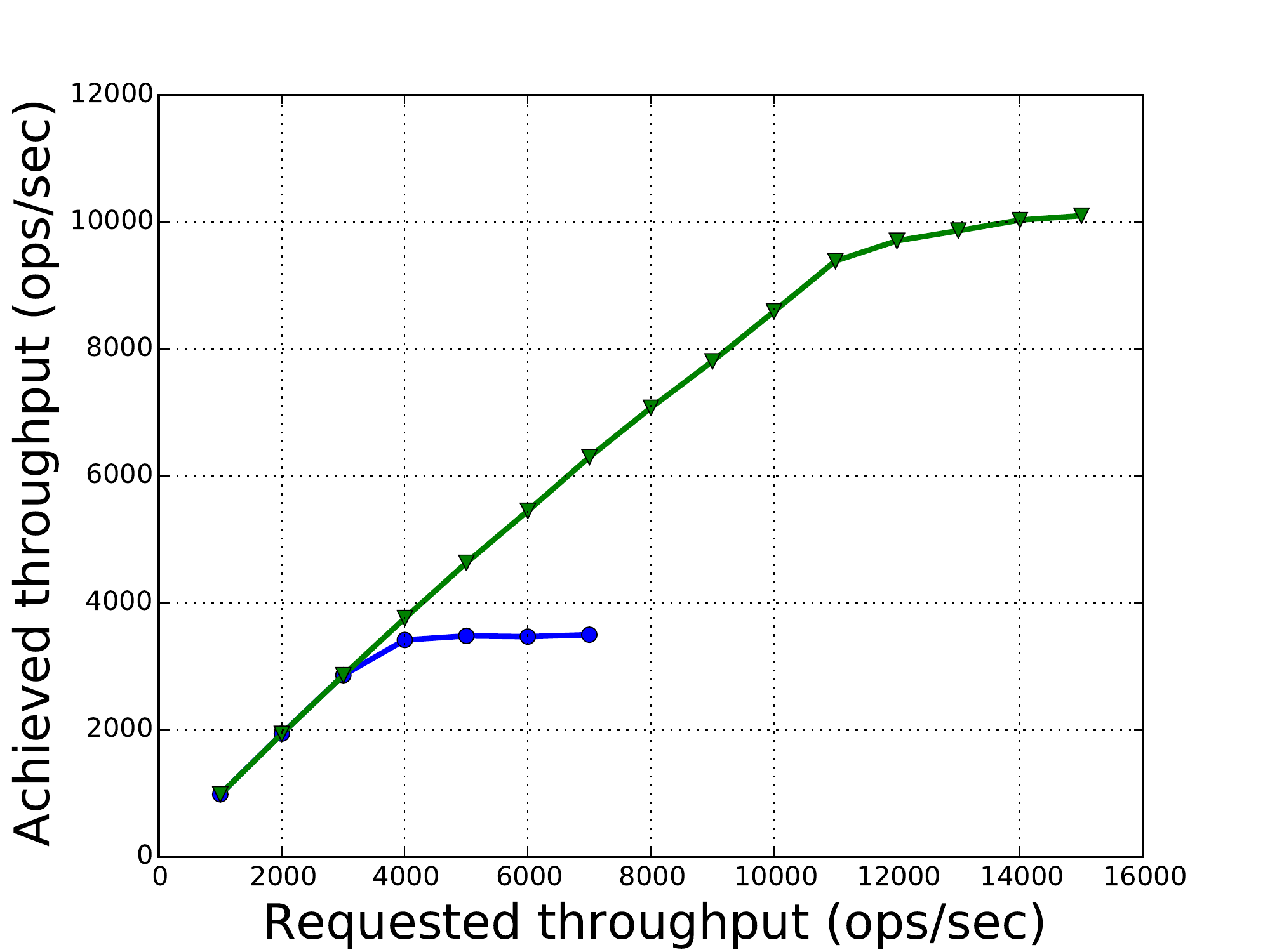}
	\end{subfigure} \\
	\begin{subfigure}{.331\textwidth}
		\centering
		\includegraphics[width=\linewidth]{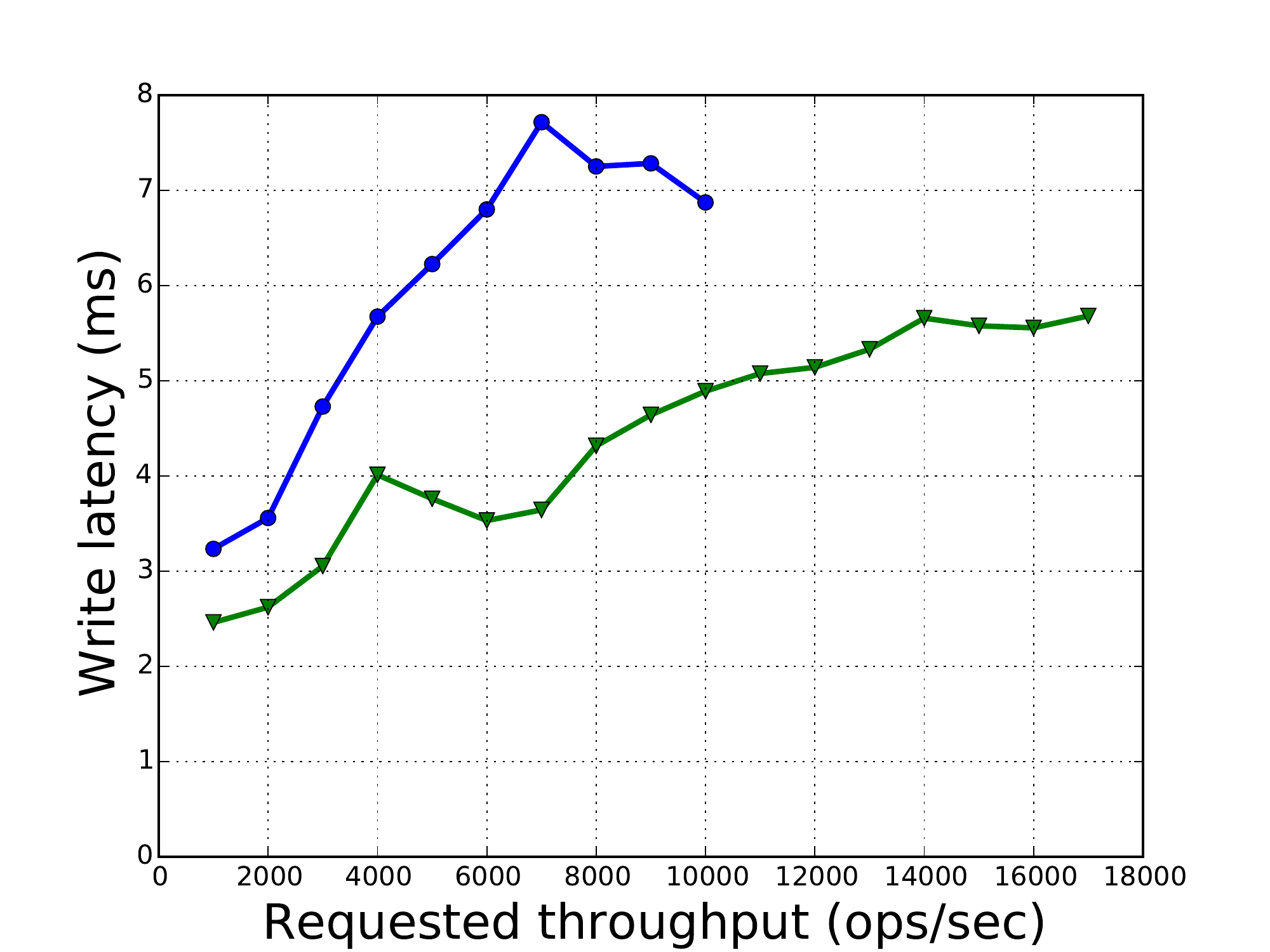}
	\end{subfigure}%
	\begin{subfigure}{.331\textwidth}
		\centering
		\includegraphics[width=\linewidth]{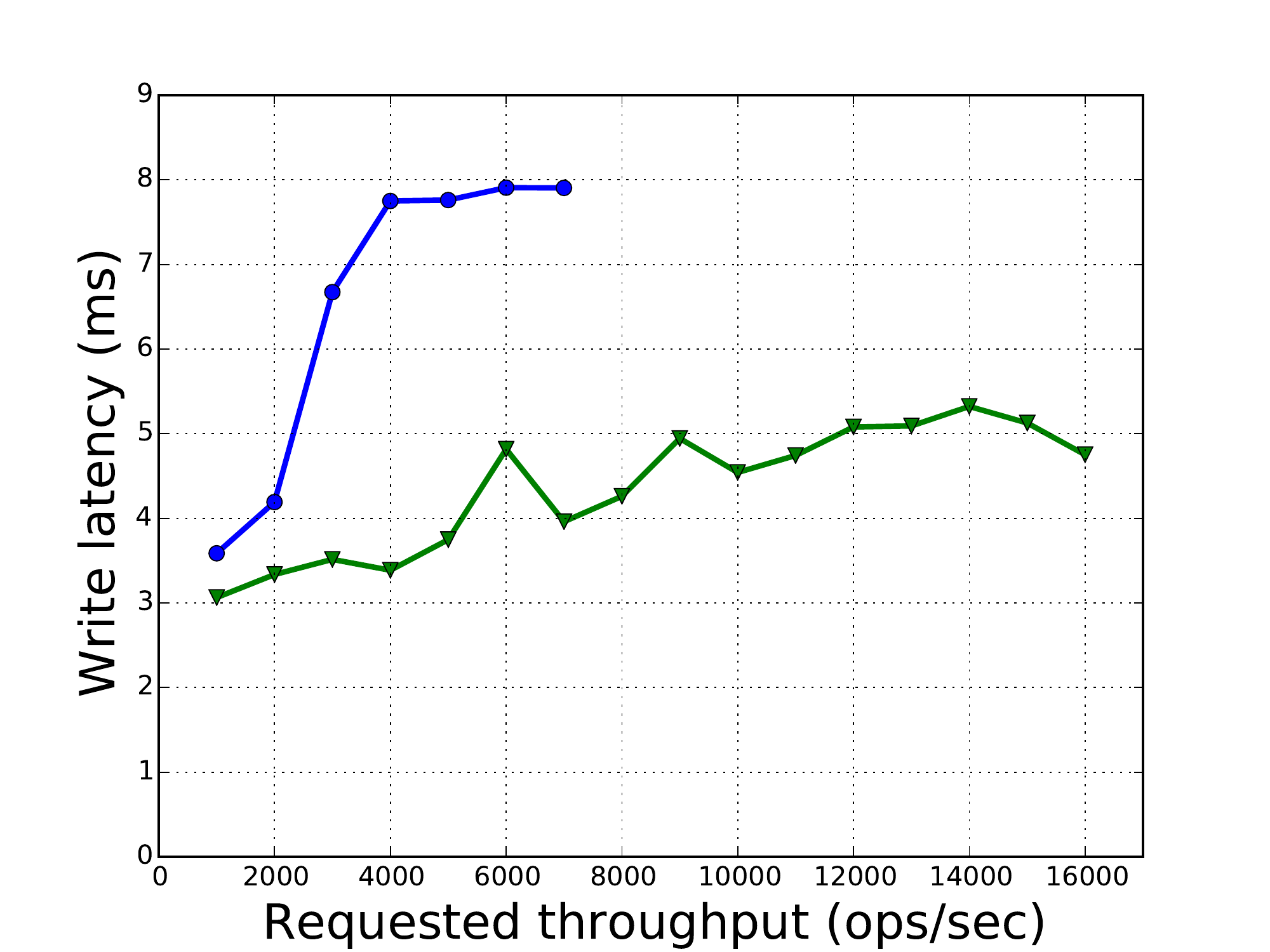}
	\end{subfigure} \\
	\begin{subfigure}{.331\textwidth}
		\centering
		\includegraphics[width=\linewidth]{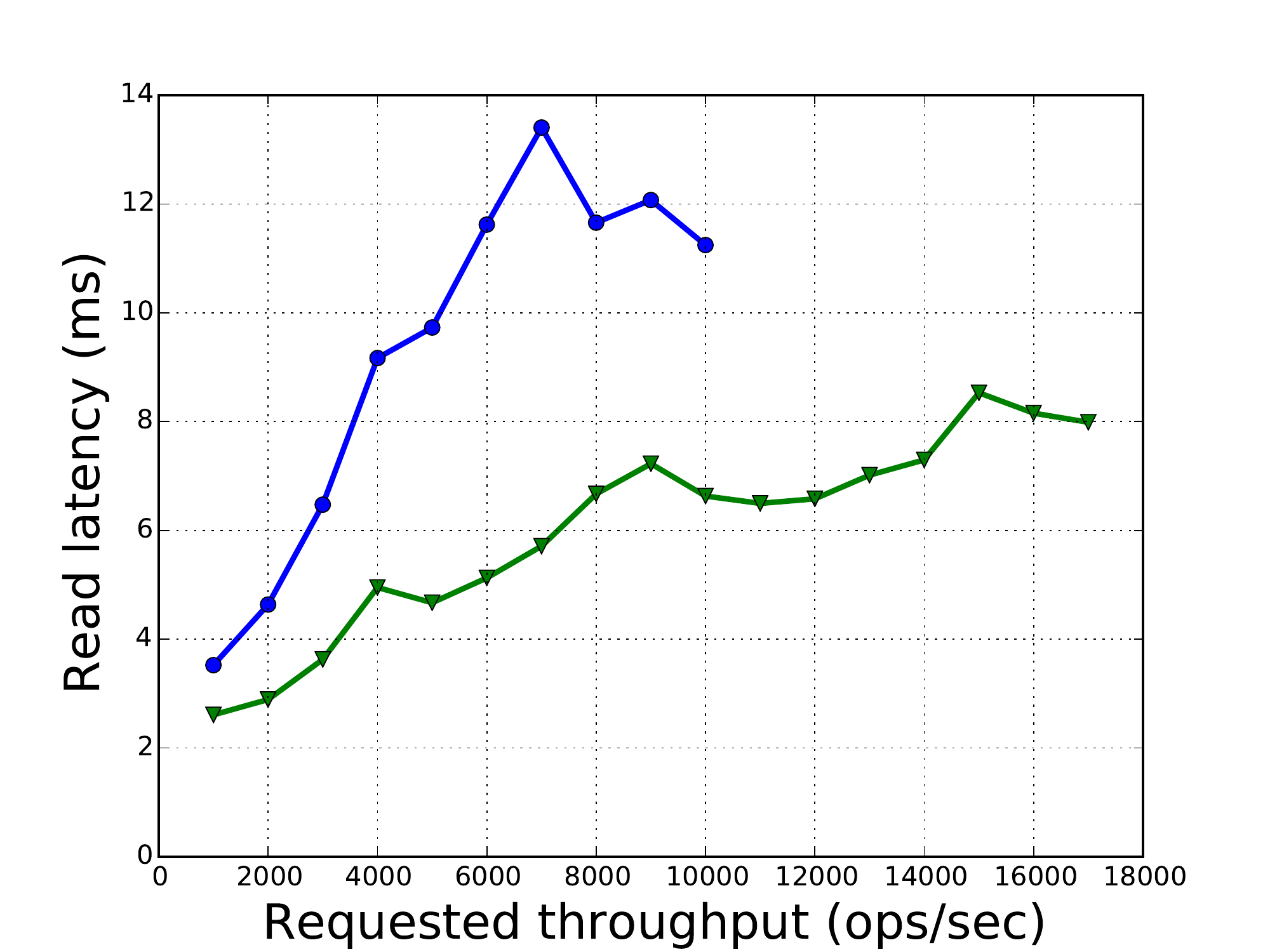}
		\caption{Workload A}
	\end{subfigure}%
	\begin{subfigure}{.331\textwidth}
		\centering
		\includegraphics[width=\linewidth]{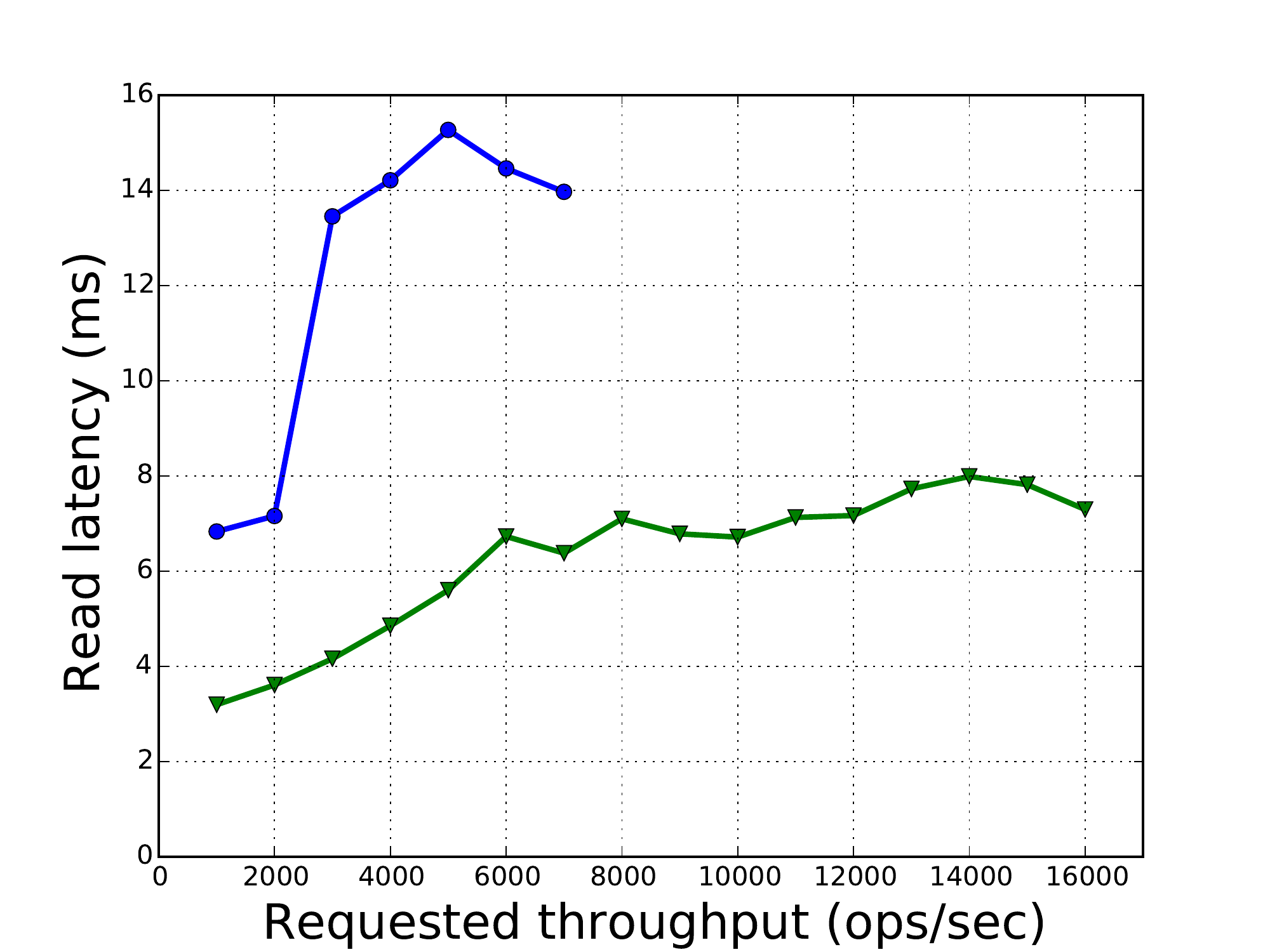}
		\caption{Workload B}
	\end{subfigure}
	\begin{subfigure}{.331\textwidth}
		\centering
		\includegraphics[width=\linewidth]{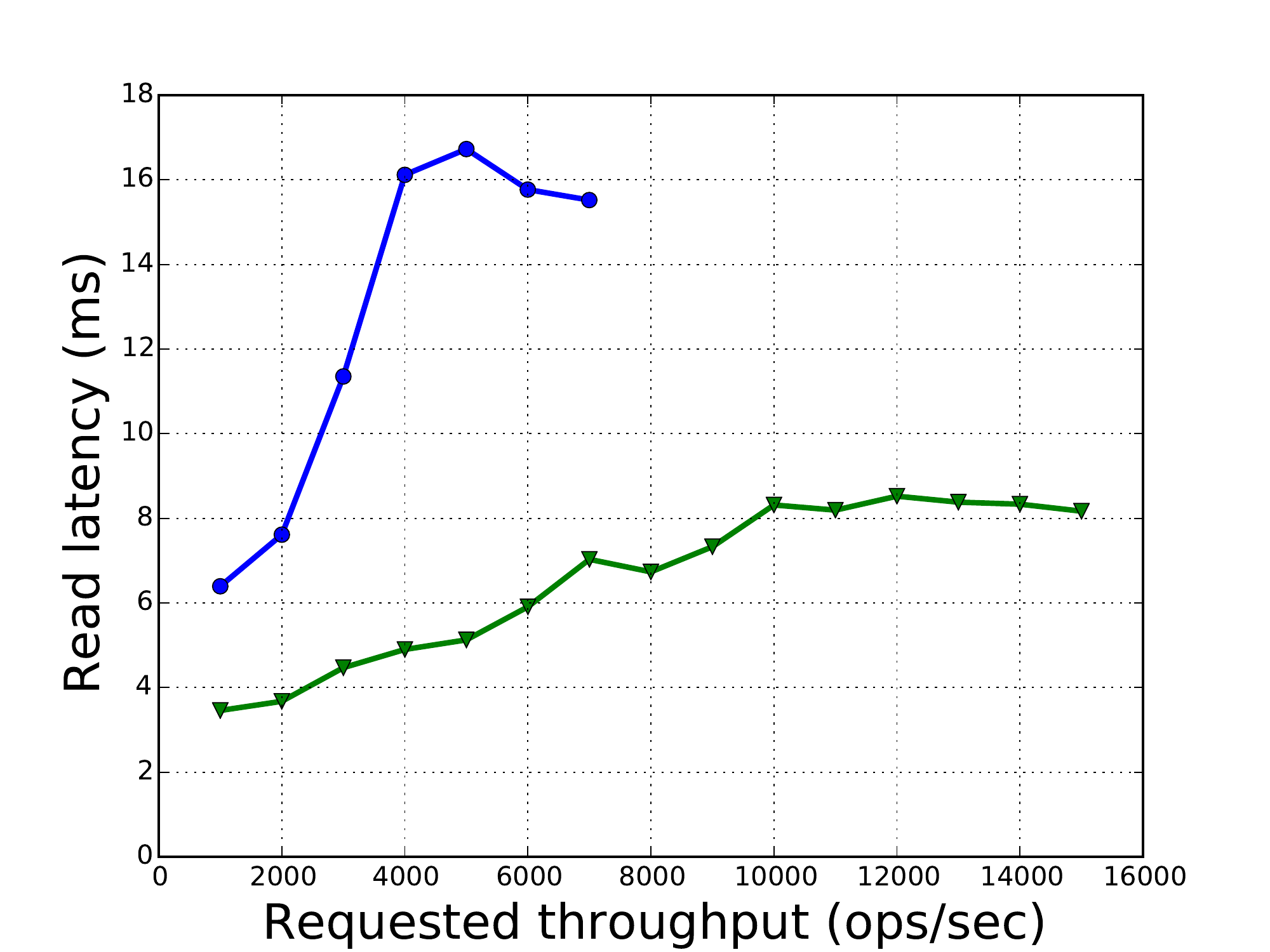}
		\caption{Workload C}
	\end{subfigure}
	\caption{Comparing the best solution in benign behavior and the scenario of one node that replies only with bad signatures.}
	\label{fig:byz_test2}
\end{figure*}

The algorithms reported here were implemented\footnote{https://github.com/ronili/HardenedCassandra} as patches to \Cassandra~2.2.4\footnote{https://github.com/apache/cassandra/tree/cassandra-2.2.4}.
We evaluated the performance of the variants of our solution and compared them to the original \Cassandra~using the standard YCSB 0.7\footnote{https://github.com/brianfrankcooper/YCSB/tree/0.7.0} benchmark~\cite{cooper2010benchmarking}, adjusted to use our BFT client library\footnote{https://github.com/ronili/HardenedCassandraYCSB}.
We used Datastax's Java driver 2.1.8\footnote{https://github.com/datastax/java-driver/tree/2.1.8} on the client side.
There are nearly 390K \emph{LOC} (lines of code) in \Cassandra. Our patch added about 3.5K LOC to the servers code and about 4K LOC to the client code (including YCSB integration), which uses the client driver as is.
Our entire code adds less than 2\% LOC.

All experiments were run on four to five machines (Ubuntu14, dual 64-bit 6 core 2.2GHz  Intel Xeon E5 CPUs, 32GB of RAM, 7200 RPM hard drive and 10Gb ethernet), one for the client and three to four for the \Cassandra{} nodes.

We pre-loaded the database with 100,000 rows and then benchmarked it with the following five YCSB workloads that vary in the ratio of the writes and reads: (1) Workload A - 50/50 reads/writes. (2) Workload B - 95/5 reads/writes. (3) Workload C - only reads. (4) Workload D - 95/5 reads/writes, where the reads are for the latest inserts (and not random). (5) Workload F - 50/50 writes/Read-Modify-Writes.
In all workloads except workload D, the write is for one column while in workload D it is for the entire row.
Every workload ran with 100 client threads that in total preformed 100,000 operations with a varying throughput target.
The tables consisted of 10 columns (default in YCSB) as well as tables consisting of one value, modeling a key-value datastore.
Each value is of size 100 bytes while the key size is randomly chosen in the range of 5-23Bytes.
Therefore, each record/line with 10 columns has an average length of 1014Bytes.

We implemented the algorithms presented in Figures~\ref{fig:write-flow} and~\ref{fig:read-flow} where the proxy authenticates the nodes acknowledgments.
We refer to these as \e{Proxy-Verifies}.
In addition, we implemented the variant where the proxy does not verify the acknowledgments and lets the client fetch more acknowledgments in case it is not satisfied, as appear in Figures~\ref{fig:write-flow-v2} and~\ref{fig:read-flow-v2}.
We will refer to it as \e{Proxy-No-Verify}.
We ran that last algorithm in two modes, one where the proxy is in charge of resolving inconsistent reads, as appears in Figure~\ref{fig:read-flow-v2}, and one where the client is, as appears in Figure~\ref{fig:read-flow-v3}.
In our work, we present only the version where the proxy resolves as it behaves similar to the client resolves version.

When using MAC tags, we analyzed it in two steps: (1) using MAC tags on messages from nodes to client, referred to as \e{Half-Sym} and (2) using it for both ways, referred to as \e{Full-Sym}.
%We refer to the last two modes with \e{Proxy-Resolves} and \e{Client-Resolves}, respectively.

We used two types of private key signatures: (1) RSA with keys of length of 2048b and (2) ECDSA with keys of length 256b and the \e{secp256r1} curve.
As for symmetric keys, we used keys of length of 128b with the \e{HMAC}~\cite{krawczyk1997hmac} MAC algorithm.
In all signatures algorithms, we have used \e{SHA256}~\cite{eastlake2006us} for the hashing process.

To evaluate the cost of our algorithm without cryptographic overhead, we ran them also without any signatures.
That is, we swapped the signing methods with a \e{base64} encoded on a single char, referred to as \e{No-Sign}.

We ran \Cassandra~with SSL support and witnessed only a marginal performance impact.
Therefor, all results presented here are without SSL support.

The YCSB tool we used is throttling the requests rate in correlation with the achieved maximum throughput. 
Given that, we run each experiment until achieving a stable throughput for several following request rates.

\subsection{Performance In A Benign Environment}

In Figures~\ref{fig:one-a_to_c}, \ref{fig:one-d_to_f}, \ref{fig:two-a_to_c}~and \ref{fig:two-d_to_f}, we present the performance results in the standard \Cassandra{} multi-column model.
%in Figures~\ref{fig:one-a_to_c} and \ref{fig:one-d_to_f}, we present the performance results in the standard \Cassandra~multi-column model.
As can be seen, our best solution is the variant where the proxy does not verify the acknowledgments, and we use ECDSA and MAC tags for both ways (ECDSA Proxy-No-Verify Full-Sym 4Nodes).
The slowdown of this solution is roughly a factor of 2-2.5 in terms of the maximum throughput, 2.5-3 in the write latency and 2-4 in the read latency.
Interestingly, for plain \Cassandra, increasing the cluster from 3 nodes to 4 nodes (while also increasing the quorum sizes from 2 to 3, respectively) actually improves the performance by about 10\%.
This is because the role of the proxy as well as the corresponding load is now split between 4 nodes rather than only 3.
%It can also be seen that ECDSA has dramatic improvement on the performance when only signing is required.
The No-Sign experiment represents the BFT algorithmic price that includes larger quorums, extra verifications and storing signatures.
The ECDSA experiment represents the cryptography price.
It can be seen that using the RSA signing algorithm has a significant negative impact on the performance.

%In Figures~\ref{fig:kv-a_to_c} and \ref{fig:kv-d_to_f}, we present a comparison of our best algorithms in a key-value model, i.e., a table with one non-key column.
We have also explored the performance in the key-value model, i.e., in a table with one non-key column.
%For lack of space, the results are not shown, but are qualitatively similar.
In Figures~\ref{fig:kv-a_to_c} and \ref{fig:kv-d_to_f}, we present a comparison of our best algorithms in a key-value model, i.e., a table with one non-key column.
As can be expected, the results show a small improvement compared to the multi-column model, as it requires a lower signatures overhead.
This implies less network traffic that mostly affect the read path and fewer public key singing operations that affect the write path.
The write latency improvement is marginal as in most of the workloads we update only one column as opposed to workload D where we update the entire row.

\subsection{Performance When Facing Byzantine Behavior}
In Figure~\ref{fig:byz_test1}, we present the performance of our best solution under the scenario of one stopped node.
We run workload A on a fully synchronized, four nodes setup, on maximum throughput. After 50 seconds, we stopped one node for 30 seconds and then restarted it.
It took the node between 20 to 30 seconds to start. Immediately after it finished loading, the other nodes started retransmitting the missed writes to the failed node.
In our best solution, the distributed retransmitting took about 250 seconds and in the plain \Cassandra, about 170 seconds.
We repeated this test with workloads B and C with one change, failing the node in t=40 instead of t=50.
From this experiment, we can see that our solution behaves as plain \Cassandra~under this scenario, and can tolerate a one node outage with an acceptable performance impact.

In Figure~\ref{fig:byz_test2}, we present the performance of our best solution under the scenario of one node that always returns a bad signature.
This impacts the entire system as on every failed signature verification, the client has to contact the proxy again.
Additionally, on every read that addresses the Byzantine node, a resolving and a write-back process is initiated.
As can be seen in the results, the performance degrades to about 40\%-50\%, still leaving the system in a workable state.

We have also explored the performance of our solution in case of a stalling proxy.
In this case, following a correct execution of an operation, the proxy waits most of the timeout duration before supplying the client with the response.
As a result, the system's performance might decrease dramatically.
Since the attack effects vary in correlation with the timeout configuration, the attack can be mitigated by lowering the timeout as low as possible.
On the contrary, a tight timeout might fail correct requests during congestion times.
The right optimization of timeouts relies on several deployment factors e.g., the application requirements, the connection path of the client to system, the network topology of the nodes and more.
Therefore, we could not deduce interesting definitive insights when facing this case.
Finally, we would like to point out that the client can be configured to contact the fastest responding nodes first and thus reduce the effect of this attack. 

%%%%%%%%%%%%%%%%%%%%%%%%% Section 7 : Discussion %%%%%%%%%%%%%%%%%%%%%%%
\section{Conclusion}
\label{sec:disscussion}
\Cassandra's wide adoption makes it a prime vehicle for analyzing and evaluating various aspects of distributed data stores.
In our work, we have studied \Cassandra's vulnerabilities to Byzantine failures and explored various design alternatives for hardening it against such failures.
Our solution addressed the use of quorums, the proxy mediated interaction between the client and replicas, conflict resolutions on the reply path, configuration changes, overcoming temporary unavailability of nodes, timestamps generation, and the use of digital signatures on stored values and messages.

We have also evaluated the attainable performance of our design alternatives using the standard YCSB benchmark.
The results of the performance evaluation indicated that our best Byzantine tolerant design yields a throughput that is only 2-2.5 times lower than plain \Cassandra~while write and read latencies are only a factor of 2-3 and 2-4 higher, respectively, than in the non-Byzantine tolerant system.
Interestingly, the performance we obtained with the Byzantine tolerant version of \Cassandra~is similar to the performance obtained for a non-Byzantine \Cassandra~in the YCSB paper from 2010~\cite{cooper2010benchmarking}.

%Performance wise, the two most significant design decisions are the specific use of cryptographic signatures and resolving all conflicts during reads only and do so at the invoking client.
Performance wise, the two most significant design decisions are the specific use of cryptographic signatures and resolving all conflicts during reads only.
Specifically, our novel design of sending a vector of MAC tags, signed by itself with the symmetric key of the client and target node, plus the ECDSA public key signature, means that the \emph{usual path} involves no public key verifications and only one elliptic curve signature.
This evades costly public key verifications and RSA signatures.

%
%We have not witnessed any work that examined and added BFT capabilities to the main \Cassandra~mechanisms. While we have relied on Malkhi \& Reiter's Byzantine Quorum Systems, we have adopted the protocol to \Cassandra~by (1) using wall clocks instead of logical timestamps and (2) tunneling the requests through a proxy node. The former decreases the number of messages to $1/3$ and the latter adds an encryption step.
%
%We have showed a new approach for switching from public key signatures to private key signatures using double signing. We have presented an alternative approach when handling Byzantine clients by repairing the system when required rather than ensuring that a specif state is preserved.

%We assume that our solution could be further improved, generating better performance that reach closer to the plain \Cassandra~abilities.
%We left for future work some of \Cassandra~interesting mechanisms that could have robustness against Byzantine failures. Such mechanisms are: optimized read and writes when using multiple data-centers, batching writes and \e{light-weight transactions}. We would also be interested in exploring quicker converging membership dissemination protocols in \Cassandra~using protocols such as Araneola~\cite{araneola}.

%\subsection{Future Work}
Looking into the future, we would like to extend our Byzantine support to the full range of CQL functionality.
Also, optimizing the protocols for multi data-center operation and supporting \e{light-weight transactions}.

Exploring batching as a performance booster~\cite{FR97} is challenging, especially given the documentation of the BATCH command in DATASTAX' reference manual:
\begin{quote}
``Using batches to optimize performance is usually not successful, as described in $\ldots$''
\end{quote}
The current batching implementation in \Cassandra{} minimizes the client-to-system traffic, but increases the in-system traffic.
We assume that \Cassandra{} could be improved also in a benign environment by using traditional batching.
Once this is resolved, fitting such a batching solution to the Byzantine environment should also lower the cryptographic overhead of our solutions.

Finally, we would like to try quicker converging membership dissemination protocols in \Cassandra, using protocols such as Araneola~\cite{araneola}.

%\chapter{Conclusion and open questions}
%\label{chap:conclusion}

%This kind of chapter can include may different things (or only some of them):
%\begin{itemize}
%\item Discussion of results
%\item Conclusions from the results or from the process in general
%\item Open questions for future research, resulting from the research performed or from the results obtained
%\end{itemize}

%But not things like the bibliography or other back matter which is generated outside of this chapter.

%\section{Some conclusion}

%Here is what I conclude.

%\section{Some open questions}

%\paragraph{A question in brief.} In \autoref{chap:firstchap} we explored a certain subject, but what about this-or-that idea? Perhaps it is worth exploring. Can one produce interesting results?

%\paragraph{A second question in brief.} A broader exposition of the question and indications of directions or ideas regarding its resolution.

\paragraph*{Acknowledgment}
This research was partially funded by the Israeli Ministry of Science and Technology grant \#3-10886.
We also thank the National Israeli Cyber Lab.

%\bibliographystyle{IEEEtran}
%\bibliography{IEEEabrv,references}

\bibliographystyle{abbrv}
\bibliography{references}

\appendix

\section{Appendix: Detailed Algorithms}
\label{appendix:detailed_algos}

In this appendix, we present in details a variety of the algorithms we described in the main part. All of them are divided into three parts: client algorithm, proxy algorithm and node algorithm.

In Figures~\ref{fig:clean-cassandra-write-flow} and~\ref{fig:clean-read-flow}, we present the write and read flows as in plain \Cassandra, as described in Section~\ref{sec:proxy}.

In Figures \ref{fig:write-flow-v2} and~\ref{fig:read-flow-v2}, we present the variant of our original solution, presented in Figures \ref{fig:write-flow}~and \ref{fig:read-flow}, where the proxy does not verify the acknowledgments, as described in Section~\ref{Proxy-Acknowledgments}. Figures~\ref{fig:write2} and~\ref{fig:read2} illustrate the run of these algorithms.

%n Figure, we present the variant of our original solution, presented in \ref{fig:read-flow}, where the proxy does not verify the acknowledgments, as described in Section~\ref{Proxy-Acknowledgments}.

In Figure~\ref{fig:read-flow-v3}, we present another variant, where the proxy does not verify the acknowledgments and the client has to resolve conflicts, as described in Section~\ref{Client-Resolving}. Figure~\ref{fig:read3} illustrates the run of this algorithm.

%%%%%%%%%%%% Extra for appendix %%%%%%%%%%%%%%

{
	\begin{figure*}[t]
		\captionsetup{font=scriptsize}
		\begin{algorithmic}[1] 		
			\scriptsize
			\Function{OnNodeToNodeWriteRequest}{$key, value, ts$}
			\State Store locally $<key, value, ts>$
			\State \Return $Success$
			%	\Else
			%	\State \Return $\bot$
			\EndFunction
			\\
			\Function{OnClientToNodeWriteRequest}{$key, value, ts$}
			\For {each node $n$ that is responsible for the $key$} \Comment{N nodes}
			\State Send write request with \e{$<$key, value, ts$>$} to $n$
			\EndFor
			\State Wait for $f+1$ acknowledgments OR timeout
			\State \Return $Success$
			\EndFunction
			\\
			\Function{ClientWriteRequest}{$key,value$}
			\State $ ts \gets $ Current timestamp
			\State $ p \gets $ Some random system node
			\State Send write request with \e{$<$key, value, ts$>$} to $p$
			\State Wait for an acknowledgment OR timeout \Comment{Retry options available}
			\State \Return $Success$
			\EndFunction
			% \fontsize{8pt}{8pt}\selectfont~
			\caption{The write flow in plain \Cassandra. ClientWriteRequest is invoked by the client for each write. OnClientToNodeWriteRequest is invoked on the proxy node by the client. OnNodeToNodeWriteRequest is invoked on a node that has the responsibility to store the value.}
			\label{fig:clean-cassandra-write-flow}
		\end{algorithmic}
	\end{figure*}
}

{
\begin{figure*}[t]
	\captionsetup{font=scriptsize}
	\begin{algorithmic}[1]
		\scriptsize
		\Function{OnNodeToNodeReadRequest}{$key$}
		\State $<value,ts> \gets $ The newest associated timestamp and value with $key$
		\If {isDigestQuery}
		\State \Return $<hash(value),ts>$
		\Else
		\State \Return $<value,ts>$
		\EndIf
		\EndFunction
		\\
		\Function{OnClientToNodeReadRequest}{$key$}
		
		\State {$target Endpoints \gets all Relevant Nodes$ for $key$ or a subset of $f+1$ fastest relevant nodes}
		\State \Comment{Optimization}
		\State {$dataEndpoind \gets$ One node from $target Endpoints$}
		\State Send read request for data to $ dataEndpoind $
		\State Send read request for digest to $target Endpoints \setminus \{dataEndpoind\} $
		\State Wait for responses from $f+1$ nodes OR timeout
		\If {timeout}
		\State \Return $\bot$
		\EndIf
		\If {got response from $ dataEndpoind $ AND all responses agree on the digest}
		\State \Return data
		\EndIf
		
		\State Send read request for data from all nodes in $ responded Nodes  \setminus \{dataEndpoind\}$
		\State Wait for responses from all $ contacted Nodes$ OR timeout
		\If {timeout}
		\State \Return $\bot$
		\EndIf
		
		\State $resolved \gets$ Latest response from $responses$
		\State Send write-back with $resolved$ to $ responded Nodes $ except those that are known to be updated
		\State Wait for responses from all $contacted Nodes$ OR timeout
		\If {timeout}
		\State \Return $\bot$
		\EndIf
		
		\State \Return resolved
		\EndFunction
		\\
		\Function{ClientReadRequest}{$key$}
		\State $ p \gets $ Some random system node
		\State Send read request with $key$ to $p$
		\State Wait for data OR timeout
		\State \Return data
		\EndFunction
		\caption{The read flow in plain \Cassandra. ClientReadRequest is invoked by the client for each read. OnClientToNodeReadRequest is invoked on the proxy node by the client. OnNodeToNodeReadRequest is invoked on a node that has the responsibility to store the value.}
		\label{fig:clean-read-flow}
	\end{algorithmic}
\end{figure*}
}

{
	\begin{figure*}[t]
		\captionsetup{font=scriptsize}
		\begin{algorithmic}[1] 		
			\scriptsize

			\Function{OnNodeToNodeWriteRequest}{$key, value, ts, clientSignature, clientID$}
				\If {clientSignature is valid}
					\State $nodeSignature \gets ComputeSignature(clientSignature) $
					\State Store locally $<key, value, ts, clientSignature, clientID>$
					\State \Return $nodeSignature$
				\EndIf
			\EndFunction
			
			\\
			\Function{OnClientToNodeWriteRequest}{$key, value, ts, clientSignature, clientID$}
				\For {each node $n$ that is responsible for the $key$} \Comment{N nodes}
					\State Send write request with \e{$<$key, value, ts, clientSignature, clientID$>$} to $n$
				\EndFor
				\State Wait for $2f+1$ acknowledgments OR timeout \blue {\Comment{Not verifying acknowledgment's signatures}}
				\State \Return responses
			\EndFunction
			\\
			\Function{ClientWriteRequest}{$key,value$}
				\State $ ts \gets $ Current timestamp
				\State $ clientSignature \gets $ ComputeSignature($key$ $||$ $value$ $||$ $ts$)
				\State $ p \gets $ Some random system node
				\State Send write request with \e{$<$key, value, ts, clientSignature, clientID$>$} to $p$ \label{lst:line:sendToP}
				\State Wait for acknowledgments OR timeout
				\If {$|valid Acknowledgments| \geq 2f+1$} \label{lst:line:checkIfFinish}
					\State \Return Success
				\EndIf

				\blue {		\If { $|valid Acknowledgments| \geq f+1$ AND $retryNumber \le f $}
						\State Send same write request to $p$ asking for $2f+1 - |valid Acknowledgments|$ from \textbf{new nodes}
						\State Wait for acknowledgments OR timeout
						\If {$|newAcknowledgments| \geq 1$}
							\State $valid Acknowledgments \gets valid Acknowledgments \cup new Valid Acknowledgments$
							\State goto line \ref{lst:line:checkIfFinish} \Comment{Fetching new acknowledgments succeeded}
						\Else
							\State goto line~\ref{lst:line:tryNewNode} \Comment{This proxy node failed to produce new acknowledgments}
						\EndIf }

				\Else \State $ p \gets $ Some random system node that was not used in this function invocation \label{lst:line:tryNewNode}
				\If {$p = \bot$ OR $contacted Nodes > f$ }
				\State \Return Failure
				\EndIf
				\State goto line~\ref{lst:line:sendToP} \Comment{Trying to contact new proxy nodes}
				\EndIf
			\EndFunction
			\caption{A variant of our hardened write algorithm. In this variant, the proxy does not verify the acknowledgments and lets the client contact it again if it is unsatisfied. ClientWriteRequest is invoked by the client for each write. OnClientToNodeWriteRequest is invoked on the proxy node by the client. OnNodeToNodeWriteRequest is invoked on a node that has the responsibility to store the value. Changes from figure~\ref{fig:write-flow} are marked in blue.}
			\label{fig:write-flow-v2}
		\end{algorithmic}
	\end{figure*}
}

{
\begin{figure*}[t]
	\captionsetup{font=scriptsize}
	\begin{algorithmic}[1]
		\scriptsize
			\Function{OnNodeToNodeReadRequest}{$key, client-ts$}
			\If {$key$ is sored in the node}
			\State $<value, ts, clientSignature, clientID> \gets $ The newest associated timestamp and value with $key$
			\Else
			\State $clientSignature \gets EMPTY$
			\EndIf
			
			\State $nodeSignature \gets ComputeSignature(key || hash(value) || clientSignature || client-ts) $
			
			\If {isDigestQuery}
			\State \Return $<hash(value), ts, clientSignature, clientID, nodeSignature>$
			\Else
			\State \Return $<value, ts, clientSignature, clientID, nodeSignature>$
			\EndIf
			\EndFunction
		
		\\
		\Function{OnClientToNodeReadRequest}{$key, client-ts$}
		
		\State {$target Endpoints \gets all Relevant Nodes$ for $key$ or a subset of $2f+1$ fastest relevant nodes}
		\State {$dataEndpoind \gets$ One node from $target Endpoints$}
		\State Send read request for data to $ dataEndpoind $
		\State Send read request for digest to $target Endpoints \setminus \{dataEndpoind\} $
		\State Wait for $2f+1$ responses or timeout \blue {\Comment{Not verifying signatures}}
		\If {timeout AND all relevant nodes were targeted at the first phase}		
		\State \Return $\bot$
		\EndIf
		\If {got response from $dataEndpoind$ AND all responses agree on the digest}
		\State \Return $<value, nodesSignatures>$
		\EndIf
		\State Send read request for data from all nodes in $ all Relevant Nodes $ \Comment {N nodes} \label{lst:line:readFull}
		\State Wait for  $2f+1$ responses OR timeout \blue {\Comment{Not verifying signatures}}
		\If {timeout}
		\State \Return $\bot$
		\EndIf
		
		\State $resolvedValue \gets$ Latest response from $responses$ that is client-signature \textbf{verified}.
		\State Send write-back with $resolvedValue$ to $ all Relevant Nodes $ except those that are known to be updated
		\State Wait for responses till we have knowledge about $2f+1$  updated nodes OR timeout
		\State \blue {\Comment{Not verifying signatures}}
		%\State \Comment{Responded before with updated data or for the write back}
		\If {timeout}
		\State \Return $\bot$
		\EndIf
		
		\State \Return $<resolvedValue, nodesSignatures, originalValuesUsedForTheResolve>$
		\EndFunction
		\\
		\Function{ClientReadRequest}{$key$}
		\State $ client-ts \gets $ Current timestamp
		\State $ p \gets $ Some random system node
		
		\State Send read request with $<key, client-ts>$ to $p$ \label{lst:line:startWithNode0}
		\State Wait for responses OR timeout
		\If { $|valid NodesSignatues| \geq 2f+1$}
		\State \Return data
		\EndIf

		\blue{	\If {$|valid NodesSignatues| \geq f+1$}
			\State $ blackList \gets$ Nodes that returned bad signatures
			\State Send same read request to $p$ asking for full read from $2f+1$ nodes that are not in $blackList$ \label{lst:line:fetchMore}
			\State Wait for responses OR timeout
			\If {$|valid NodesSignatues| \geq 2f+1$}
			\State \Return data
			\EndIf
		} \blue{
		\State $blackList \gets blackList$ $\cup $ new nodes that returned bad signatures
		\If {$blackList$ size increased AND $retryNumber \le f $}
		\State goto line~\ref{lst:line:fetchMore} \Comment{Try again without the bad nodes}
		\EndIf		}	
	\blue{			
		\EndIf }

	\State $ p \gets $ Some random system node that was not used in this function invocation \Comment{Failed reading from $p$}
	\If {$p = \bot$ OR $contacted Nodes > f$ }
	\State \Return Failure
	\EndIf
	\State goto line~\ref{lst:line:startWithNode0}
	
	\EndFunction	
	\caption{A variant of our hardened read algorithm. In this variant, the proxy does not verify the answers and lets the client contact it again if it is unsatisfied. ClientReadRequest is invoked by the client for each read. OnClientToNodeReadRequest is invoked on the proxy node by the client. OnNodeToNodeReadRequest is invoked on a node that has the responsibility to store the value. Changes from figure~\ref{fig:read-flow} are marked in blue.}
	\label{fig:read-flow-v2}
\end{algorithmic}
\end{figure*}
}

{
\begin{figure*}[t]
	\captionsetup{font=scriptsize}
	\begin{algorithmic}[1]
			\tiny
			\Function{OnNodeToNodeReadRequest}{$key, client-ts$}
			\If {$key$ is sored in the node}
			\State $<value, ts, clientSignature, clientID> \gets $ The newest associated timestamp and value with $key$
			\Else
			\State $clientSignature \gets EMPTY$
			\EndIf
			
			\State $nodeSignature \gets ComputeSignature(key || hash(value) || clientSignature || client-ts) $
			
			\If {isDigestQuery}
			\State \Return $<hash(value), ts, clientSignature, clientID, nodeSignature>$
			\Else
			\State \Return $<value, ts, clientSignature, clientID, nodeSignature>$
			\EndIf
			\EndFunction
		
		\\
		\Function{OnClientToNodeReadRequest}{$key, client-ts$}
		
		\State {$target Endpoints \gets all Relevant Nodes$ for $key$ or a subset of $2f+1$ fastest relevant nodes}
		\State {$dataEndpoind \gets$ One node from $target Endpoints$}
		\State Send read request for data to $ dataEndpoind $
		\State Send read request for digest to $target Endpoints \setminus \{dataEndpoind\} $
		\State Wait for $2f+1$ responses or timeout
		\If {timeout AND all relevant nodes were targeted at the first phase}		
		\State \Return $\bot$
		\EndIf
		\If {got response from $dataEndpoind$ AND all responses agree on the digest}
		\State \Return $<value, nodesSignatures>$
		\EndIf
		\State Send read request for data from all nodes in $ all Relevant Nodes $  \label{lst:line:readFull1}
		\State Wait for $2f+1$ responses OR timeout
		\If {timeout}
		\State \Return $\bot$
		\EndIf
		\State \Return $2f+1$ data versions \blue{ \Comment{Resolving responsibility has moved to the client}}
		
		\EndFunction
		\\
		
		\Function{ClientReadRequest}{$key$}
		\State $ client-ts \gets $ Current timestamp
		\State $ p \gets $ Some random system node
			
		\State Send read request with $<key, client-ts>$ to $p$ \label{lst:line:startWithNode2}
		
		\State Wait for responses OR timeout
		\If {\blue{got one version} AND $|valid NodesSignatues| \geq 2f+1$}
		\State \Return data
		\EndIf
		
		\blue{
			\If {got $2f+1$ versions with valid nodes signatures}
			\State goto line~\ref{lst:line:resolve} \Comment{Resolve data}
			\EndIf
		}
		
		\If {$|valid NodesSignatues| \geq f+1$}
		\State $ blackList \gets$ Nodes that returned bad signatures
		\State Send same read request to $p$ asking for full read from $2f+1$ nodes that are not in $blackList$ \label{lst:line:fetchMore1}
		\State Wait for responses OR timeout
		\If {$|valid NodesSignatues| \geq 2f+1$}
		\State \Return data
		\EndIf
		
		\State {Add to $blackList$ new nodes that returned bad signatures}
		\If {$blackList$ size increased AND $retryNumber \le f $}
		\State goto line~\ref{lst:line:fetchMore1}
		\EndIf			
		\EndIf
		
		\blue{
			\If {got $2f+1$ versions with valid nodes signatures}
			\State $resolved \gets$ Latest response from $responses$ that is client-signature \textbf{verified}.  \label{lst:line:resolve}
			\State Send write-back with $resolved$ to all stalled nodes
			\State 		{\Comment{Using one of our write protocols, the client contacts one system node}}
			\If {write success}
			\State \Return resolved data
			\EndIf
			\EndIf
		}
		
		\State $ p \gets $ Some random system node that was not used in this function invocation \Comment{Failed reading from $p$}
		\If {$p = \bot$ OR $contacted Nodes > f$ }
		\State \Return Failure
		\EndIf
		\State goto line~\ref{lst:line:startWithNode2}
		
		\EndFunction
	   \caption{A variant of our hardened read algorithm. In this variant, the proxy does not verify the answers and the client is responsible to resolve conflicts. ClientReadRequest is invoked by the client for each read. OnClientToNodeReadRequest is invoked on the proxy node by the client. OnNodeToNodeReadRequest is invoked on a node that has the responsibility to store the value. Changes from Figure~\ref{fig:read-flow-v2} are marked in blue.}
        \label{fig:read-flow-v3}	
	\end{algorithmic}
\end{figure*}
}

\begin{figure}[t]
	\centering
	\includegraphics[width=0.90\columnwidth]{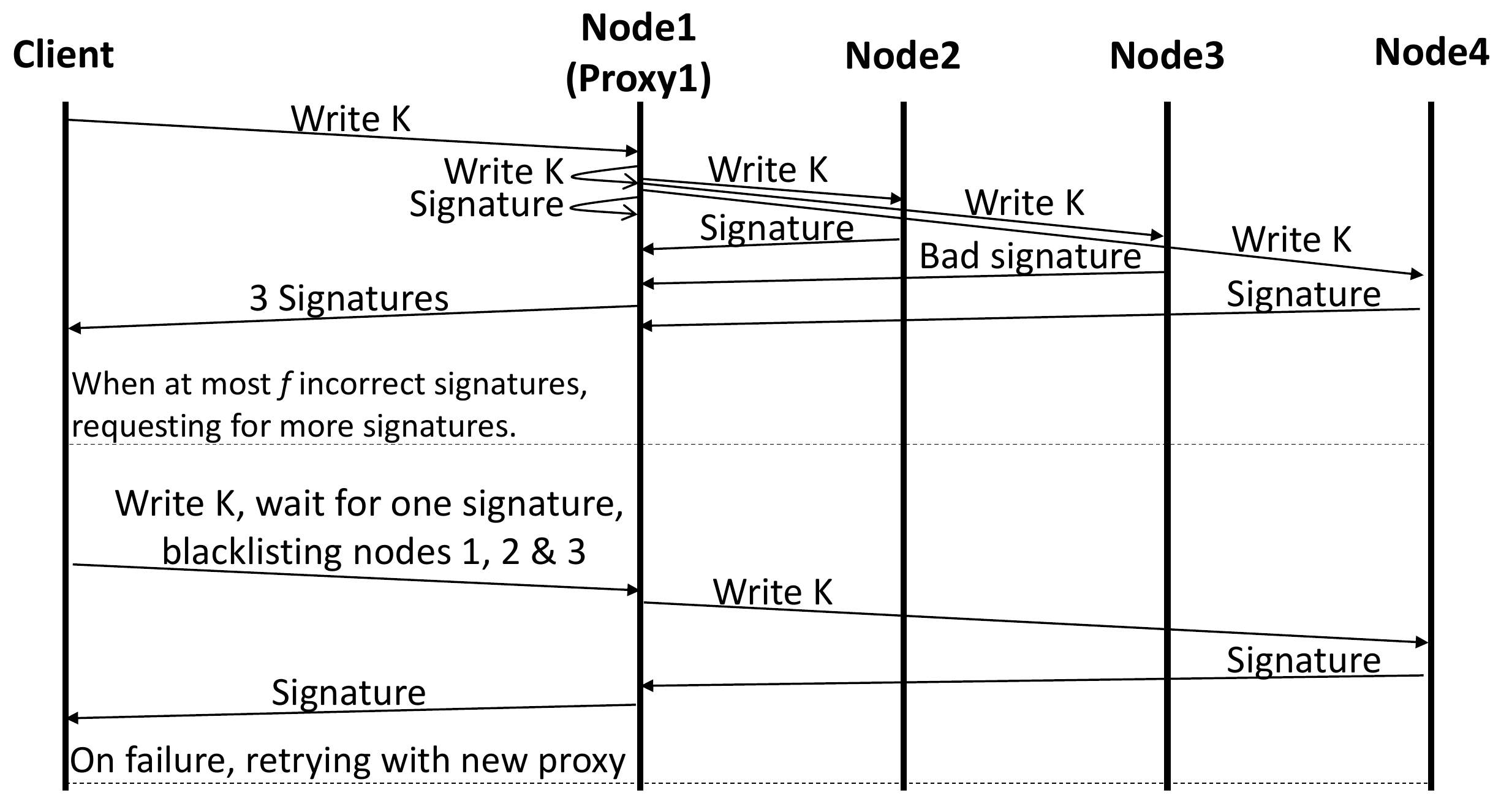}
	%\caption{Diagram for the algorithm from Figure~\ref{fig:clean-cassandra-write-flow}, this is the current write algorithm. Configuration:  N=4 and W=3.}
	\caption{Illustrating our write algorithm from Figure~\ref{fig:write-flow-v2} where the proxy does not verify the store acknowledgments. Configuration: N=4 and W=3.}
	\label{fig:write2}
\end{figure}

\begin{figure}[t]
	\centering
	\includegraphics[width=0.90\columnwidth]{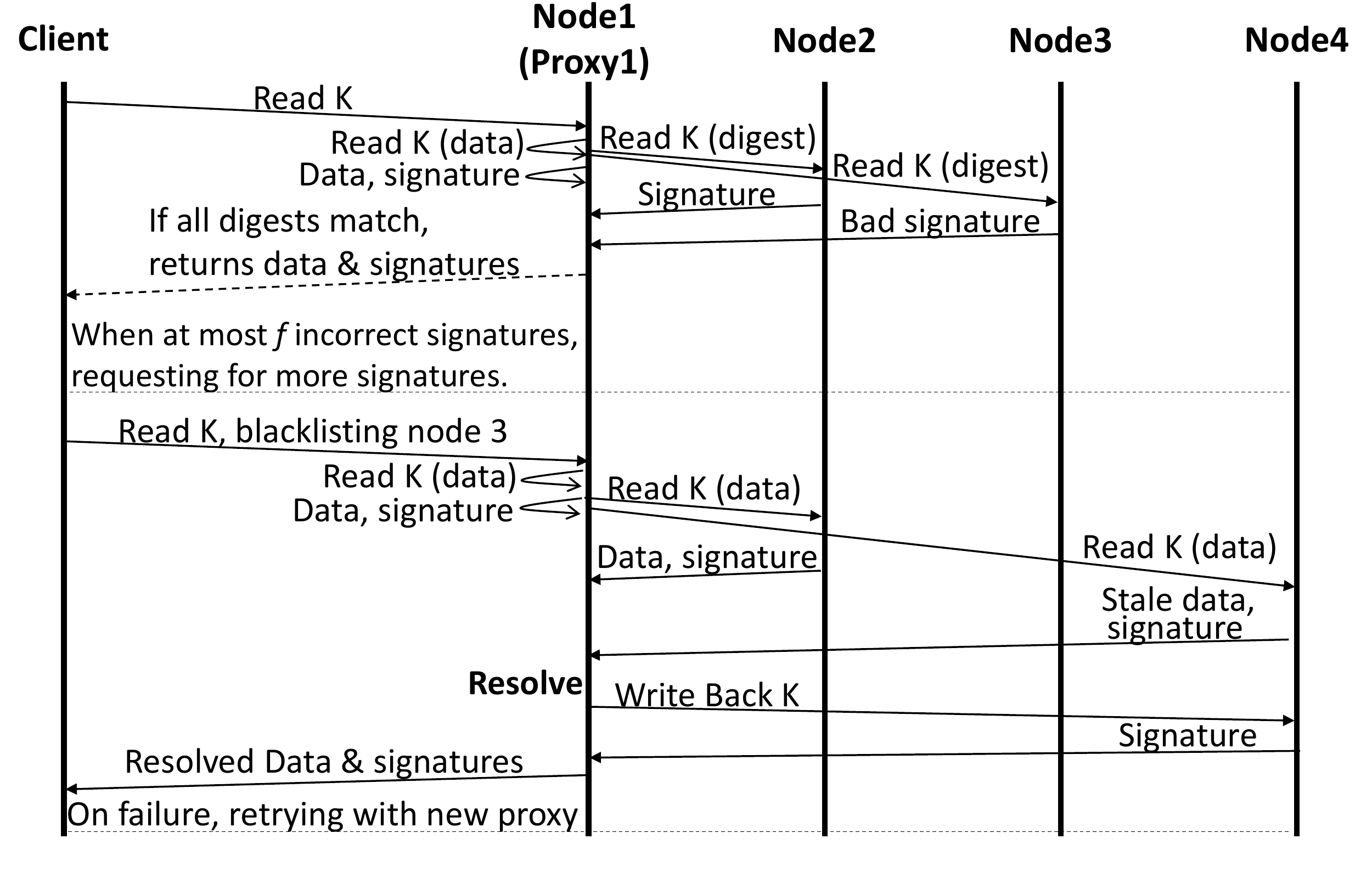}
	\caption{Illustrating our read algorithm from Figure~\ref{fig:read-flow-v2}~where the proxy does not verify the answers. Configuration: N=4 and R=3.}
	\label{fig:read2}
\end{figure}

\begin{figure}[t]
	\centering
	\includegraphics[width=\columnwidth]{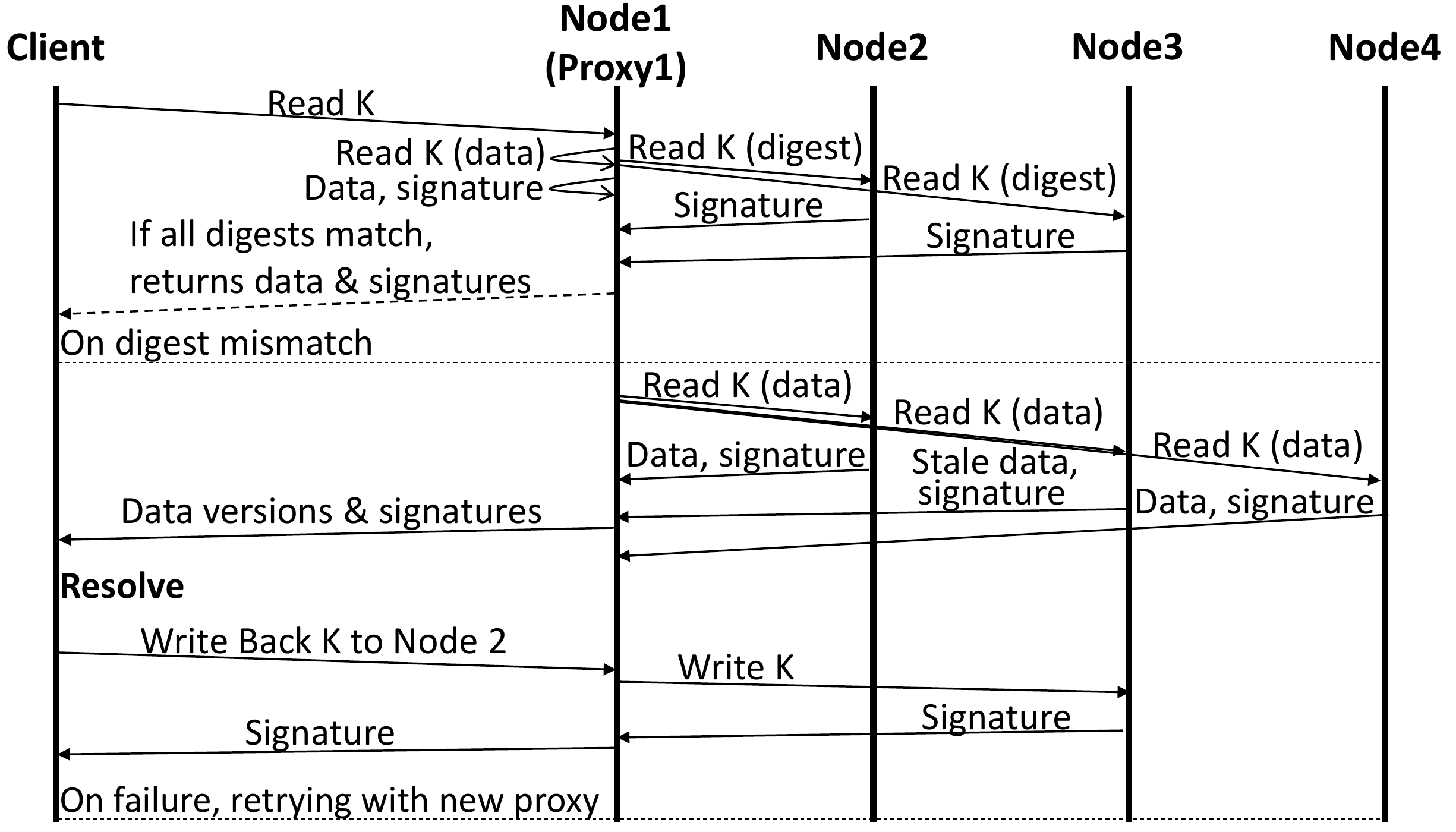}
	\caption{Illustrating our read algorithm from Figure~\ref{fig:read-flow-v3} where the proxy does not verify the answers and the client is responsible to resolve conflicts. Configuration: N=4 and R=3.}
	\label{fig:read3}
\end{figure} 

\end{document}